\documentclass[a4paper,9pt]{article}
\pdfoutput=1 % if your are submitting a pdflatex (i.e. if you have
\usepackage{jheppub} % for details on the use of the package, please
\usepackage[T1]{fontenc} % if needed

\usepackage{graphicx, epsfig}
\usepackage{amsmath,amssymb}

\usepackage{color}

\title{\boldmath Electroweak phase transition triggered by fermion sector}

\author[a,b,c]{Qing-Hong~Cao}
\author[b]{, Katsuya~Hashino}
\author[a]{, Xu-Xiang~Li}
\author[d,e]{, Zhe~Ren}
\author[d,e,b,f,g]{, Jiang-Hao~Yu}

\affiliation[a]{Department of Physics and State Key Laboratory of Nuclear Physics and Technology, Peking University, Beijing 100871, China}
\affiliation[b]{Center for High Energy Physics, Peking University, Beijing 100871, China}
\affiliation[c]{Collaborative Innovation Center of Quantum Matter, Beijing 100871, China}
\affiliation[d]{CAS Key Laboratory of Theoretical Physics, Institute of Theoretical Physics, Chinese Academy of Sciences, Beijing 100190, P. R. China}
\affiliation[e]{School of Physical Sciences, University of Chinese Academy of Sciences, Beijing 100049, P.R. China}
\affiliation[f]{School of Fundamental Physics and Mathematical Sciences, Hangzhou Institute for Advanced Study,  University of Chinese Academy of Sciences, Hangzhou 310024, China} 
\affiliation[g]{International Center for Theoretical Physics Asia-Pacific, Beijing/Hangzhou, China}

\emailAdd{qinghongcao@pku.edu.cn}
\emailAdd{hashino@pku.edu.cn}
\emailAdd{jhyu@itp.ac.cn}

\abstract{  
To realize first-order electroweak phase transition, it is necessary to generate a barrier in the thermal Higgs potential, which is usually triggered by scalar degree of freedom. 
We instead investigate phase transition patterns in pure fermion extensions of the standard model, and 
find that additional fermions with mass hierarchy and mixing could develop such barrier and realize strongly first-order phase transition in such models. 
In the Higgs potential with polynomial parametrization, the barrier can be generated in the following two patterns:  (I) positive quadratic term, negative cubic term and positive quartic term or (II) positive quadratic term, negative quartic term and positive higher dimensional term, such as dimensional 6 operator. 
}

\begin{document} 

\maketitle
\flushbottom

%%%%%%%%%%%%%%%%%%%%%%%%%%%%%%%%%%%%%%%%%%%%%%%%%%%%%%%%%%%%
%%%%%%%%%%%%%%%%%%%%%%%%%%%%%%%%%%%%%%%%%%%%%%%%%%%%%%%%%%%%
%%%%%%%%%%%%%%%%%%%%%%%%%%%%%%%%%%%%%%%%%%%%%%%%%%%%%%%%%%%%
%%%%%%%%%%%%%%%%%%%%%%%%%%%%%%%%%%%%%%%%%%%%%%%%%%%%%%%%%%%%
%%%%%%%%%%%%%%%%%%%%%%%%%%%%%%%%%%%%%%%%%%%%%%%%%%%%%%%%%%%%
%%%%%%%%%%%%%%%%%%%%%%%%%%%%%%%%%%%%%%%%%%%%%%%%%%%%%%%%%%%%
\section{Introduction}
%%%%%%%%%%%%%%%%%%%%%%%%%%%%%%%%%%%%%%%%%%%%%%%%%%%%%%%%%%%%
%%%%%%%%%%%%%%%%%%%%%%%%%%%%%%%%%%%%%%%%%%%%%%%%%%%%%%%%%%%%
%%%%%%%%%%%%%%%%%%%%%%%%%%%%%%%%%%%%%%%%%%%%%%%%%%%%%%%%%%%%
%%%%%%%%%%%%%%%%%%%%%%%%%%%%%%%%%%%%%%%%%%%%%%%%%%%%%%%%%%%%
%%%%%%%%%%%%%%%%%%%%%%%%%%%%%%%%%%%%%%%%%%%%%%%%%%%%%%%%%%%%
%%%%%%%%%%%%%%%%%%%%%%%%%%%%%%%%%%%%%%%%%%%%%%%%%%%%%%%%%%%%

Although the standard model~(SM) like Higgs boson has been discovered at the large hadron collider~\cite{Aad:2012tfa}, currently the shape of the Higgs potential is still undetermined~\cite{Agrawal:2019bpm}.
It is well known that dynamics of the electroweak phase transition~(EWPT), governed by the shape of the Higgs potential, is quite important to describe the early universe at the electroweak scale.
In particular, the first-order EWPT is very interesting, because it is one of the three necessary conditions in the electroweak baryogenesis scenario to explain baryon asymmetry of the universe~\cite{Kuzmin:1985mm}, and furthermore, shape of the potential could be tested by the gravitational waves which are produced by the collision of bubbles via the first-order EWPT~\cite{Kosowsky:1991ua}.

To realize first-order EWPT, it is necessary to develop a sizable potential barrier in the thermal Higgs potential. 
In the SM, although there are thermal loop corrections from the gauge bosons, the Higgs boson mass (125 GeV) is too large to realize the first-order phase transition. 
The thermal Higgs potential in the SM, parametrized by various polynomial terms, reads
%%%%%%%
	\begin{align}
	\label{SMfinite}
	V_{\rm eff}^{SM}(\varphi, T) \simeq D(T^2-T_0^2) \varphi^2 -E T \varphi^3 +\frac{\lambda_T}{4} \varphi^4,
	\end{align}
%%%%%%%
where $\varphi$ is the classical background in the SM Higgs field and $D$, $T_0$ and $E$ are parameters independent of temperature $T$ and $\lambda_T$  depends on $T$ through logarithmic term. 
The coefficients $DT_0^2$ and $\lambda_T$ have dominant contributions from the tree-level potential, while $E$ can only be generated from the loop effect of gauge bosons:
%%%%%%%
	\begin{align}
	\label{thermalbosons}
	E = \frac{1}{4\pi v^3} \left(2m_W^3+m_Z^3 \right)\simeq 10^{-2}.
	\end{align}
%%%%%%% 
This term $E$ has much smaller contribution on the potential than other terms at $T\sim\varphi$, and thus the cubic term could be ignored. 
In such a case, the potential only has quadratic and quartic term of $\varphi$:
%%%%%%%
	\begin{align}
	V_{\rm eff}^{SM}(\varphi, T) \simeq D(T^2-T_0^2) \varphi^2 +\frac{\lambda_T}{4} \varphi^4.
	\end{align}
%%%%%%%
Thus, the first-order EWPT cannot be realized in the SM~\cite{Dine:1992vs}.  
One needs to increase size of the cubic term in the potential to obtain first-order EWPT. 
Usually, the cubic effect can be enhanced by degrees of freedom of additional bosons, because the cubic term typically comes from thermal loop effects of bosons~\cite{Dolan:1973qd} and furthermore tree-level effects of scalar bosons~\cite{Pietroni:1992in}. 
Thus it motivates various bosonic extensions of the SM, such as extended scalar boson models, and extended gauge sectors.
On the contrary, it is usually believed that the fermion could not contribute to the cubic term. 
This could be understood from the one-loop finite temperature bosonic and fermionic corrections on the Higgs potential in the high temperature limit:
\begin{align}
	\Delta V_T \simeq \begin{cases} -\frac{\pi^2}{90} T^4 + \frac{M^2(\varphi)}{24} T^2  -  \frac{M^3(\varphi)}{12\pi} T -  \frac{M^4(\varphi)}{64}\ln\left(\frac{M^2(\varphi)}{\alpha_B T^2}\right)+\dots, & \,\,\textrm{boson}\\
	\frac{7\pi^2}{720} T^4 - \frac{M^2(\varphi)}{48} T^2 -  \frac{M^4(\varphi)}{32}\ln\left(\frac{M^2(\varphi)}{\alpha_F T^2}\right)+\dots,& \,\,\textrm{fermion}
	\end{cases}
\end{align}	
where the cubic term $M^3(\varphi)$ does not appear in the fermionic thermal potential.

In this work, we consider phase transition patterns in fermionic extensions of the SM, and investigate whether the first-order phase transition could be triggered by pure fermion effects. 
In this case, although the cubic term cannot be enhanced, it is still possible to realize first-order EWPT by decreasing the quadratic and quartic contributions or adding higher dimensional terms in the Higgs potential.
If there is only one additional fermion added to the SM, it is very hard to satisfy such conditions in the Higgs potential. 
However, if there are several additional fermions contributed to the Higgs potential, the situation might be different.
This motivates us to consider extended models with several fermions included, and the simplest one is adding two fermions in the SM.
Although phase transition with several fermions has been considered in Refs.~\cite{Carena:2004ha, Davoudiasl:2012tu, Angelescu:2018dkk, Matsedonskyi:2020mlz},
we would like to explore various mass hierarchy regions and obtain the most generic phase transition patterns by utilizing the multi-scale effective Higgs potential in the effective field theory framework. 
In particular, we take an effective field theory approach to study the effective Higgs potential with matching and running procedure systematically with avoiding artificial large logarithm in the potential.
With the complete loop and running effects encoded in the Higgs potential, we find not only the scenario considered in literature~\cite{Davoudiasl:2012tu, Angelescu:2018dkk} but also new scenario could realize first-order EWPT.

To be specific, we consider a simple extended fermion model with two new fermions, such as vector like leptons, one SU(2) doublet fermion and one neutral singlet fermion, and systematically discuss various possibilities of how to generate a sizable barrier in the potential.
The mass hierarchy patterns of two new fermions can be divided into three categories as following: 
 
\paragraph{(A) both fermions are at the EW scale}
In this mass region only the EW scale physics is involved and there is no need to perform integrating-out and matching procedure. 
Therefore, to obtain a sizable barrier, it is necessary to decrease the size of quadratic and quartic terms in the potential Eq.~(\ref{SMfinite}) and thus make the quadratic and quartic terms comparable to the small cubic term due to the SM gauge boson mentioned above. 
However, it is impossible to realize such small quadratic term because new fermion contribution to the quadratic term has the same sign with the SM contribution and cancellation cannot happen. 
Therefore in this case the quadratic term $\varphi^2$ cannot be decreased to the same size as the cubic term to generate a sizable barrier and thus additional degree of freedom of additional bosons or more complex fermion sector is required~\cite{Carena:2004ha}.

\paragraph{(B) both fermions are at the TeV scale}
In order to analyze the potential at the EW scale, heavy fields with the TeV scale mass should be integrated out, otherwise large logarithmic terms could be developed in the potential and thus the perturbation expansion might become invalid. 
Ref.~\cite{Angelescu:2018dkk} considers a similar model with heavy fermions at TeV scale, however, they do not take into account such a treatment and thus the barrier might be appeared due to artificially large logarithm.  
In this work, we utilize a systematical method to get the effective potential with matching and running procedure~\cite{Davoudiasl:2012tu}.
In this case, integrating out heavy particles only contributes to higher dimensional operators and cannot give rise to large negative quartic coupling in the potential. 
Therefore although positive higher dimensional contributions are obtained by integrating out new fermions, there is no enough negative quartic contribution to develop a barrier. 

\paragraph{(C) one fermion is at the TeV scale and another is at the EW scale}
Similar to the case (B), one integrates out the TeV scale heavy fermion and obtain the higher dimension operators. 
However, different from the case (B), the EW scale new fermion cannot be integrated out, and thus the mixing effects between heavy and light fermions have new contribution on the potential at the EW scale. 
Typically the larger mixing effect causes the more negative quartic coupling in the Higgs potential. 
In this case there are both high dimensional~operators from TeV scale fermion contribution and thermal effects from EW scale fermion contribution~\cite{Davoudiasl:2012tu,Matsedonskyi:2020mlz}, which add up to generate positive quadratic, negative quartic and high dimensional operator in the potential. 
It could develop a sizable barrier and a first-order EWPT is realized.

On the other hand, when the mixing effect is not so large, a new scenario appears to realize first-order phase transition. 
In this case, the quartic coupling is small but still positive, and mixing effects would not cause negative quartic coupling, and thus the quartic term is comparable to the cubic term in the SM. 
At the same time, the quadratic term receives a positive contribution from integrating out heavy fermion, which is opposite to the SM tree-level contribution. 
Thus, there are cancellation between the positive new contribution and the SM one, and thus the quadratic term is decreased to smaller size which is comparable to the cubic term in the SM. 
In total, the effective potential with comparable positive quadratic, negative cubic and positive quartic terms could realize first-order EWPT.

Among the above cases, only the case (C) could realize first-order EWPT, and as discussed above a sizable barrier could be developed under the following two scenarios:
\begin{itemize}
	\item 
  (I)  First scenario: positive $\varphi^2$ term, negative $\varphi^3$ term and positive $\varphi^4$ term
%%%%%%%
	\begin{align}
	\label{firstpotform}
	V_{\rm eff}(\varphi, T)\simeq  \frac{1}{2}\mu^2 \varphi^2 -\frac{1}{3}\lambda_{3} \varphi^3 +\frac{1}{4}\lambda \varphi^4\quad (\mu^2,\lambda_{3}, \lambda>0 ),  
	\end{align}
%%%%%%%
where negative cubic term only comes from the SM as shown Eq.~(\ref{thermalbosons}). 
 The potential of this scenario looks like one of the SM in Eq.~(\ref{SMfinite}), however, the fermion model has additional reductions in $\varphi^2$ and $\varphi^4$ terms through new fermion effects. 
Then a barrier can be generated in the effective potential, because the quadratic and quartic terms can be almost same size as cubic term.  
\item
(II) Second scenario: positive $\varphi^2$ term, negative $\varphi^4$ term and positive high dimensional~term of heavy fermion effect
 %%%%%%%%%%%%%%%%%%%%%%%%%%%%%%%%%%%%%%%%%%%%%%%%%%%%%%%%%%%%
	\begin{align}
	\label{secondproc}
	V_{\rm eff}(\phi, T) \simeq \frac{1}{2}\mu^2\varphi^2-\frac{1}{4}\lambda\varphi^4+\frac{1}{6}\gamma\varphi^6\quad (\mu^2,\lambda, \gamma>0 ).
	\end{align}
%%%%%%%%%%%%%%%%%%%%%%%%%%%%%%%%%%%%%%%%%%%%%%%%%%%%%%%%%%%%
This scenario has larger fermion effects than scenario (I), and then the quartic term becomes negative. 
 Thus a large barrier can be generated in the potential without cubic term in scenario (II). 
 \end{itemize}

In the following, we investigate the multi-scale effective potential using the matching and running procedure in section 2.
 The detail of extended fermion model is shown in section 3 and we check the temperature dependence of coefficients of $\varphi^n$ in order to discuss the detail of generation of a barrier in the model.
 In section 4, we explore the parameter region where a barrier appears in the potential.
The conclusion is drawn in section 5.

%%%%%%%%%%%%%%%%%%%%%%%%%%%%%%%%%%%%%%%%%%%%%%%%%%%%%%%%%%%%
%%%%%%%%%%%%%%%%%%%%%%%%%%%%%%%%%%%%%%%%%%%%%%%%%%%%%%%%%%%%
%%%%%%%%%%%%%%%%%%%%%%%%%%%%%%%%%%%%%%%%%%%%%%%%%%%%%%%%%%%%
%%%%%%%%%%%%%%%%%%%%%%%%%%%%%%%%%%%%%%%%%%%%%%%%%%%%%%%%%%%%
%%%%%%%%%%%%%%%%%%%%%%%%%%%%%%%%%%%%%%%%%%%%%%%%%%%%%%%%%%%%
%%%%%%%%%%%%%%%%%%%%%%%%%%%%%%%%%%%%%%%%%%%%%%%%%%%%%%%%%%%%
\section{Multi-scale effective potential}
%%%%%%%%%%%%%%%%%%%%%%%%%%%%%%%%%%%%%%%%%%%%%%%%%%%%%%%%%%%%
%%%%%%%%%%%%%%%%%%%%%%%%%%%%%%%%%%%%%%%%%%%%%%%%%%%%%%%%%%%%
%%%%%%%%%%%%%%%%%%%%%%%%%%%%%%%%%%%%%%%%%%%%%%%%%%%%%%%%%%%%
%%%%%%%%%%%%%%%%%%%%%%%%%%%%%%%%%%%%%%%%%%%%%%%%%%%%%%%%%%%%
%%%%%%%%%%%%%%%%%%%%%%%%%%%%%%%%%%%%%%%%%%%%%%%%%%%%%%%%%%%%
%%%%%%%%%%%%%%%%%%%%%%%%%%%%%%%%%%%%%%%%%%%%%%%%%%%%%%%%%%%%

Here we consider a general ultraviolet theory containing several fields: the SM particles, new particles at the EW scale and new heavy particles at the TeV scale.
According to Coleman-Weinberg's treatment~\cite{Coleman:1973jx}, the one-loop effective potential develops the logarithmic term $\log\frac{m^2}{Q^2}$ where $m^2$ is the mass-squared running over all the relevant particles, and $Q$ is the renormalization scale taken to be the EW scale in the effective potential.
Since there are new heavy particles, the effective potential would contain at least the large logarithmic term $\log\frac{M_{\rm heavy}^2}{Q^2}$, in which $M_{\rm heavy}$ is the mass of heavy particles and $Q$ stays at the EW scale.
According to the loop expansion, the $n$-th order loop contribution contains the logarithmic term $\left(\lambda \times \log\frac{M_{\rm heavy}^2}{Q^2}\right)^N$, so if the mass hierarchy in logarithmic term between $M_{\rm heavy}$ and $Q$ is large, the perturbation will break down.

\begin{figure}[tb]
  \begin{center}
\includegraphics[width=0.9\textwidth]{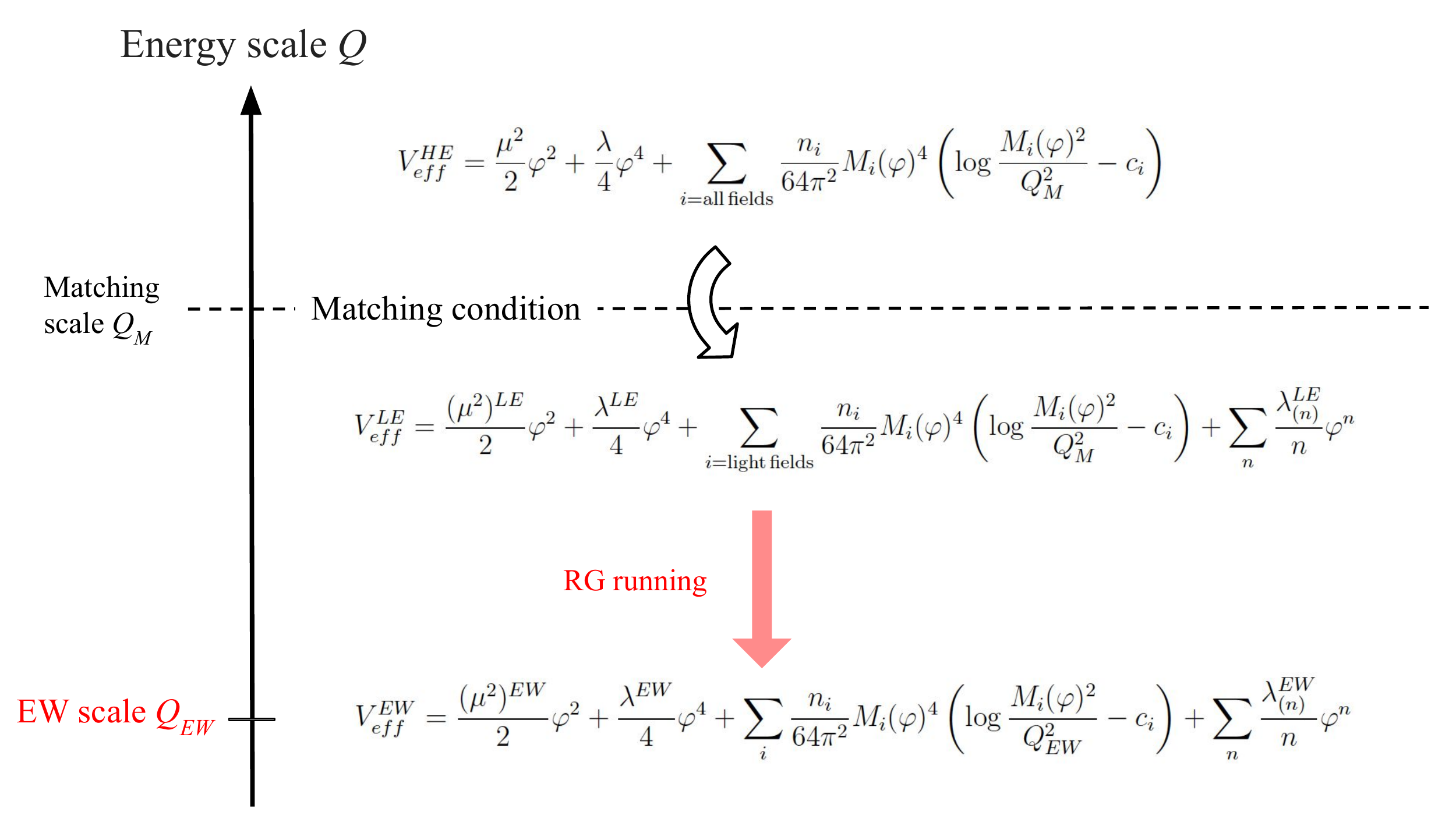}
\caption{The procedure of obtaining the effective potential at the EW scale.
The effective potential contains the classical background $\varphi$ of the SM Higgs field and the divergence in it is absorbed by the $\overline{\rm MS}$ scheme.
At the HE scale, the effective potential $V_{eff}^{HE}$, which is calculated at the matching scale $Q_M$, contains all fields.
At the LE scale, the effective potential $V_{eff}^{LE}$ containing light fields and high dimensional~operators of heavy fields can be obtained by the matching condition Eq.~(\ref{GEM}).
After performing the RG running from $Q_M$ to $Q_{EW}$, we obtain the effective potential at the EW scale $V_{eff}^{EW}$. }
\label{Flow}
  \end{center}
\end{figure}

To solve this problem, several kinds of methods, such as the renormalization group (RG) improved effective potential~\cite{Coleman:1973jx}, the integrate out and matching method~\cite{Weinberg:1980wa, Masina:2015ixa}, are used to get the effective potential at the EW scale. In this work, we utilize the matching method in Ref.~\cite{Masina:2015ixa} and generalize the formalism to cases with several new particles. The following procedure is used to obtain the effective potential at the EW scale: 
 \begin{enumerate}
 \item Define the ultraviolet (UV) Lagrangian containing all fields,
  \item Calculate the one-loop effective potential at high energy (HE) scale,
  \item Absorb the divergence in the calculation by the renormalization and tadpole scheme,
  \item  Match the potential at the HE scale with the one containing light fields and high dimensional~operators at low energy (LE) scale at matching scale $Q_M$ by the matching condition,
  \item Evaluate RG runnings of parameters in the effective potential from the matching scale $Q_M$ to the EW scale $Q_{EW}$.
    \item Obtain the effective potential at the EW scale.
 \end{enumerate}
 This procedure on calculation of the effective potential at the EW scale is summarized in Fig.~\ref{Flow}. 
%%%%%%%%%%%%%%%%%%%%%%%%%%%%%%%%%%%%
%%%%%%%%%%%%%%%%%%%%%%%%%%%%%%%%%%%%
%%%%%%%%%%%%%%%%%%%%%%%%%%%%%%%%%%%%
%%%%%%%%%%%%%%%%%%%%%%%%%%%%%%%%%%%%
The general UV Lagrangian containing massive bosonic and fermionic degree of freedom at the HE scale could be written as
%%%%%%%%%%%%%%%%%%%%%%%%%%%%%%%%%%%%
	\begin{align}
	{\cal L} &=  \frac{1}{2}\left|D_\mu\phi_{(i)}\right|^2 +  \frac{1}{2}\left|D_\mu\Phi_{(i)}\right|^2 + i \bar{f}_{(i)}\hspace{-0.15cm}\not\hspace{-0.15cm}D f_{(i)}+ i \bar{F}_{(i)}\hspace{-0.15cm}\not\hspace{-0.15cm}D F_{(i)} \nonumber\\
	& - \lambda^{(a,b,c,d), \{e,f,g,h\}}_{(i,j,k,l), \{m,n,o,p\}} \phi_{(i)}^m \phi_{(j)}^n \phi_{(k)}^o \phi_{(l)}^p  \Phi_{(a)}^e \Phi_{(b)}^f \Phi_{(c)}^g \Phi_{(d)}^h \nonumber\\
	&  - Y^{(a,b,c), \{g,h, i\}}_{(i,j,k), \{m,n,o\}} \phi_{(i)}^m \Phi_{(a)}^g \bar{f}_{(j)}^n f_{(k)}^o \bar{F}_{(b)}^h F_{(c)}^i ,
	\end{align}
%%%%%%%%%%%%%%%%%%%%%%%%%%%%%%%%%%%%
where $\phi^m_{(i)}$ ($f_{(i)}$) and $\Phi_{(i)}$ ($F_{(i)}$) are light and heavy boson (fermion) fields, superscript like $m$ is the power of field and subscript like $(i)$ is the kind of field. 
For simplicity, kinetic mixing term is not included in the Lagrangian and we assume only the SM-like Higgs boson has the vacuum expectation value.

First let us consider the HE effective potential at the TeV scale.
The effective potential can be obtained by the functional trace of action regarding the Lagrangian, such as $\log \det\left(\frac{\delta^2 {\cal L}}{\delta \phi_{(i)}\delta \phi_{(j)} }\right)$~\cite{Jackiw:1974cv}.
%%%%%%%%%%%%%%%%%%%%%%%%%%%%%%%%%%%%%%%%%%%%%%%%%%%%%%%%%%%%
\begin{equation}
V^{HE} =\frac{\mu^2}{2}\varphi^2 + \frac{\lambda}{4}\varphi^4  +  \sum_{i={\rm all\,fields}}\frac{n_i}{64\pi^2} M_i(\varphi)^4\left(\log\frac{M_i(\varphi)^2 }{Q^2}-c_i\right) +  \frac{\delta\mu^2}{2}\varphi^2+ \frac{\delta\lambda}{4}\varphi^4 +V_{div}^{HE}  , 
\end{equation}
%%%%%%%%%%%%%%%%%%%%%%%%%%%%%%%%%%%%%%%%%%%%%%%%%%%%%%%%%%%%
where $n_i$ is degree of freedom of the $i$-th field, $M_i(\varphi)^2$ is its field dependent mass, $c_{i}=\frac{3}{2}$ (for scalar bosons and fermions), $\frac{5}{6}$ (for gauge bosons), $\delta\mu^2$ and $\delta\lambda$ are respectively the counter terms of the quadratic and quartic $\varphi$ terms and 
 %%%%%%%%%%%%%%%%%%%%%%%%%%%%%%%%%%%%
\begin{align}
 V_{div}^{HE} = \frac{Q^{2\epsilon}}{\bar{\epsilon}} \sum_{i}\frac{n_i}{64\pi^2} (M_i(\varphi)^2)^{2-\epsilon},
\end{align}
%%%%%%%%%%%%%%%%%%%%%%%%%%%%%%%%%%%%
where $\frac{1}{\bar{\epsilon}}=\frac{1}{\epsilon}-\gamma +\log 4\pi$.
Then the $\overline{\rm MS}$ scheme is utilized to absorb the UV divergence as follows 
 %%%%%%%%%%%%%%%%%%%%%%%%%%%%%%%%%%%%
\begin{align}
\left. \frac{\partial^2 }{\partial \varphi^{2}} \left( V_{div}^{HE} + \frac{\delta\mu^2}{2}\varphi^2+ \frac{\delta\lambda}{4}\varphi^4\right) \right|_{\varphi=0} &= 0,\quad \left. \frac{\partial^4 }{\partial \varphi^{4}} \left( V_{div}^{HE} + \frac{\delta\mu^2}{2}\varphi^2+ \frac{\delta\lambda}{4}\varphi^4 \right) \right|_{\varphi=0} &= 0,
\end{align}
%%%%%%%%%%%%%%%%%%%%%%%%%%%%%%%%%%%%
The effective potential under the $\overline{\rm MS}$ scheme is
%%%%%%%%%%%%%%%%%%%%%%%%%%%%%%%%%%%%
\begin{align}
V_{eff}^{\overline{\rm MS}} &= \frac{\mu^2}{2}\varphi^2 + \frac{\lambda}{4}\varphi^4  + \sum_{i}\frac{n_i}{64\pi^2} M_i(\varphi)^4\left(\log\frac{M_i(\varphi)^2 }{Q^2}-c_i\right).
\end{align}
%%%%%%%%%%%%%%%%%%%%%%%%%%%%%%%%%%%%

In order to get the HE effective potential, one also need to fix the tadpole schemes~\cite{Fleischer:1980ub}. 
  In the tadpole schemes, the shift of scalar field with $v_0$ ($\varphi\to\varphi + v_0$) shows up, where the subscript represents the bare parameter. 
  This shift evolves a linear term in the effective potential: $-(\mu_0^2+\lambda_0 v_0^2)v_0\varphi$, and the form of effective potential containing tree-level and counter term is
    %%%%%%%%%%%%%%%%%%%%%%%%%%%%%%%%%%%%
  \begin{align}
    V_{eff, \,0, c}^{Tad} &= v (\mu^2+\lambda v^2) \varphi + \frac{\mu^2+3\lambda v^2}{2}\varphi^2 + \lambda v\varphi^3 + \frac{\lambda}{4}\varphi^4 + \left(v\delta\mu^2+\delta\lambda v^3 + \delta v \left(\mu^2+3\lambda v^2\right) \right) \varphi  \nonumber\\
    &\quad +   \frac{\delta\mu^2 + 3v \left(\delta\lambda v+2\lambda \delta v\right) }{2}\varphi^2 +  \left(\delta\lambda v+ \lambda \delta v\right)\varphi^3+ \frac{\delta\lambda}{4}\varphi^4 + O(\lambda^3).
   \end{align}
    %%%%%%%%%%%%%%%%%%%%%%%%%%%%%%%%%%%%
  The tadpole condition for fixing $v$ and $\delta v$ is
    %%%%%%%%%%%%%%%%%%%%%%%%%%%%%%%%%%%%
  \begin{align}
    \langle 0 | \varphi | 0 \rangle &= 0 \Rightarrow -i\delta t + i \Gamma_\varphi^{(1)} = 0,
  \end{align}
    %%%%%%%%%%%%%%%%%%%%%%%%%%%%%%%%%%%%
  where $\delta t$ appears in the effective potential as the linear term of $\varphi$, $V_{eff,0}^{Tad} \supset \delta t \varphi$, and $\Gamma_\varphi$ is the 1-point function of $\varphi$. 
  The tadpole conditions at the tree-level and the one-loop level are
    %%%%%%%%%%%%%%%%%%%%%%%%%%%%%%%%%%%%
  \begin{align}
    v=\sqrt{-\frac{\mu^2}{\lambda}},\quad-i(v\delta\mu^2+\delta\lambda v^3 + \delta v \left(\mu^2+3\lambda v^2\right)) +i \Gamma_\varphi^{(1),1-loop}=0.
  \end{align}
     %%%%%%%%%%%%%%%%%%%%%%%%%%%%%%%%%%%%
 Other counter terms $\delta\mu^2$, $\delta\lambda$ are fixed by 2-point and 4-point function of $\varphi$.
  If the divergence parts are absorbed by the counter terms, such as the $\overline{\rm MS}$ scheme, the renormalization conditions are 
  %%%%%%%%%%%%%%%%%%%%%%%%%%%%%%%%%%%%
  \begin{align}
  -i\left(\delta\mu^2 + 3v \left(\delta\lambda v+2\lambda \delta v^{div}\right)\right)+ i \Gamma_\varphi^{(2), div}=0, \quad -i 6\delta\lambda  + i \Gamma_\varphi^{(4), div}=0.
  \end{align}
    %%%%%%%%%%%%%%%%%%%%%%%%%%%%%%%%%%%%
  This tadpole scheme is called Fleischer-Jegerlehner tadpole scheme (FJTS).
  Note that $\delta v$ contains finite term while $\delta\mu^2$ and $\delta\lambda$ have terms with $1/\overline{\epsilon}$. Then the HE effective potential is
    %%%%%%%%%%%%%%%%%%%%%%%%%%%%%%%%%%%%
  \begin{align}
    V_{eff}^{FJTS} &= \frac{\mu^2}{2}\varphi^2 + \frac{\lambda}{4}\varphi^4 + \sum_i \frac{n_i}{64\pi^2}M_i^4(\varphi) \left(\log\frac{M_i^2(\varphi)}{Q^2}-c_i\right)\nonumber\\
    & + \left[\frac{\delta\mu^2}{2}\varphi^2 + \frac{\delta\lambda}{4}\varphi^4 + \sum_i \frac{n_i}{64\pi^2} \frac{M_i^4(\varphi)}{\bar{\epsilon}}\right].
	\label{eq:VHEfjts}
  \end{align} 
   For spontaneously symmetry breaking $\lambda\phi^4$ theory
  \begin{align}
    \mathcal{L} = \frac{1}{2}\left(\partial_\mu \varphi\right)^2 - \frac{1}{2}\mu^2\varphi^2 - \frac{1}{4}\lambda \varphi^4, \quad \mu^2 <0, ~ \lambda >0
  \end{align}
    %%%%%%%%%%%%%%%%%%%%%%%%%%%%%%%%%%%%
  all divergences in the effective potential cancel with each other and $V_{eff}^{FJTS} = V_{eff}^{\overline{\rm MS}}$.
In the following analysis, we use the effective potential with the $\overline{\rm MS}$ scheme.

After obtaining the HE effective potential, we start to consider the matching between the HE and LE effective potentials. 
The HE effective potential is given in Eq.~(\ref{eq:VHEfjts}), while the LE effective potential at the same UV scale can be written similarly with less field degree of freedom.
Therefore, the HE and LE effective potential at the matching scale $Q_{M}$ are
%%%%%%%%%%%%%%%%%%%%%%%%%%%%%%%%%%%%
\begin{align}
\label{HEP}
V_{eff}^{HE} &= \frac{\mu^2}{2}\varphi^2 + \frac{\lambda}{4}\varphi^4  + \sum_{i={\rm all\,fields}}\frac{n_i}{64\pi^2} M_i(\varphi)^4\left(\log\frac{M_i(\varphi)^2 }{Q_{M}^2}-c_i\right) ,\\
\label{LEP}
V_{eff}^{LE} &= \frac{(\mu^2)^{LE}}{2}\varphi^2 + \frac{\lambda^{LE}}{4}\varphi^4 + \sum_{i={\rm light\,fields}}\frac{n_i}{64\pi^2} M_i(\varphi)^4\left(\log\frac{M_i(\varphi)^2 }{Q_{M}^2}-c_i\right) + \sum_n \frac{\lambda^{LE}_{(n)}}{n} \varphi^n.
\end{align}
%%%%%%%%%%%%%%%%%%%%%%%%%%%%%%%%%%%%
  The coefficient of $\varphi^n$ due to heavy fields, $\lambda_{(n)}$, is fixed by the matching conditions at the matching scale $Q_{M}$:
 %%%%%%%%%%%%%%%%%%%%%%%%%%%%%%%%%%%%
\begin{align}
\label{GEM}
(\mu^2)^{LE}=\mu^2 + \left. \frac{\partial^2 }{\partial \varphi^{2}} \sum_{i={\rm heavy\,fields}}\frac{n_i}{64\pi^2} M_i(\varphi)^4\left(\log\frac{M_i(\varphi)^2 }{Q_{M}^2}-c_i\right)\right|_{\varphi=0}  , \nonumber\\
\lambda^{LE}= \lambda + \frac{1}{6} \left. \frac{\partial^4 }{\partial \varphi^{4}}  \sum_{i={\rm heavy\,fields}}\frac{n_i}{64\pi^2} M_i(\varphi)^4\left(\log\frac{M_i(\varphi)^2 }{Q_{M}^2}-c_i\right) \right|_{\varphi=0} , \nonumber\\
\lambda^{LE}_{(n)} =  \frac{1}{(n-1)!} \left.  \frac{\partial^n }{\partial \varphi^{n}}   \sum_{i={\rm heavy\,fields}}\frac{n_i}{64\pi^2} M_i(\varphi)^4\left(\log\frac{M_i(\varphi)^2 }{Q_{M}^2}-c_i\right) \right|_{\varphi=0}.
\end{align}
%%%%%%%%%%%%%%%%%%%%%%%%%%%%%%%%%%%%
 Differential terms represent the threshold effects of heavy fields, because such a loop effect in the effective potential corresponds to the 1PI Feynman diagrams with $n$ external lines of $\varphi$.

After obtaining the LE effective potential at the matching scale $Q_M$, we could obtain the effective potential at the EW scale $Q_{EW}$ via the RG running.
 The beta function can be obtained through the RG equation of the effective potential at the LE scale:
 %%%%%%%%%%%%%%%%%%%%%%%%%%%%%%%%%%%%
\begin{align}
\label{CSE}
\sum_a\left( \beta_a\frac{\partial}{\partial\lambda_a} - \gamma_\varphi \varphi \frac{\partial}{\partial\varphi} - \frac{\partial}{\partial Q} \right) V_{eff}^{LE} = 0.
\end{align}
%%%%%%%%%%%%%%%%%%%%%%%%%%%%%%%%%%%%
By using the beta functions, the effective potential at the EW scale is given by
%%%%%%%%%%%%%%%%%%%%%%%%%%%%%%%%%%%%
\begin{align}
\label{LEscalepotential}
V^{EW}_{eff} &=\frac{(\mu^2)^{EW}}{2}\varphi^2 + \frac{\lambda^{EW}}{4}\varphi^4  + \sum_{i}\frac{n_i}{64\pi^2} M_i(\varphi)^4\left(\log\frac{M_i(\varphi)^2 }{Q_{EW}^2}-c_i\right) + \sum_n \frac{\lambda^{EW}_{(n)}}{n} \varphi^n , 
\end{align}
%%%%%%%%%%%%%%%%%%%%%%%%%%%%%%%%%%%%
where $(\mu^2)^{EW}$, $\lambda^{EW}$ and $\lambda^{EW}_{(n)}$ terms are the $LE$ terms containing the running effects.

The above effective potential at the EW scale only contains the zero temperature effects. 
The finite temperature effects $\Delta V_T$ and $V_T^{\rm ring}$ should be added to the effective potential, in order to describe the phase transition.
 The 1-loop finite temperature effect $\Delta V_T$ can be written as
%%%%%%%%%%%%%%
	\begin{align} % requires amsmath; align* for no eq. number
	\label{LEscalepotentialT}
	 \Delta V_T&= \frac{T^4}{2\pi^2} 
  \Biggl\{ \sum_{i={\rm bosons}} 
  n_i  \int_0^\infty d x x^2\ln 
  \left[ 1- \exp \left( -\sqrt{x^2+(M_i(\varphi)/T)^2}\right) \right]\nonumber\\
 & + \sum_{i = {\rm fermions}} n_i  \int_0^\infty d x x^2\ln 
  \left[ 1+ \exp \left( -\sqrt{x^2+(M_i(\varphi)/T)^2}\right) \right] \Biggl\}.
	\end{align}
%%%%%%%%%%%%%%
 In high temperature limit, the finite temperature effects are given as following forms:
%%%%%%%%%%%%%%
	\begin{align} % requires amsmath; align* for no eq. number
 &\int_0^\infty d x x^2\ln \left[ 1- \exp \left( -\sqrt{x^2+a^2}\right) \right] \to -\frac{\pi^4}{45} + \frac{\pi^2}{12} a^2 -  \frac{\pi}{6} (a^2)^{3/2} -  \frac{a^4}{32}\ln\left(\frac{a^2}{\alpha_B}\right)+\dots,\nonumber\\
 & \int_0^\infty d x x^2\ln \left[ 1+ \exp \left( -\sqrt{x^2+a^2}\right) \right] \Bigl\} \to \frac{7\pi^4}{360} - \frac{\pi^2}{24} a^2 -  \frac{a^4}{32}\ln\left(\frac{a^2}{\alpha_F}\right)+\dots, \nonumber
	\end{align}
%%%%%%%%%%%%%%
 where $a = M(\varphi)/T$, $\log \alpha_B=2\log4\pi -2\gamma_E+3/2$, $\log\alpha_F=2\log\pi -2\gamma_E+3/2$ and $\gamma_E$ is Euler constant. 
One could also include the daisy ring effect with the resummation, $V_T^{\rm ring}$, which is obtained to be
%%%%%%%
	\begin{align}
	V_T^{\rm ring}=\frac{T}{12\pi}\sum_{\rm i= bosons}n_i \left( (M_i(\varphi)^2)^{3/2} - (M_i(\varphi,T)^2 )^{3/2}\right),\quad M_i(\varphi,T)^2=M_i(\varphi)^2 + \Pi_i,
	\end{align}
%%%%%%%
where $\Pi_i$ is the thermal self-energy.
Since we expect the self-energy effect to be smaller than other contributions, here we neglect this daisy resummation effect in the following discussions.  
 
In this section, the key result is the Eq.~(\ref{LEscalepotential}), the effective potential at the EW scale, which serves as the starting point to describe how a sizable barrier can be generated in a model with heavy fields \footnote{  We will discuss the different multi-scale effective potential among other renormalization schemes and several ways to get the effective potential at the LE scale in future work~\cite{future}. }. 
In the following analysis, we will focus on the extended fermion model and explore the phase transition pattern in such models.

%%%%%%%%%%%%%%%%%%%%%%%%%%%%%%%%%%%%%%%%%%%%%%%%%%%%%%%%%%%%
%%%%%%%%%%%%%%%%%%%%%%%%%%%%%%%%%%%%%%%%%%%%%%%%%%%%%%%%%%%%
%%%%%%%%%%%%%%%%%%%%%%%%%%%%%%%%%%%%%%%%%%%%%%%%%%%%%%%%%%%%
%%%%%%%%%%%%%%%%%%%%%%%%%%%%%%%%%%%%%%%%%%%%%%%%%%%%%%%%%%%%
%%%%%%%%%%%%%%%%%%%%%%%%%%%%%%%%%%%%%%%%%%%%%%%%%%%%%%%%%%%%
%%%%%%%%%%%%%%%%%%%%%%%%%%%%%%%%%%%%%%%%%%%%%%%%%%%%%%%%%%%%
\section{Extended fermion model}
%%%%%%%%%%%%%%%%%%%%%%%%%%%%%%%%%%%%%%%%%%%%%%%%%%%%%%%%%%%%
%%%%%%%%%%%%%%%%%%%%%%%%%%%%%%%%%%%%%%%%%%%%%%%%%%%%%%%%%%%%
%%%%%%%%%%%%%%%%%%%%%%%%%%%%%%%%%%%%%%%%%%%%%%%%%%%%%%%%%%%%
%%%%%%%%%%%%%%%%%%%%%%%%%%%%%%%%%%%%%%%%%%%%%%%%%%%%%%%%%%%%
%%%%%%%%%%%%%%%%%%%%%%%%%%%%%%%%%%%%%%%%%%%%%%%%%%%%%%%%%%%%
%%%%%%%%%%%%%%%%%%%%%%%%%%%%%%%%%%%%%%%%%%%%%%%%%%%%%%%%%%%%

In this section, we apply the general treatment of the multi-scale effective potential to the extended fermion model, more specifically, the model with one isospin doublet and one neutral isospin singlet fermions which are vector like leptons.
The Lagrangian with the doublet fermion $L^T=(N, E)$ and the neutral singlet fermions $N'$ is 
%%%%%%%%%%%%%%%%%%%%%%%%%%%%%%%%%%%%
\begin{align}
\label{Lagrang}
-{\cal L}_{fermion}\,\,\supset \,\,& y_{N}( \overline{L} \tilde{H} N^\prime  +  h.c.) +m_L \overline{L} L +m_N \overline{N}^\prime N^\prime.
\end{align}
%%%%%%%%%%%%%%%%%%%%%%%%%%%%%%%%%%%%
where new model parameters are $y_N$, $m_N$ and $m_L$. 
In this Lagrangian, the mass matrix for the neutral fermion $N, N^\prime$ reads 
%%%%%%%%%%%%%%%%%%%%%%%%%%%%%%%%%%%%%%%%%%%%%%%%%%%%%%%%%%%%
\begin{equation}
\mathcal{M}_N = \begin{pmatrix}
m_L &\frac{\varphi}{\sqrt{2}}  \, y_{N} \\ \frac{\varphi}{\sqrt{2}}  \, y_{N} & m_N
\end{pmatrix}.
\end{equation}
%%%%%%%%%%%%%%%%%%%%%%%%%%%%%%%%%%%%%%%%%%%%%%%%%%%%%%%%%%%%
After diagonalization, the field dependent masses for the mass eigenstates of the new fermions $N_1$ and $N_2$ are given by
%%%%%%%%%%%%%%%%%%%%%%%%%%%%%%%%%%%%%%%%%%%%%%%%%%%%%%%%%%%%
\begin{align}
\label{fieldMASS}
 M_{N_1, N_2} ^2 (\phi)= \frac{1}{2} \Bigg( m_L^2 + m_N^2 + y_{N}^2 \phi^2 \mp(m_L^2-m_N^2) \sqrt{1 +  \frac{2y_{N}^2 \phi^2 }{ \left( m_L - m_N  \right)^2}}\Bigg). 
\end{align}
%%%%%%%%%%%%%%%%%%%%%%%%%%%%%%%%%%%%%%%%%%%%%%%%%%%%%%%%%%%%

In order to analyze behaviour of effective potential under different parameter region, we parameterize the potential into a following polynomial form:
   %%%%%%%%%%%%%%%%%%%%%%%%%%%%%%%%%%%%
  \begin{align}
  V_{eff}^{P} &=  \lambda_{1,\, eff}\varphi + \frac{\mu^2_{eff}}{2}\varphi^2+ \frac{\lambda_{3,\, eff}}{3}\varphi^3 + \frac{\lambda_{eff}}{4}\varphi^4 + \frac{\lambda_{5,\, eff}}{5}\varphi^5+ \frac{\gamma_{eff}}{6}\varphi^6 + \frac{\lambda_{7,\, eff}}{7}\varphi^7 + \frac{\delta_{eff}}{8}\varphi^8 \nonumber\\
  &\quad  + \frac{\lambda_{9,\, eff}}{9}\varphi^9 + \frac{\epsilon_{eff}}{10}\varphi^{10} + {\cal O}(\varphi^{11}).
  \end{align}
  %%%%%%%%%%%%%%%%%%%%%%%%%%%%%%%%%%%%
These coefficients, recognized as the effective couplings, are calculated by differentials of the effective potential with the finite temperature effect:
  %%%%%%%%%%%%%%%%%%%%%%%%%%%%%%%%%%%%%%%%%%%%%%%%%%%%%%%%%%%%
  \begin{align}
  \label{effT}
  \mu^2_{eff}&\equiv \left. \frac{\partial^2 }{\partial \varphi^2} \left(V_{eff} + \Delta V_T \right)\right|_{\varphi=0},\quad \lambda_{eff}\equiv \frac{1}{3!} \left. \frac{\partial^4 }{\partial \varphi^4}  \left(V_{eff} + \Delta V_T \right)\right|_{\varphi=0},\dots \\\nonumber 
 \lambda_{n, eff}&\equiv \frac{1}{(n-1)!} \left. \frac{\partial^{n} }{\partial \varphi^{n}}  \left(V_{eff} + \Delta V_T \right)\right|_{\varphi=0}.
  \end{align}
  %%%%%%%%%%%%%%%%%%%%%%%%%%%%%%%%%%%%%%%%%%%%%%%%%%%%%%%%%%%%
Among them the coefficient of the $\varphi^4$ term ($\lambda_{eff}$) is important to generate a barrier in the following two scenarios:   
In scenario (I) where positive $\mu^2_{eff}$, negative $\lambda_{3,\, eff}$ and positive $\lambda_{eff}$ terms, a barrier may show up when the positive $\lambda_{eff}$ is closed to $E$ in Eq.~(\ref{thermalbosons}) by cancellation between the SM effect and the new fermion effect.
On the other hand, a barrier may also be generated by scenario (II) where negative $\lambda_{eff}$ term has larger negative fermion effect than scenario (I). 
Before discussing the $T$ dependence of the effective couplings, we first estimate behaviour of $\lambda_{eff}^{T=0}$ both analytically and numerically. 
Such an effective coupling $\lambda_{eff}^{T=0}$ is roughly given by 
%%%%%%%%%%%%%%%%%%%%%%%%%%%%%%%%%%%%
 \begin{align}
 \label{lameffgene}
 \lambda_{eff}^{T=0}&\sim \lambda_{eff}^{SM}  - 2\lambda_{(6)} v^2  + \frac{1}{6}\left. \frac{\partial^4V_{1-loop}^{VLL}}{\partial \varphi^4} \right|_{\varphi=0},\quad V_{1-loop}^{VLL} = \sum_{VLL}\frac{-4}{64\pi^2} M_i(\varphi)^4\left(\log\frac{M_i(\varphi)^2 }{v^2}-\frac{3}{2}\right),
 \end{align}
%%%%%%%%%%%%%%%%%%%%%%%%%%%%%%%%%%%% 
where $\lambda_{eff}^{SM}$ is the SM contribution containing the loop effects.
 Under finite temperature, we use the high temperature approximation for the SM effects in order to avoid the divergence of logarithmic term in the SM effects at $\varphi=0$.
We note that this approximation becomes invalid at low temperature $m / T \sim 1$.

 In the following, we will discuss the effective couplings in the potential in three mass parameter cases (A) $m_L \sim m_N \sim y_N v$, (B) $m_L\gg m_N \gg y_N v$, $m_L\sim m_N \gg y_N v$ and (C) $m_L\gg m_N \sim y_N v$.
In some cases, mass parameters are chosen to be at the TeV scale, which could come from physics at even higher energy scale, such as new symmetry breaking, composite Higgs model and so on.
But in this work we do not consider such UV completion above the TeV scale and only introduce the heavy mass parameter by hand.

%%%%%%%%%%%%%%%%%%%%%%%%%%%%%%%%%%%%%%%%%%%%%%%%%%%%%%%%%%%%
%%%%%%%%%%%%%%%%%%%%%%%%%%%%%%%%%%%%%%%%%%%%%%%%%%%%%%%%%%%%
%%%%%%%%%%%%%%%%%%%%%%%%%%%%%%%%%%%%%%%%%%%%%%%%%%%%%%%%%%%%
\subsection{Both fermions are at EW scale}
%%%%%%%%%%%%%%%%%%%%%%%%%%%%%%%%%%%%%%%%%%%%%%%%%%%%%%%%%%%%
%%%%%%%%%%%%%%%%%%%%%%%%%%%%%%%%%%%%%%%%%%%%%%%%%%%%%%%%%%%%
%%%%%%%%%%%%%%%%%%%%%%%%%%%%%%%%%%%%%%%%%%%%%%%%%%%%%%%%%%%%
In the case that both new fermions are at the EW scale $m_L \sim m_N\sim y_N v$, the effective potential does not encounter the multi-scale problem and thus there is no need to introduce high dimensional operators into the effective potential, and the field dependent masses for new fermions in Eq.~(\ref{fieldMASS}) are approximated as
%%%%%%%%%%%%%%%%%%%%%%%%%%%%%%%%%%%%%%%%%%%%%%%%%%%%%%%%%%%%
\begin{align}
\label{MassescaseEW}
 M_{N_1, N_2} ^2 (\phi)\sim m_L^2 + \frac{ y_{N}^2 \phi^2 }{2} \mp \sqrt{2} m_L  y_{N} \phi. 
\end{align}
%%%%%%%%%%%%%%%%%%%%%%%%%%%%%%%%%%%%%%%%%%%%%%%%%%%%%%%%%%%%
The effective potential at the EW scale is
 %%%%%%%%%%%%%%%%%%%%%%%%%%%%%%%%%%%%
\begin{align}
V_{eff}^{EW} &= \frac{\mu^2}{2}\varphi^2 + \frac{\lambda}{4}\varphi^4  + \sum_{i=N_1, N_2, SM}\frac{n_i}{64\pi^2} M_i(\varphi)^4\left(\log\frac{M_i(\varphi)^2 }{v^2}-c_i\right), 
\end{align}
%%%%%%%%%%%%%%%%%%%%%%%%%%%%%%%%%%%%
where $n_{t}=-12, n_{W}=6, n_{Z}=3, n_{h}=1, n_{NGB}=3$, $n_{N_1, N_2}=-4$.
The quartic coupling from new fermion effects is
%%%%%%%%%%%%%%%%%%%%%%%%%%%%%%%%%%%%
 \begin{align}
 \lambda_{eff}^{T=0}&\sim \lambda_{eff}^{SM}  -  \frac{y_N^4}{8\pi^2}\left(\log\frac{m_L^2}{v^2}+\frac{8}{3}\right), 
 \end{align}
%%%%%%%%%%%%%%%%%%%%%%%%%%%%%%%%%%%% 
where the second term may reduce the quartic coupling to the same level with $\lambda_{3,eff}$.

%%%%%%%%%%%%%%%%%%%%%%%%%%%%%%%%%%%%%%%%%%%%%%%%%%%%%%%%%%%%
%%%%%%%%%%%%%%%%%%%%%%%%%%%%%%%%%%%%%%%%%%%%%%%%%%%%%%%%%%%%
%%%%%%%%%%%%%%%%%%%%%%%%%%%%%%%%%%%%%%%%%%%%%%%%%%%%%%%%%%%%
\paragraph*{\textbf{Temperature dependence}}
%%%%%%%%%%%%%%%%%%%%%%%%%%%%%%%%%%%%%%%%%%%%%%%%%%%%%%%%%%%%
%%%%%%%%%%%%%%%%%%%%%%%%%%%%%%%%%%%%%%%%%%%%%%%%%%%%%%%%%%%%
%%%%%%%%%%%%%%%%%%%%%%%%%%%%%%%%%%%%%%%%%%%%%%%%%%%%%%%%%%%%
From Eqs.~(\ref{LEscalepotential}), (\ref{LEscalepotentialT}) and (\ref{effT}), the quadratic and quartic terms originated from new fermions, including the $T$ dependence,  are  
%%%%%%%%%%%%%%
	\begin{align} % requires amsmath; align* for no eq. number
(\mu^{\rm new \,fermions}_{eff})^2 &\simeq  \frac{y_{N}^2}{ 12 }  \left(  2 T^2 - \frac{9 m_L^2}{\pi^2} \left(\ln\frac{\alpha_FT^2}{v^2}-\frac{3}{2}\right) \right),\label{eq:mu2-caseA}\\
\lambda^{\rm new\, fermions}_{eff} &\simeq - \frac{y_N^4}{8\pi^2} \left(\ln\frac{\alpha_FT^2}{v^2}-\frac{3}{2}\right) ,
	\end{align}
%%%%%%%%%%%%%% 
where the high temperature approximation is taken. 
In Eq.~(\ref{eq:mu2-caseA}), the first term proportional to the $T^2$ is dominant at high temperature while the second term with thermal logarithm is dominant at the temperature $T < m_L$.
In total, the quadratic term from new fermion contribution is negative around the electroweak phase transition temperature.
The new quadratic term has the same sign as the tree-level $\mu^2$ term in the effective potential,
and  cancellation between two terms cannot happen to obtain smaller overall quadratic term which is comparable to the cubic term from the SM gauge bosons. 
Therefore in this case there is no comparable contributions of positive quadratic, negative cubic and positive quartic terms in the potential to generate a sizable barrier.  
This is the reason why the first-order EWPT can not be realized in the simple fermion model~\cite{Carena:2004ha}. 

 %%%%%%%%%%%%%%%%%%%%%%%%%%%%%%%%%%%%%%%%%%%%%%%%%%%%%%%%%%%%
%%%%%%%%%%%%%%%%%%%%%%%%%%%%%%%%%%%%%%%%%%%%%%%%%%%%%%%%%%%%
%%%%%%%%%%%%%%%%%%%%%%%%%%%%%%%%%%%%%%%%%%%%%%%%%%%%%%%%%%%%
\subsection{Both fermions are at TeV scale}
%%%%%%%%%%%%%%%%%%%%%%%%%%%%%%%%%%%%%%%%%%%%%%%%%%%%%%%%%%%%
%%%%%%%%%%%%%%%%%%%%%%%%%%%%%%%%%%%%%%%%%%%%%%%%%%%%%%%%%%%%
%%%%%%%%%%%%%%%%%%%%%%%%%%%%%%%%%%%%%%%%%%%%%%%%%%%%%%%%%%%%
Since both fermions are at the TeV scale, the effective potential at the EW scale would develop large logarithmic term, and thus we use the matching method in section 2 to avoid such term. 
There are two parameter regions to be considered: $m_L \gg m_N\gg y_N v$ or $m_L \sim m_N\gg y_N v$.
The detail of treatment of the multi-scale effective potential in case (B) is given in appendix A, and here we only list the main results.
 
  %%%%%%%%%%%%%%%%%%%%%%%%%%%%%%%%%%%%%%%%%%%%%%%%%%%%%%%%%%%%
%%%%%%%%%%%%%%%%%%%%%%%%%%%%%%%%%%%%%%%%%%%%%%%%%%%%%%%%%%%%
%%%%%%%%%%%%%%%%%%%%%%%%%%%%%%%%%%%%%%%%%%%%%%%%%%%%%%%%%%%%
\subsubsection{$m_L \gg m_N\gg y_N v$ case}
%%%%%%%%%%%%%%%%%%%%%%%%%%%%%%%%%%%%%%%%%%%%%%%%%%%%%%%%%%%%
%%%%%%%%%%%%%%%%%%%%%%%%%%%%%%%%%%%%%%%%%%%%%%%%%%%%%%%%%%%%
%%%%%%%%%%%%%%%%%%%%%%%%%%%%%%%%%%%%%%%%%%%%%%%%%%%%%%%%%%%%
In this case, we consider a two-step matching because there are two heavy scales: very HE scale $m_L$ and HE scale $m_N$. 
We will consider the high dimensional~operators up to dimensional~10 in this work. 
After the matching and running procedure, the effective potential at the EW scale is
 %%%%%%%%%%%%%%%%%%%%%%%%%%%%%%%%%%%%
\begin{align}
\label{caseBpotentialEW1}
V_{eff}^{EW} &= \frac{\mu^2}{2}\varphi^2 + \frac{\lambda}{4}\varphi^4  + \sum_{i=SM}\frac{n_i}{64\pi^2} M_i(\varphi)^4\left(\log\frac{M_i(\varphi)^2 }{v^2}-c_i\right)\nonumber\\
&\quad+\frac{\lambda_{(6)}}{6}\varphi^6+\frac{\lambda_{(8)}}{8}\varphi^8+\frac{\lambda_{(10)}}{10}\varphi^{10},
\end{align}
%%%%%%%%%%%%%%%%%%%%%%%%%%%%%%%%%%%%
 where $\lambda_{(n)}$ is the $n$-point coupling at the EW scale:
  %%%%%%%%%%%%%%%%%%%%%%%%%%%%%%%%%%%%
\begin{align}
\label{couplingA}
\lambda_{(6)} =  \gamma^{LE} -\beta_{6\varphi}^{LE} \log\frac{m_N}{v},\quad
\lambda_{(8)} =  \delta^{LE} -\beta_{8\varphi}^{LE} \log\frac{m_N}{v},\quad\lambda_{(10)}  = \epsilon^{LE} -\beta_{10\varphi}^{LE} \log\frac{m_N}{v}. 
\end{align}
 %%%%%%%%%%%%%%%%%%%%%%%%%%%%%%%%%%%
The first and second terms in the right hand side are respectively the threshold effect in Eq.~(\ref{B1thEW}) and the RG running term in Eq.~(\ref{betaLNheavyLE}).
If we exchange $m_N$ and $m_L$, the above calculations correspond to the case of $m_N \gg m_L\gg y_N v$.

%%%%%%%%%%%%%%%%%%%%%%%%%%%%%%%%%%%%%%%%%%%%%%%%%%%%%%%%%%%%
%%%%%%%%%%%%%%%%%%%%%%%%%%%%%%%%%%%%%%%%%%%%%%%%%%%%%%%%%%%%
%%%%%%%%%%%%%%%%%%%%%%%%%%%%%%%%%%%%%%%%%%%%%%%%%%%%%%%%%%%%
\subsubsection{$m_L\sim m_N\gg y_N v$ case}
%%%%%%%%%%%%%%%%%%%%%%%%%%%%%%%%%%%%%%%%%%%%%%%%%%%%%%%%%%%%
%%%%%%%%%%%%%%%%%%%%%%%%%%%%%%%%%%%%%%%%%%%%%%%%%%%%%%%%%%%%
%%%%%%%%%%%%%%%%%%%%%%%%%%%%%%%%%%%%%%%%%%%%%%%%%%%%%%%%%%%%

In this case, the field dependent masses of the new fermions are Eq.~(\ref{MassescaseEW}). 
 The form of the effective potential at the EW scale is the same as Eq.~(\ref{caseBpotentialEW1}) and dimensional~6, 8 and 10 operators at the EW scale are roughly 
 %%%%%%%%%%%%%%%%%%%%%%%%%%%%%%%%%%%%
\begin{align}
\label{couplingmLnmN}
\lambda_{(6)} &\simeq  \frac{y_N^6}{160\pi^2m_L^2} \left(  1 +  \frac{3}{8\pi^2} \left(-3 y_t^2  + \frac{3}{4}g_1^2+ \frac{9}{4}g_2^2 \right) \log\frac{m_L}{v}\right),\nonumber\\
\lambda_{(8)} &\simeq \frac{y_N^8}{2240\pi^2m_L^4}  \left(  1 +  \frac{1}{2\pi^2} \left(-3 y_t^2  + \frac{3}{4}g_1^2+ \frac{9}{4}g_2^2 \right) \log\frac{m_L}{v}\right),\nonumber\\
\lambda_{(10)} &\simeq \frac{y_N^{10}}{16128\pi^2m_L^6} \left(  1 +  \frac{5}{8\pi^2} \left(-3 y_t^2  + \frac{3}{4}g_1^2+ \frac{9}{4}g_2^2 \right) \log\frac{m_L}{v}\right),
\end{align}
 %%%%%%%%%%%%%%%%%%%%%%%%%%%%%%%%%%
where $y_t$ is the top Yukawa coupling and $g_1$ and $g_2$ are the gauge couplings of U(1)$_Y$ and SU(2)$_I$ symmetry, respectively.

The form of $\lambda_{(n)}$ differs between cases (B-1) and (B-2).  
For example, the dimensional~6 operators are $\lambda_{(6)}^{\rm (B\mathchar`-1)}\sim \frac{y_N^6}{16\pi^2m_L^2}$ and $\lambda_{(6)}^{\rm (B\mathchar`-2)}\sim \frac{y_N^6}{160\pi^2m_L^2}$.
Here the 1/10 factor difference comes from different effective interactions between $\varphi$ and new fermion at the electroweak scale after integrating out TeV fermions.
The Lagrangian of case (B-1) has $\varphi\varphi \bar{N} N$ interaction, while case (B-2) has $\varphi \bar{N} N$ interaction at the EW scale. 
So the same high dimensional~operator could originate from different UV topology with number of propagators from heavy fermions  between cases (B-1) and (B-2).
 
 %%%%%%%%%%%%%%%%%%%%%%%%%%%%%%%%%%%%%%%%%%%%%%%%%%%%%%%%%%%%
%%%%%%%%%%%%%%%%%%%%%%%%%%%%%%%%%%%%%%%%%%%%%%%%%%%%%%%%%%%%
%%%%%%%%%%%%%%%%%%%%%%%%%%%%%%%%%%%%%%%%%%%%%%%%%%%%%%%%%%%%
\paragraph*{\textbf{Temperature dependence}}
%%%%%%%%%%%%%%%%%%%%%%%%%%%%%%%%%%%%%%%%%%%%%%%%%%%%%%%%%%%%
%%%%%%%%%%%%%%%%%%%%%%%%%%%%%%%%%%%%%%%%%%%%%%%%%%%%%%%%%%%%

In these cases (B-1) and (B-2), new fermions only contribute to the high dimensional~operators with positive size, but cannot give rise to large negative quartic coupling in the potential. 
Therefore it is difficult to generate a barrier in the potential with this case (B), since the high dimensional~operator term is typically much smaller than tree-level ones, and there is no enough negative quartic contribution.

%%%%%%%%%%%%%%%%%%%%%%%%%%%%%%%%%%%%%%%%%%%%%%%%%%%%%%%%%%%%
%%%%%%%%%%%%%%%%%%%%%%%%%%%%%%%%%%%%%%%%%%%%%%%%%%%%%%%%%%%%
%%%%%%%%%%%%%%%%%%%%%%%%%%%%%%%%%%%%%%%%%%%%%%%%%%%%%%%%%%%%
\subsection{One fermion is at TeV scale and another is at EW scale }
%%%%%%%%%%%%%%%%%%%%%%%%%%%%%%%%%%%%%%%%%%%%%%%%%%%%%%%%%%%%
%%%%%%%%%%%%%%%%%%%%%%%%%%%%%%%%%%%%%%%%%%%%%%%%%%%%%%%%%%%%
%%%%%%%%%%%%%%%%%%%%%%%%%%%%%%%%%%%%%%%%%%%%%%%%%%%%%%%%%%%%
In this parameter region ($m_L \gg m_N \sim y_N v$), $m_N$ is at the EW scale and $m_L$ is at the TeV scale.
The detail of matching and running in the model with this parameter is also in the appendix A.
In this case, the effective potential at the EW scale is obtained to be
 %%%%%%%%%%%%%%%%%%%%%%%%%%%%%%%%%%%%
\begin{align}
\label{caseBpotentialEW}
V_{eff}^{EW} &= \frac{\mu^2}{2}\varphi^2 + \frac{\lambda}{4}\varphi^4  + \sum_{i=SM, N_1}\frac{n_i}{64\pi^2} M_i(\varphi)^4\left(\log\frac{M_i(\varphi)^2 }{v^2}-c_i\right)\nonumber\\
&\quad+\frac{\lambda_{(6)}}{6}\varphi^6+\frac{\lambda_{(8)}}{8}\varphi^8+\frac{\lambda_{(10)}}{10}\varphi^{10},
\end{align}
%%%%%%%%%%%%%%%%%%%%%%%%%%%%%%%%%%%%
where the dimensional~6, 8 and 10 operators at the EW scale are roughly given by
 %%%%%%%%%%%%%%%%%%%%%%%%%%%%%%%%%%%%
\begin{align}
\label{dim6810ope}
\lambda_{(6)} &\simeq  \frac{m_L y_N^6}{16\pi^2(m_L-m_N)^5}(m_L^2+7m_Nm_L - 2m_N^2)  - \beta_{6\varphi} \log\frac{m_L}{v},\nonumber\\
\lambda_{(8)} &\simeq  - \frac{y_N^8}{48\pi^2(m_L-m_N)^7}(7m_L^3+27m_Nm_L^2 - 4m_N^3) - \beta_{8\varphi} \log\frac{m_L}{v},\nonumber\\
\lambda_{(10)} &\simeq  \frac{y^{10}}{384\pi^2m_L(m_L-m_N)^9}(107m_L^4+342m_L^3m_N+42m_L^2m_N^2 - 68 m_Lm_N^3 - 3m_N^4)  - \beta_{10\varphi} \log\frac{m_L}{v},
\end{align}
 %%%%%%%%%%%%%%%%%%%%%%%%%%%%%%%%%%%%
where the first term presents the threshold effect and the second term the running effect.
The higher dimensional~operators, such as the dimensional~6 operator $\lambda_{(6)}$, in $m_L \gg m_N \sim y_N v$ case are typically larger than other cases in Eqs.~(\ref{couplingA}, \ref{couplingmLnmN}) due to mass splitting from the mixing effects.

The quartic effective coupling $\lambda_{eff}^{T=0}$ in this parameter region (C) is roughly given by
%%%%%%%%%%%%%%%%%%%%%%%%%%%%%%%%%%%%
\begin{align}
\label{lambdaval}
\lambda_{eff}^{T=0}  &\sim \lambda_{eff}^{SM} - 2\gamma_{eff}^{T=0}v^2- 3\lambda_{(8)}v^4- 4\lambda_{(10)}v^6  -  \frac{y_N^4m_N^2}{8\pi^2m_L^2} \log\frac{m_N^2}{v^2},
\end{align}
%%%%%%%%%%%%%%%%%%%%%%%%%%%%%%%%%%%% 
 where $\gamma_{eff}^{T=0}$ is $\lambda_{(6)}$ in Eq.~(\ref{dim6810ope}) and logarithmic term comes from the loop effect of the new light fermion.
 %%%%%%%%%%%%%%%%%%%%%%%%%%%%%%
 %%%%%%%%%%%% new %%%%%%%%%%%%%%%
 %%%%%%%%%%%%%%%%%%%%%%%%%%%%%%
 On the other hand, the quadratic effective coupling $(\mu_{eff}^{T=0})^2$ in the case (C) is roughly given as
%%%%%%%%%%%%%%%%%%%%%%%%%%%%%%%%%%%%
\begin{align}
\label{mueff}
(\mu_{eff}^{T=0})^2  &\sim \lambda_{eff}^{SM} + \gamma_{eff}^{T=0}v^4 + 2\lambda_{(8)}v^6 + 3 \lambda_{(10)}v^8  + \frac{y_N^2m_N^3}{4\pi^2m_L} \log\frac{m_N^2}{v^2},
\end{align}
%%%%%%%%%%%%%%%%%%%%%%%%%%%%%%%%%%%% 
which has the almost same terms as quartic coupling $\lambda_{eff}^{T=0}$. 
 The quadratic and quartic terms contain not only the light fermion loop effect but also the heavy fermion effect, which in sum have larger positive and negative fermion effects than cases (A) and (B), respectively.

\begin{figure}[tb]
  \begin{center}
\includegraphics[width=0.45\textwidth]{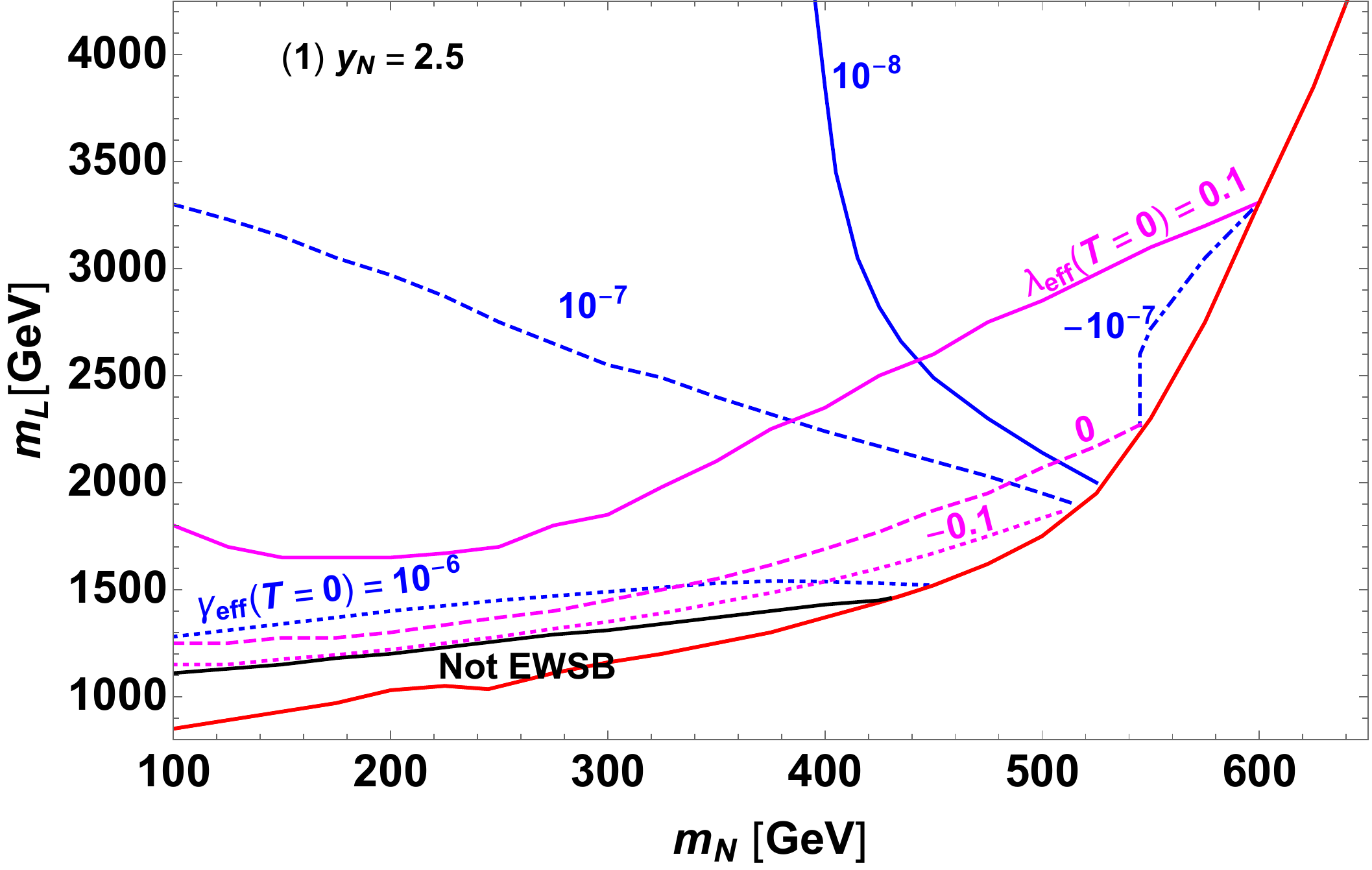}\\
\includegraphics[width=0.45\textwidth]{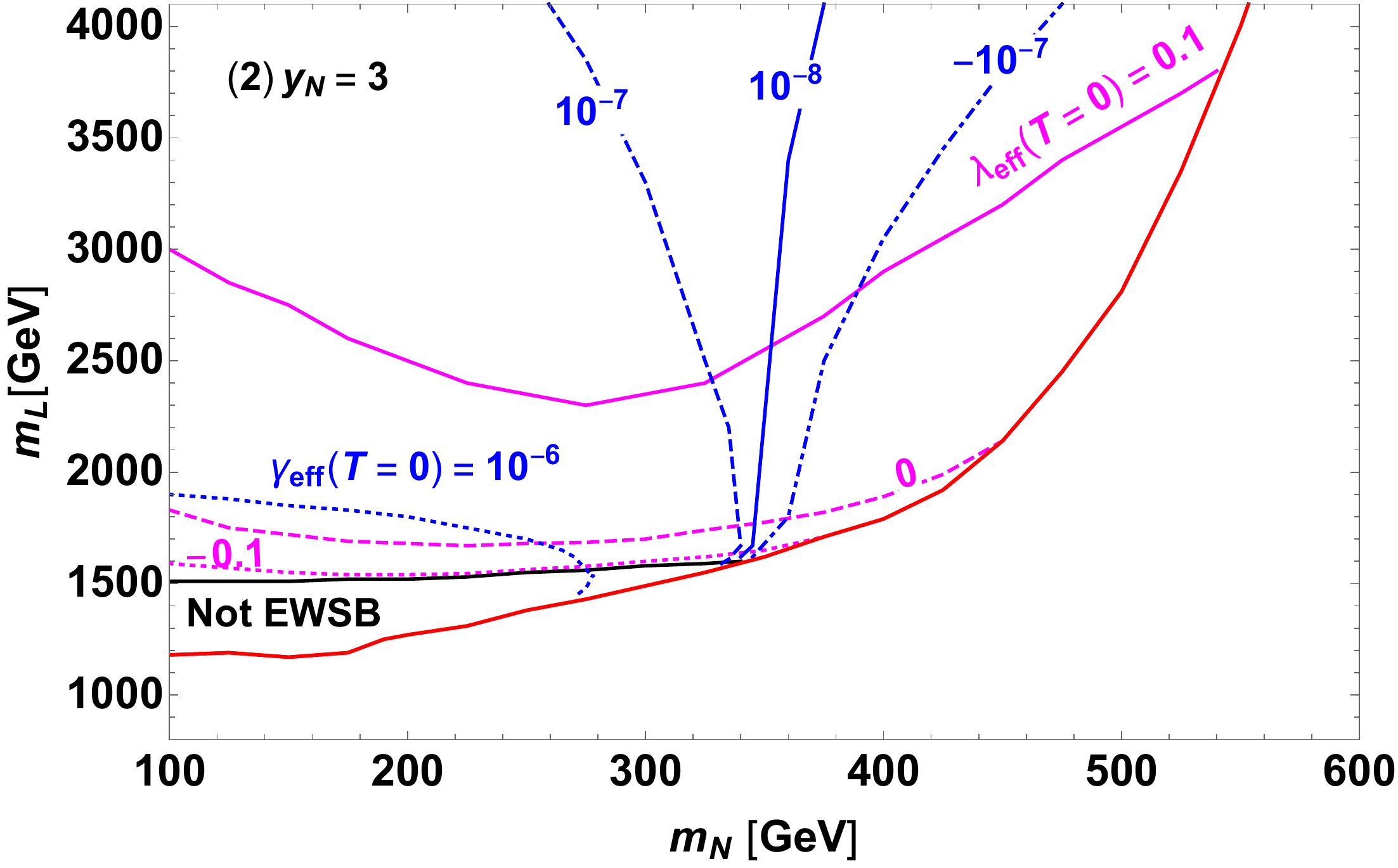}
\includegraphics[width=0.45\textwidth]{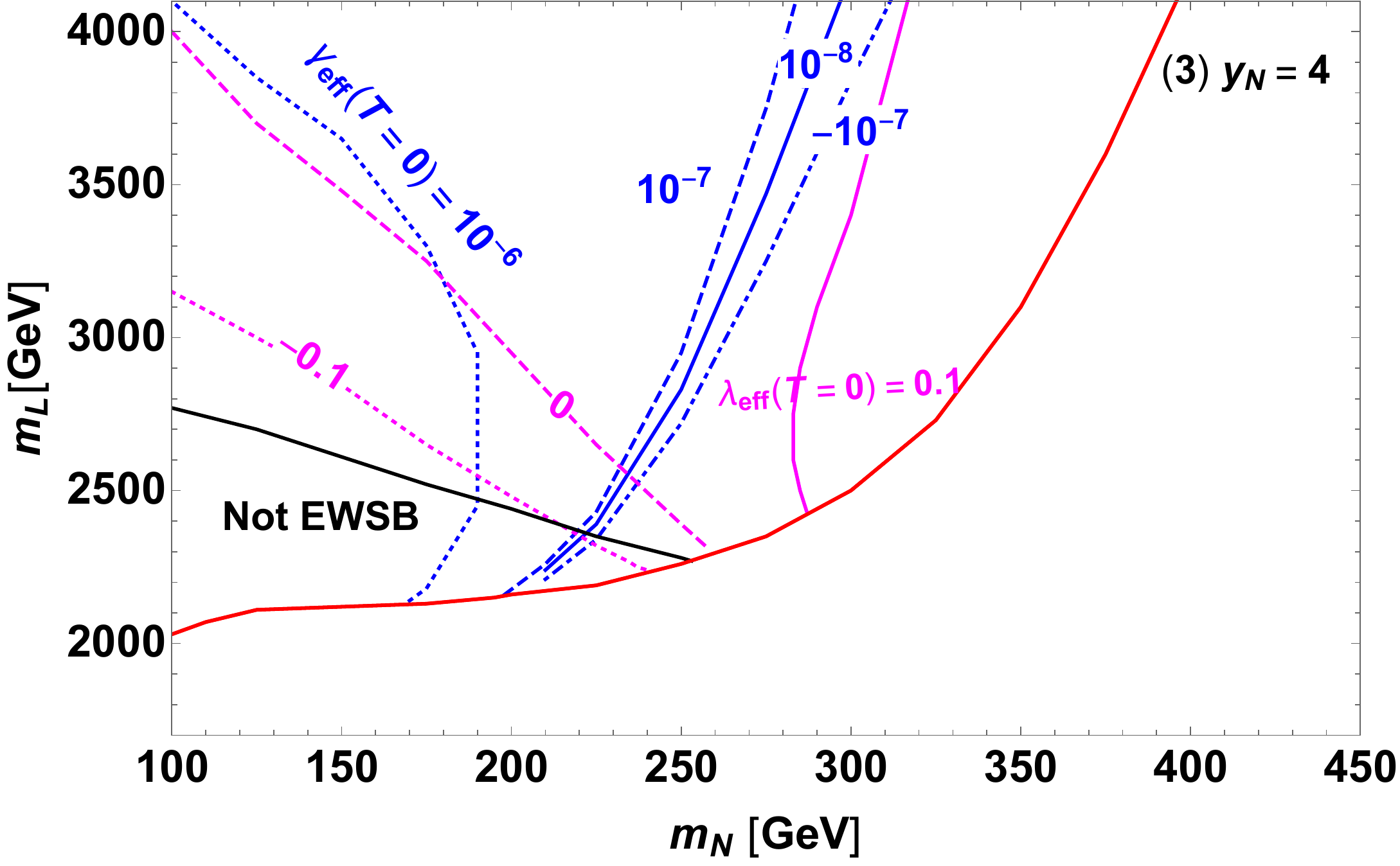}
\caption{The contours of $\lambda_{eff}$ and $\gamma_{eff}$ at zero temperature with $y_N=2.5,$ 3 and 4 in Figs.~\ref{gamlamyN}-(1), \ref{gamlamyN}-(2) and \ref{gamlamyN}-(3), respectively. 
Magenta solid, dashed, dotted lines are $\lambda_{eff}^{T=0}=0.1, 0, -0.1$, respectively.
Also, blue dotted, dashed, solid, dashed dotted lines are $\gamma_{eff}^{(T=0)}=10^{-6}, 10^{-7}, 10^{-8}, -10^{-7}$, respectively.
The red line represents the criterion that indicates the RG running analysis is appropriate or not, e.g. the condition $(\lambda_{(6)}(Q_M) - \gamma_{th})/\gamma_{th}<10^{-3}$ (above red line) or $>10^{-3}$ (below red line).
Also, symmetry breaking does not occur below the black line.}
\label{gamlamyN}
  \end{center}
\end{figure}

We will focus on this case and perform numerical analyses. Since the terms in the zero temperature serve as the starting point of our analysis, we will first investigate various terms in the zero temperature potential. 
The numerical results of $\lambda_{eff}^{T=0}$ and $\gamma_{eff}^{T=0}$ in $(m_N, m_L)$ plane with $y_N=2.5,$ 3 and 4 are shown in Figs.~\ref{gamlamyN}-(1), \ref{gamlamyN}-(2) and \ref{gamlamyN}-(3), respectively. 
%%%%%%%%%%%%%%%%%%%%%%%%%%%%%%%%%%%%
%%%%%%%%%%%%%%%%%%%%%%%%%%%%%%%%%%%%
%%%%%%%%%%%%%%%%%%%%%%%%%%%%%%%%%%%%
%%%%%%%%%%%%%%%%%%%%%%%%%%%%%%%%%%%%
In the parameter space below red lines, the result of RG running analysis cannot explain the input parameter, for example, the threshold effect at high dimensional operator $\gamma_{th}$ in Eq.~(\ref{darac}) is not the same as a result of RG running analysis at the matching scale $\lambda_{(6)}(Q_M)$ in Eq.~(\ref{dim6810ope}).
The cause of that is excessively large loop effects of new fermion and these effects disturb the RG running analysis. 
We clarify the parameter region, where $\gamma_{th} \sim \lambda_{(6)}(Q_M)$, by $(\lambda_{(6)}(Q_M) - \gamma_{th})/\gamma_{th}<10^{-3}$.
In the parameter space below the black line, the spontaneous symmetry breaking cannot occur because the minimum of the potential is at origin.
Blue lines correspond to contours about the value of $\gamma_{eff}$ in Eq.~(\ref{dim6810ope}). 
On the other hand, magenta lines represent contour lines of $\lambda_{eff}$ at zero temperature in Eq.~(\ref{lambdaval}) which have $\gamma_{eff}$ term. 
\paragraph*{\textbf{Behaviour of blue contour \mbox{\boldmath $\gamma_{eff}$}}} 
  According to Eq.~(\ref{dim6810ope}), the dimensional 6 operator is roughly given as
    %%%%%%%%%%%%%%%%%%%%%%%%%%%%%%%%%%%%
\begin{align}
\label{rgamma}
\gamma_{eff}^{T=0} &\simeq  \frac{ y_N^6 }{16\pi^2m_L^2}\left(1 - 3\left( \frac{3y_t^2 }{8\pi^2}  + 2\frac{m_N}{m_L}\right)\log\frac{m_L}{v} \right).
\end{align}
 %%%%%%%%%%%%%%%%%%%%%%%%%%%%%%%%%%%%
   From this equation, the $\gamma_{eff}^{T=0}$ could be negative when $m_N/m_L$ is large, as shown dashed dotted lines in the Fig.~\ref{gamlamyN}. 
  The negative effect is the source of different behaviours of $\gamma_{eff}^{T=0}$, especially blue dashed and solid lines, between Fig.~\ref{gamlamyN}-(1) and Figs.~\ref{gamlamyN}-(2), -(3).
   Since small $m_L$ is prohibited by the RG running in Figs.~\ref{gamlamyN}-(2), -(3), the ratio of $m_N/m_L$ can only be increased by large $m_N$ value.
    From that, the slope of blue solid line, for example, change from Fig.~\ref{gamlamyN}-(1) to Figs.~\ref{gamlamyN}-(3).   
  \paragraph*{\textbf{Behaviour of magenta contour \mbox{\boldmath $\lambda_{eff}$} }} The quartic term $\lambda_{eff}^{T=0}$ in Eq.~(\ref{lambdaval}) has not only $\gamma_{eff}^{T=0}$ but also negative loop effect of light fermion.
  Since the negative light fermion loop effect in Eq. (\ref{lambdaval}) is proportional to $(m_N/m_L)^2$, the $\lambda_{eff}^{T=0}$ could be small positive or negative value even if $\gamma_{eff}^{T=0}$ has a negative contribution from RG running term $(m_N/m_L) \log m_L/ v $.
  Therefore the behaviour of $\lambda_{eff}^{T=0}$ can have downward convex shapes in Fig.~\ref{gamlamyN}-(1) and \ref{gamlamyN}-(2).
  In the Fig.~\ref{gamlamyN}-(3), the behaviour of magenta line could be similar to blue line, because the $\gamma_{eff}^{T=0}$ and the one-loop effect of light fermion are respectively proportional to $y_N^6$ and $y_N^4$.

According to the Fig.~\ref{gamlamyN}, the quartic term at the zero temperature potential determines which cases would happen: if the quartic term is positive and comparable to the cubic term, it would be possible to satisfy the scenario (I) discussed in the introduction; if the quartic term is negative, then it  is likely to realise the scenario (II). 
 Also, in this parameter region, new fermion effects could cause the potential to be unstable.
  We discuss the detail of the stability and Landau pole in the appendix B.

%%%%%%%%%%%%%%%%%%%%%%%%%%%%%%%%%%%%%%%%%%%%%%%%%%%%%%%%%%%%
%%%%%%%%%%%%%%%%%%%%%%%%%%%%%%%%%%%%%%%%%%%%%%%%%%%%%%%%%%%%
%%%%%%%%%%%%%%%%%%%%%%%%%%%%%%%%%%%%%%%%%%%%%%%%%%%%%%%%%%%%
\paragraph*{\textbf{Temperature dependence}}
%%%%%%%%%%%%%%%%%%%%%%%%%%%%%%%%%%%%%%%%%%%%%%%%%%%%%%%%%%%%
%%%%%%%%%%%%%%%%%%%%%%%%%%%%%%%%%%%%%%%%%%%%%%%%%%%%%%%%%%%%
%%%%%%%%%%%%%%%%%%%%%%%%%%%%%%%%%%%%%%%%%%%%%%%%%%%%%%%%%%%%

From Eq.~(\ref{effT}), new fermion contributions with the $T$ effect in the quadratic and the quartic terms are obtained to be 
%%%%%%%%%%%%%%
	\begin{align} % requires amsmath; align* for no eq. number
	\label{Tdepmu}
(\mu^{\rm new \,fermion}_{eff})^2 &\simeq \gamma_{eff}^{T=0}v^4  + \frac{y_{N}^2X}{ 2(1- X)  }  \left( - \frac{ T^2}{3} + \frac{X^2m_L^2}{2\pi^2} \left(\ln\frac{\alpha_FT^2}{v^2}-\frac{3}{2}\right) \right),\\
	\label{Tdeplam}
\lambda^{\rm new\, fermion}_{eff} &\simeq  - 2\gamma_{eff}^{T=0}v^2  + \frac{4y_{N}^4(1+X)}{ 16m_L^2(1 - X)^3  }  \left( \frac{ T^2}{3} - \frac{X^2m_L^2(3-X)}{2\pi^2(1+X)} \left(\ln\frac{\alpha_FT^2}{v^2}-\frac{3}{2}\right) \right),
	\end{align}
%%%%%%%%%%%%%%
where $X \equiv m_N/m_L$. 
From the above equation, we note that the larger the mixing effect, which is proportional to $y_N X$, the more negative quartic coupling in the Higgs potential. 
At the same time, in the quadratic term, if the mixing effect becomes larger, the positive logarithmic term dominates over the negative $T^2$ term, and verse vice.
The finite temperature effect on the dimensional~6 term is
%%%%%%%%%%%%%%
	\begin{align} % requires amsmath; align* for no eq. number
	\label{gamTfermi}
\gamma^{\rm T, fermion}_{eff} &\simeq\frac{y_{N}^6(1+X)}{ 16 m_L^4 (1 - X)^5  }  \left(- 2 T^2 + \frac{3m_L^2 X }{\pi^2} \left(\ln\frac{\alpha_FT^2}{v^2}-\frac{3}{2}\right) \right).
	\end{align}
%%%%%%%%%%%%%%
Similarly the larger the mixing effects, the larger the higher dimensional term contribution. 

So beside the size and sign of the quartic term at the zero temperature, the mixing effects, parametrized by the $y_N X$ also control whether the sizable barrier could be realized. 
\begin{itemize}
	\item 
In the scenario (II), it needs negative quartic term and large mixing angle. 
In this scenario, at low temperature comparable to the EW scale, these logarithmic terms dominate the effective couplings in the quadratic and quartic terms: the quadratic term is enhanced by the new fermion positive logarithmic term besides the SM quadratic $T^2$ contribution, while the quartic coupling is further decreased by the new fermion negative logarithmic term. 
A large barrier could be generated through the positive quadratic and sextic effective couplings ($\mu_{eff}^2$, $\gamma_{eff}$) and the negative quartic one ($\lambda_{eff}$). 
\item
On the other hand, if the mixing effects are not so large, and the quartic coupling at zero temperature is small but still positive, the scenario (I) could be realized.  
In this scenario, since the mixing is not so large the negative $T^2$ term dominates over the logarithmic term in the quadratic term. Thus there is cancellation between the negative $T^2$ term from the fermions and the SM thermal contribution, and thus the quadratic term is decreased to smaller size which is comparable to the cubic term in the SM.   
In total, the effective potential with comparable positive quadratic, negative cubic and positive quartic terms could realize first-order EWPT.
\end{itemize}

In summary, the extended fermion model with this mass region could generate a barrier through the (I) positive $\varphi^2$ term, negative $\varphi^3$ term and positive $\varphi^4$ term and the (II) positive $\varphi^2$ term, negative $\varphi^4$ term and positive higher dimensional~term scenarios.

%%%%%%%%%%%%%%%%%%%%%%%%%%%%%%%%%%%%%%%%%%%%%%%%%%%%%%%%%%%%
%%%%%%%%%%%%%%%%%%%%%%%%%%%%%%%%%%%%%%%%%%%%%%%%%%%%%%%%%%%%
%%%%%%%%%%%%%%%%%%%%%%%%%%%%%%%%%%%%%%%%%%%%%%%%%%%%%%%%%%%%
%%%%%%%%%%%%%%%%%%%%%%%%%%%%%%%%%%%%%%%%%%%%%%%%%%%%%%%%%%%%
%%%%%%%%%%%%%%%%%%%%%%%%%%%%%%%%%%%%%%%%%%%%%%%%%%%%%%%%%%%%
%%%%%%%%%%%%%%%%%%%%%%%%%%%%%%%%%%%%%%%%%%%%%%%%%%%%%%%%%%%%
\section{Numerical results on phase transition pattern}
%%%%%%%%%%%%%%%%%%%%%%%%%%%%%%%%%%%%%%%%%%%%%%%%%%%%%%%%%%%%
%%%%%%%%%%%%%%%%%%%%%%%%%%%%%%%%%%%%%%%%%%%%%%%%%%%%%%%%%%%%
%%%%%%%%%%%%%%%%%%%%%%%%%%%%%%%%%%%%%%%%%%%%%%%%%%%%%%%%%%%%
%%%%%%%%%%%%%%%%%%%%%%%%%%%%%%%%%%%%%%%%%%%%%%%%%%%%%%%%%%%%
%%%%%%%%%%%%%%%%%%%%%%%%%%%%%%%%%%%%%%%%%%%%%%%%%%%%%%%%%%%%
%%%%%%%%%%%%%%%%%%%%%%%%%%%%%%%%%%%%%%%%%%%%%%%%%%%%%%%%%%%%
As discussed above, in cases (A) and (B), it is difficult to generate a sizable barrier and thus cannot realize first-order phase transition.
In the following, we will focus on the case (C) to explore how to generate a barrier under the scenarios (I) and (II). 
We will numerically analyze the phase transition pattern by using both the high temperature approximation and exact thermal effective potential.
There are three new parameters ($y_N$, $m_L$, $m_N$) in the model. 
Before we perform a parameter scan over all the parameters, we first analyze the phase transition scenarios with chosen benchmark points.

\begin{figure}[htb]
  \begin{center}
\includegraphics[width=0.45\textwidth]{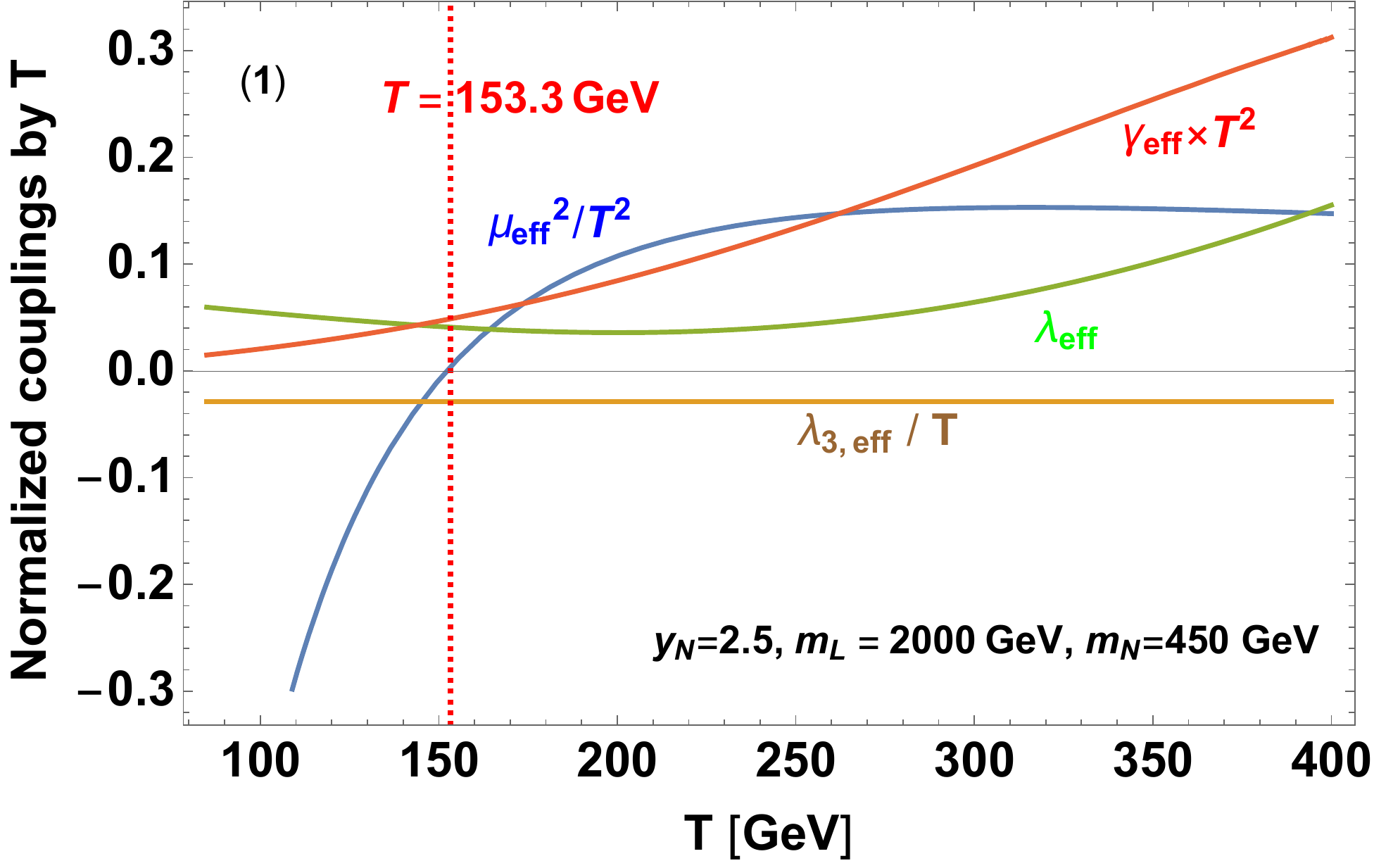}\,\,\,\,%\\
\includegraphics[width=0.4\textwidth]{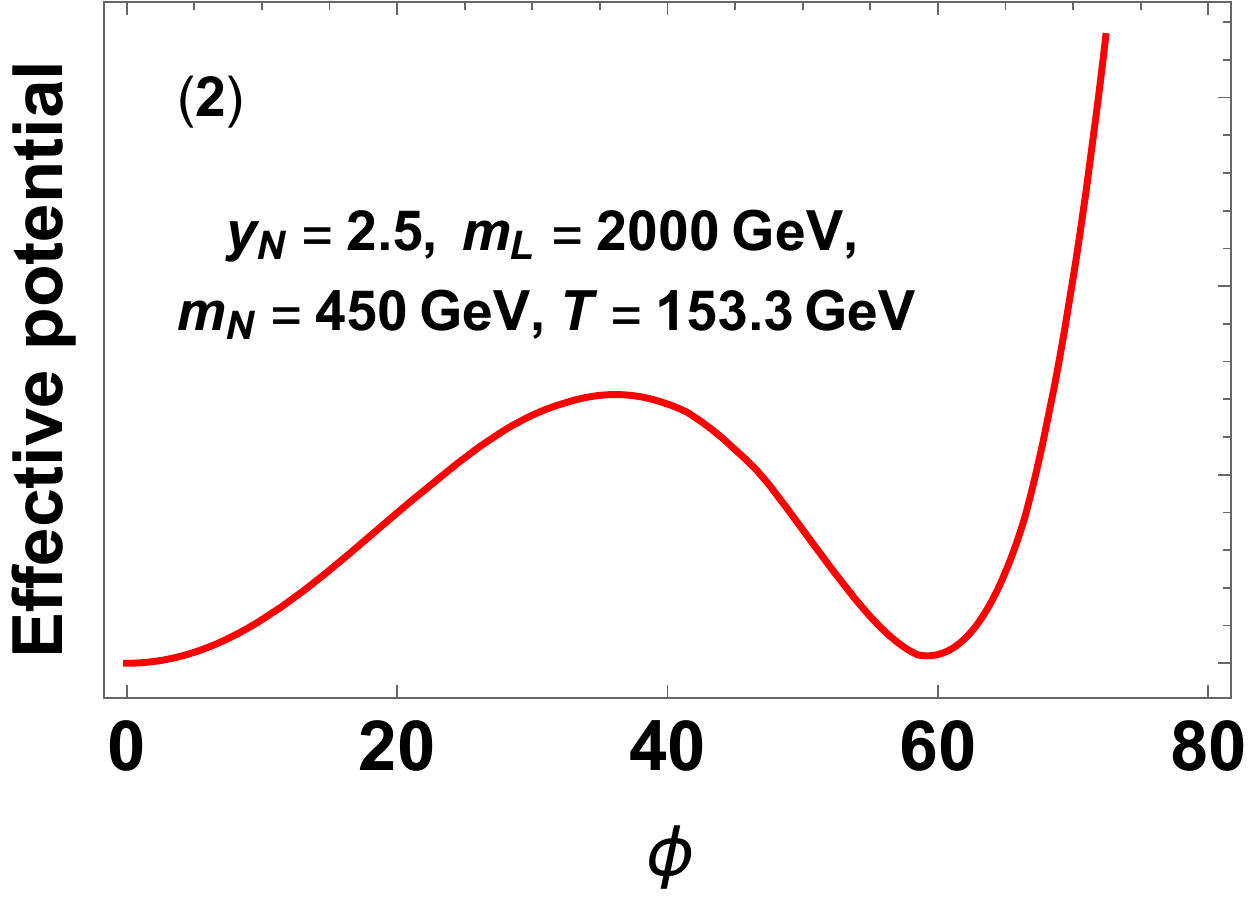}\\
\includegraphics[width=0.45\textwidth]{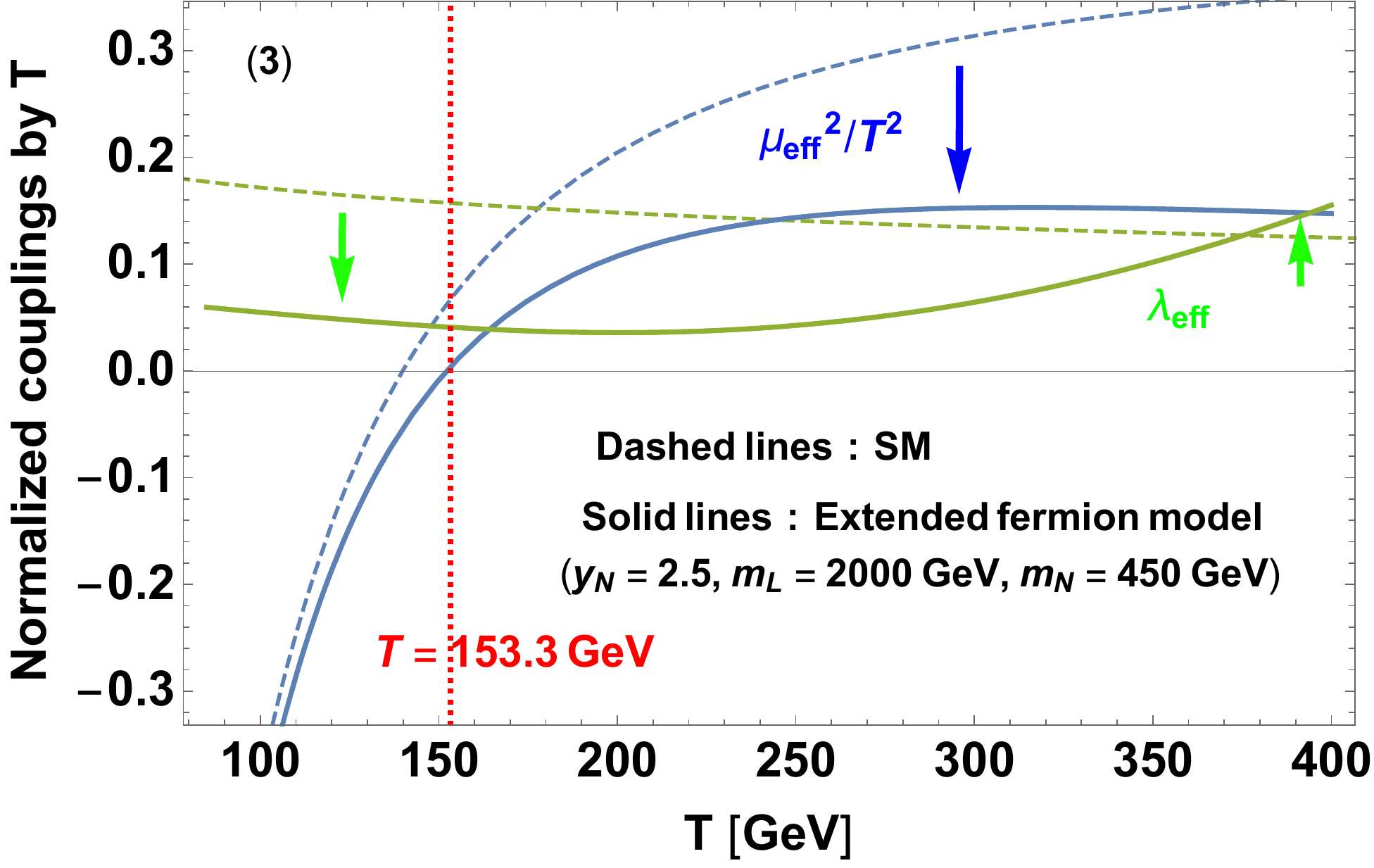}
\caption{ Upper left figure \ref{TdepI}-(1) represents the $T$ dependence of normalized effective couplings of the model with $y_N=2.5$, $m_L=2000$ GeV and $m_N=450$ GeV: blue $\mu_{eff}^2/T^2$, brawn $\lambda_{3,eff}/T$, green $\lambda_{eff}$ and red $\gamma_{eff} T^2$.
The red dotted lines in figure \ref{TdepI}-(1) and lower one \ref{TdepI}-(3) correspond to the critical temperature of the model with this benchmark point: $T=153.3$ GeV.
 Upper right figure \ref{TdepI}-(2) describe the shape of potential at the critical temperature. 
 Lower figure \ref{TdepI}-(3) represents differences of quadratic (blue) and quartic (green) couplings between in the extended fermion model with this benchmark point (solid lines) and in the SM (dashed line).
}
\label{TdepI}
  \end{center}
\end{figure}

First, let us discuss the scenario (I) with the benchmark point: $y_N=2.5$, $m_L=2000$ GeV and $m_N=450$ GeV.
The $T$ dependence of normalized effective couplings in Eq.~(\ref{effT}) with this benchmark point are shown in Fig.~\ref{TdepI}-(1).
%%%%%%%%%%%%%%%%%%%%%%%%%%%%%%%%%%%%
%%%%%%%%%%%%%%%%%%%%%%%%%%%%%%%%%%%%
%%%%%%%%%%%%%%%%%%%%%%%%%%%%%%%%%%%%
%%%%%%%%%%%%%%%%%%%%%%%%%%%%%%%%%%%%
The blue, brawn, green and red lines in the figure correspond to the $T$ dependences of $\mu_{eff}^2/T^2$, $\lambda_{3,eff}/T$, $\lambda_{eff}$ and $\gamma_{eff} T^2$, respectively.
In Fig.~\ref{TdepI}-(2), the shape of the effective potential is described at critical temperature, at which there is degenerate vacuum in the potential.
The red dotted lines in upper right Fig.~\ref{TdepI}-(1) and in lower Fig.~\ref{TdepI}-(3) represents the critical temperature of the model with this benchmark point, $T=153.3$ GeV. 
From Fig.~\ref{TdepI}-(1), we note that the normalized effective couplings are almost the same size around the critical temperature (red dotted line) and then a sizable barrier is developed.
Lower Fig.~\ref{TdepI}-(3) shows the difference in $T$ dependence of normalized effective couplings $\mu_{eff}^2/T^2$ and $\lambda_{eff}$ between the extended fermion model with the benchmark point (solid lines) and the SM (dashed lines).
Compared to the SM, the normalized effective couplings are quite different: $\mu_{eff}^2/T^2$ and $\lambda_{eff}$ are decreased compared to the SM values.
 Arrows in Fig.~\ref{TdepI}-(3) represent the amount of change from the SM to the extended fermion model with this benchmark point.
 The directions of arrows change at low temperature, because then the logarithmic term of $T$ becomes dominant contribution.

\begin{figure}[htb]
  \begin{center}
\includegraphics[width=0.45\textwidth]{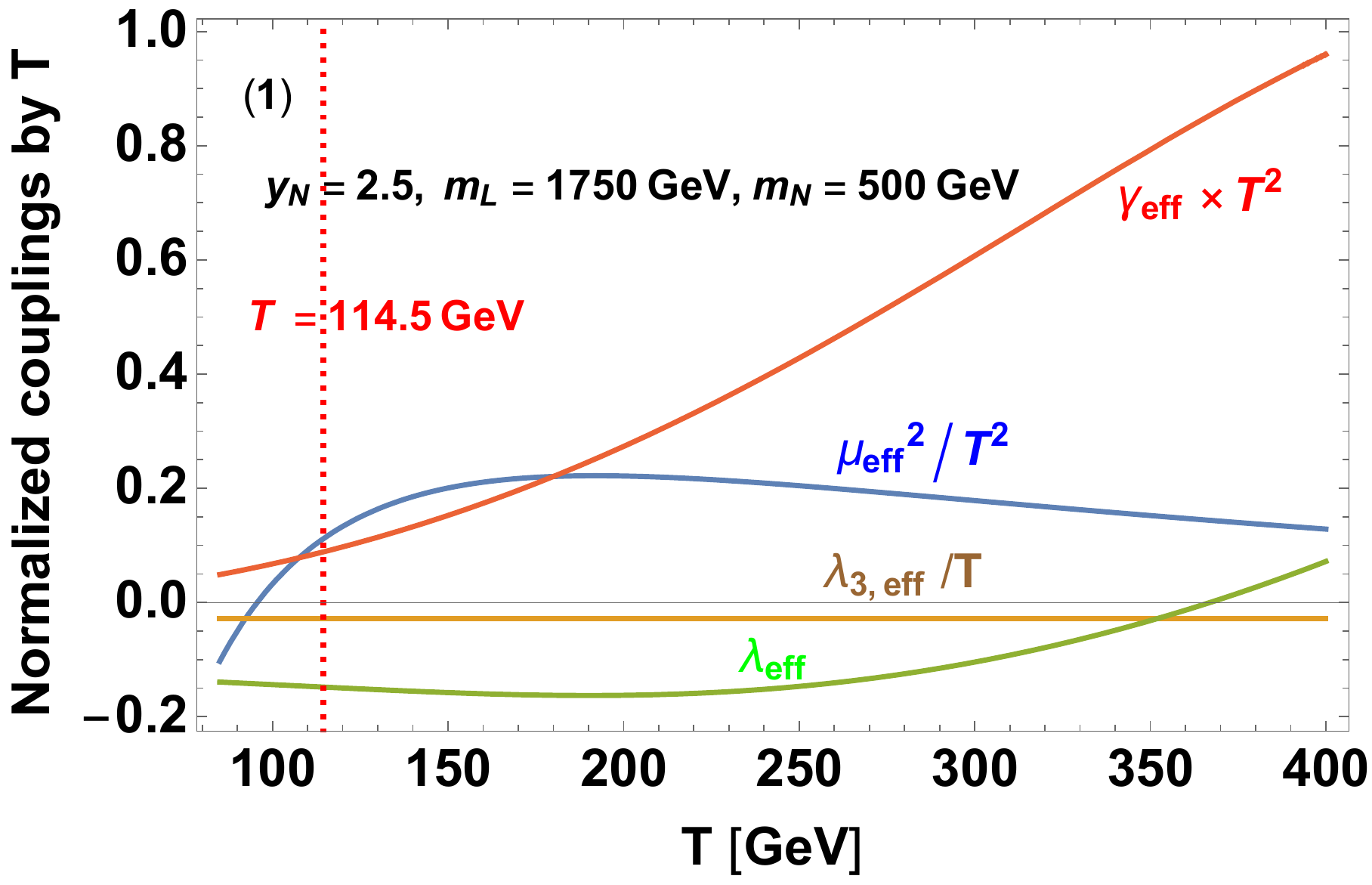}\,\,\,\,
\includegraphics[width=0.4\textwidth]{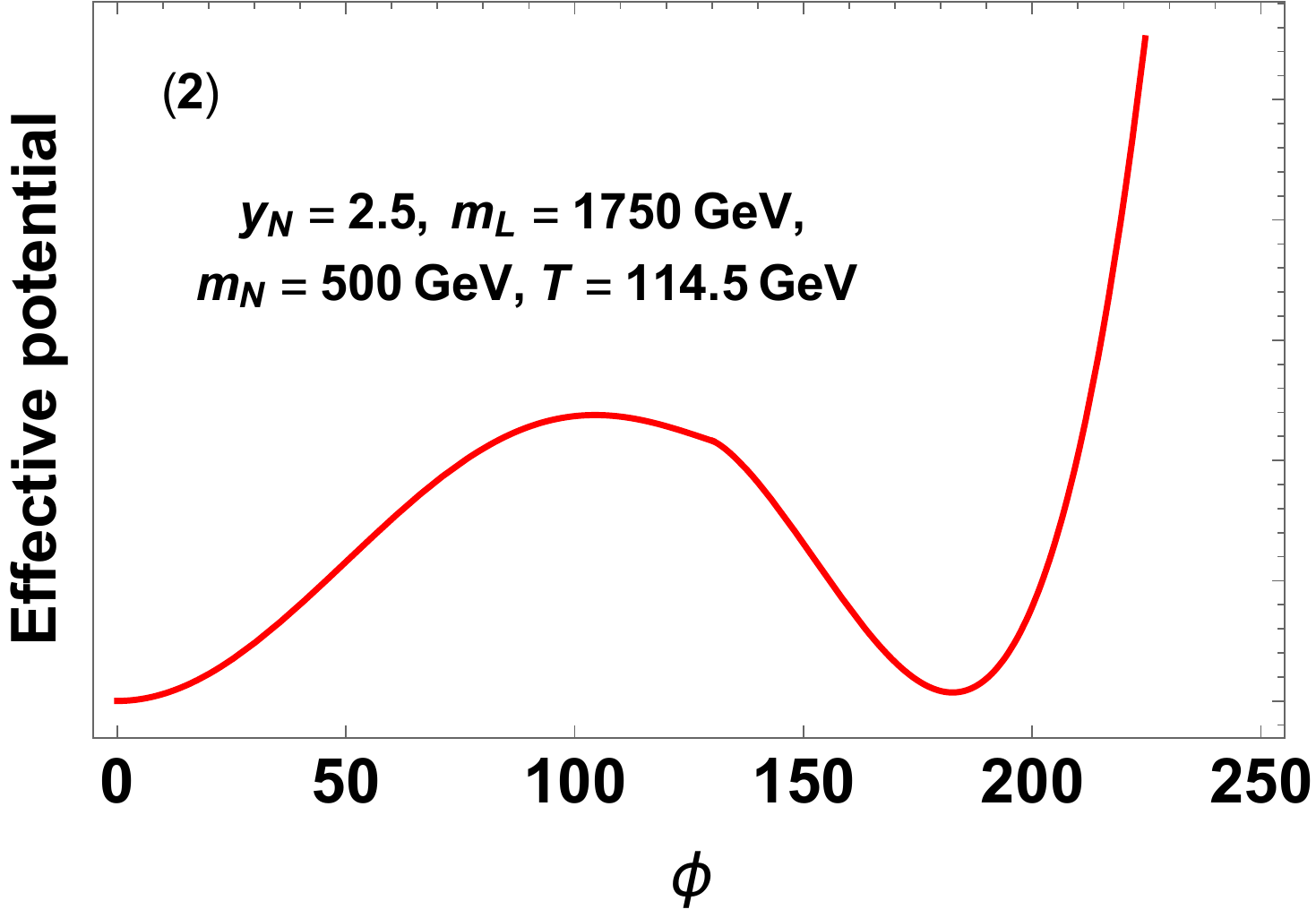}
\includegraphics[width=0.45\textwidth]{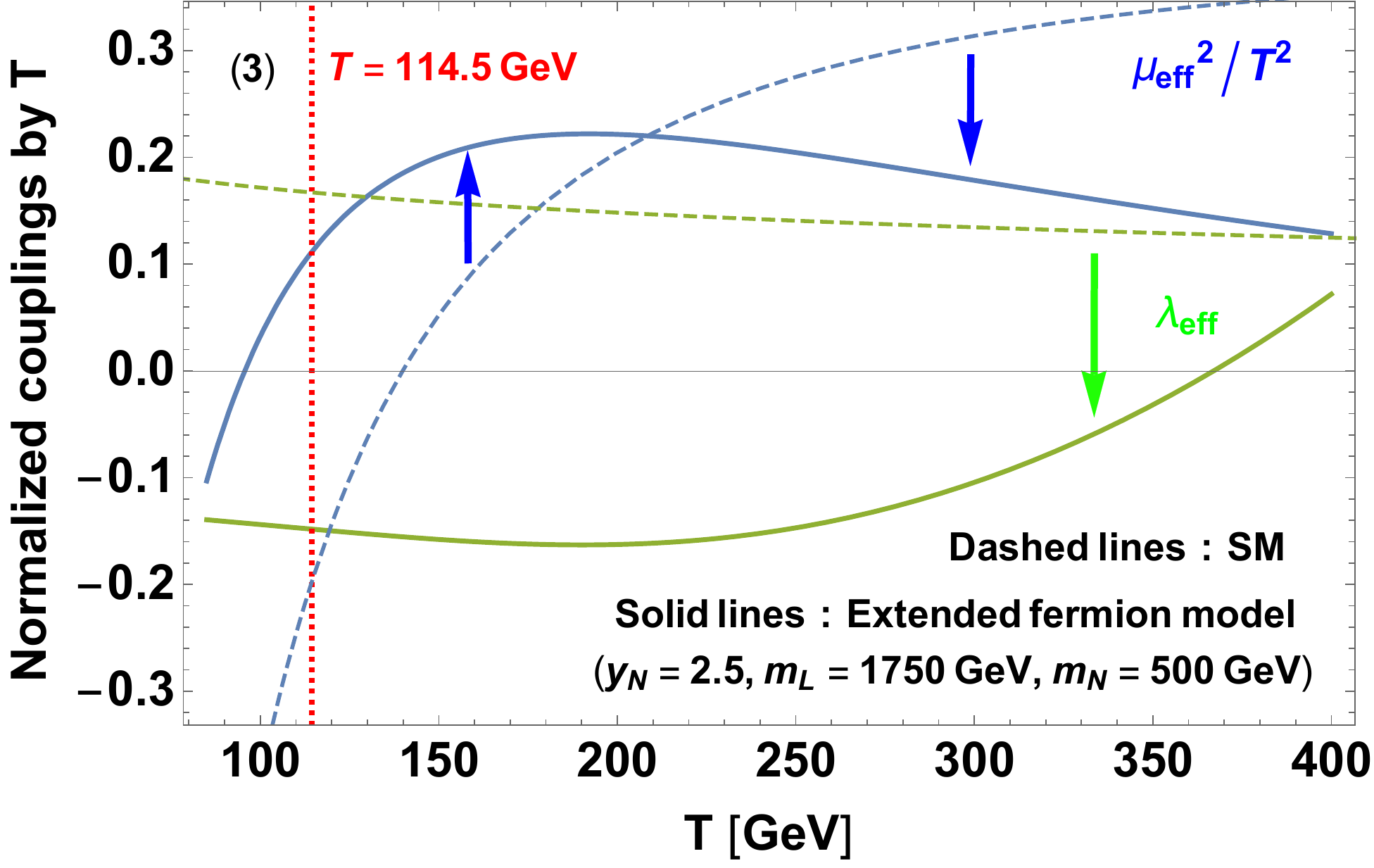}
\caption{ The case of extended fermion model with $y_N=2.5$, $m_L=1750$ GeV and $m_N=500$ GeV.
The model with this benchmark point could generate a barrier by scenario (II).
The critical temperature about this benchmark point is $T=114.5$ GeV.
Otherwise the same as Fig.~\ref{TdepI}.}
\label{TdepII}
  \end{center}
\end{figure}

 Next, we discuss the scenario (II) with another benchmark point $y_N=2.5$, $m_L=1750$ GeV and $m_N=500$ GeV. 
  Fig.~\ref{TdepII} represents results for this benchmark point by the same analyses as Fig.~\ref{TdepI}.
%%%%%%%%%%%%%%%%%%%%%%%%%%%%%%%%%%%%
%%%%%%%%%%%%%%%%%%%%%%%%%%%%%%%%%%%%
%%%%%%%%%%%%%%%%%%%%%%%%%%%%%%%%%%%%
%%%%%%%%%%%%%%%%%%%%%%%%%%%%%%%%%%%%
 This benchmark point has larger a value of $m_N/m_L$ than one for the scenario (I) in Fig.~\ref{TdepI}, and then the fermionic reduction contributions of last terms in Eqs.~(\ref{Tdepmu}--\ref{gamTfermi}) become large.
Therefore the behaviours of normalized effective couplings except for $\lambda_{eff,3}/T$ in Fig.~\ref{TdepII} are different from ones in Fig.~\ref{TdepI}.
 Since the normalized quartic coupling $\lambda_{eff}$ becomes negative around a critical temperature $T=114.5$ GeV for the model with this benchmark point, a barrier could be developed through the scenario (II).

%%%%%%%%%%%%%%%%%%%%%%%%%%%%%%%%%%%%
%%%%%%%%%%%%%%%%%%%%%%%%%%%%%%%%%%%%
\begin{figure}[htb]
  \begin{center}
\includegraphics[width=0.5\textwidth]{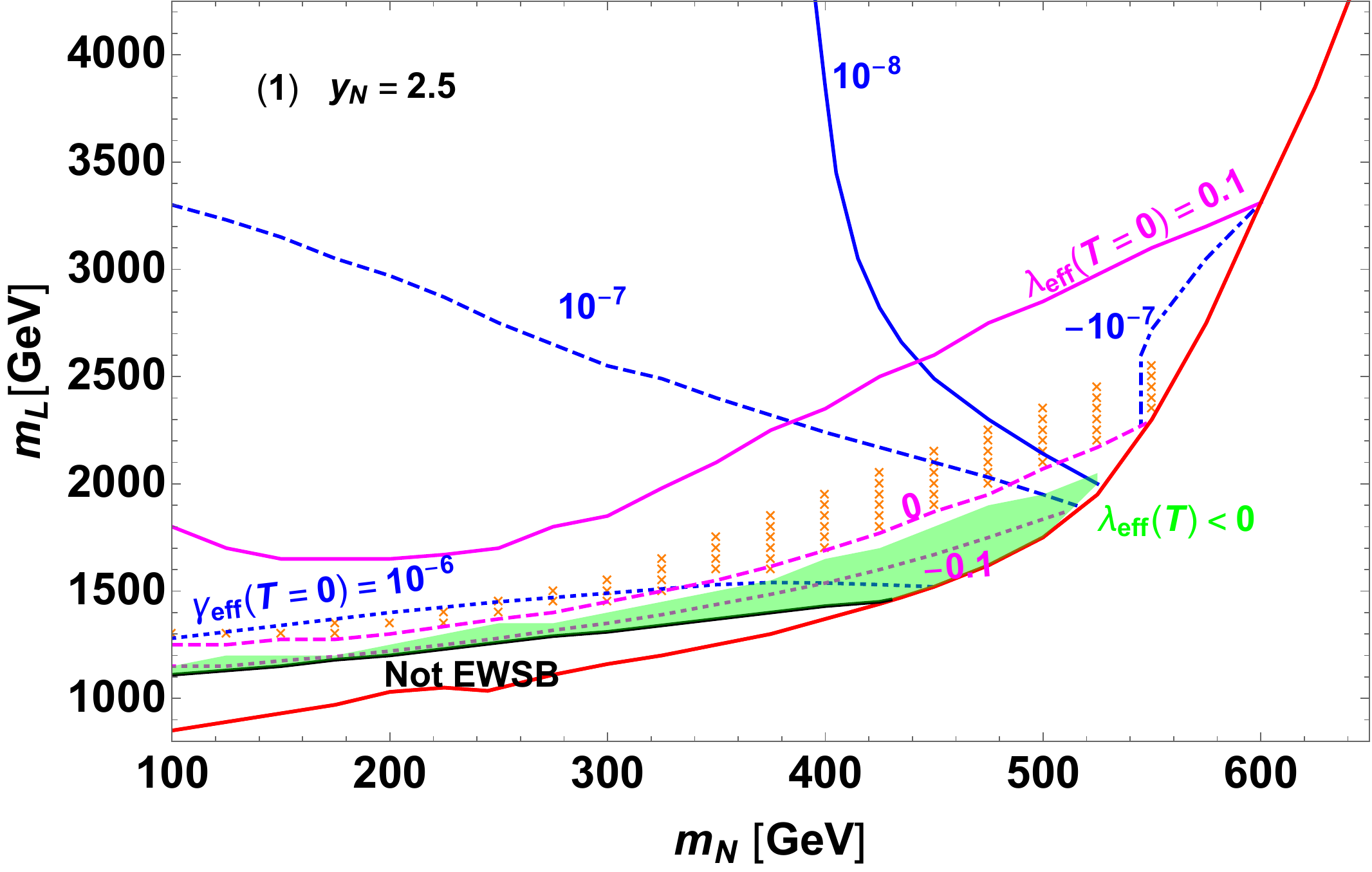}\\
\includegraphics[width=0.48\textwidth]{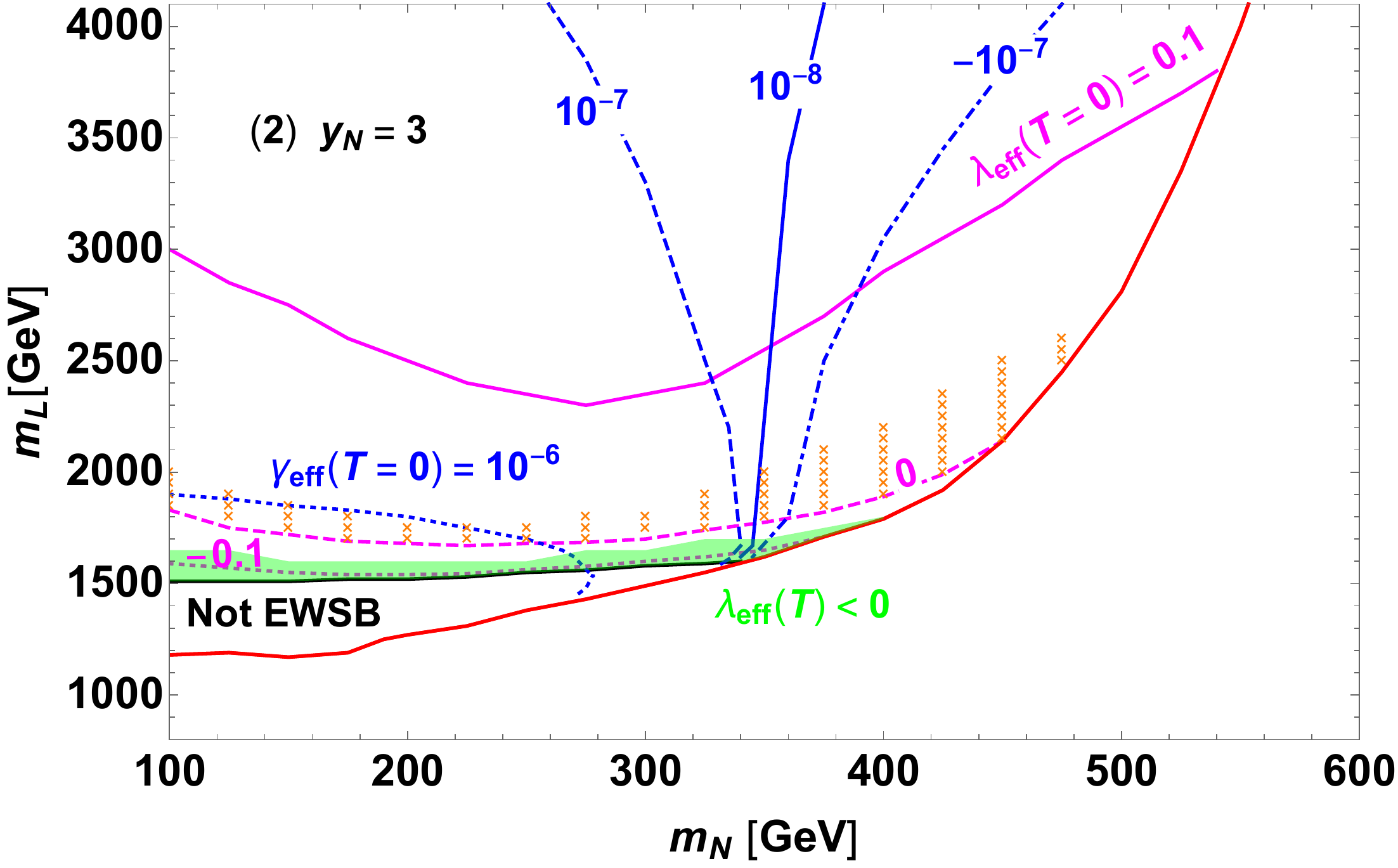}
\includegraphics[width=0.48\textwidth]{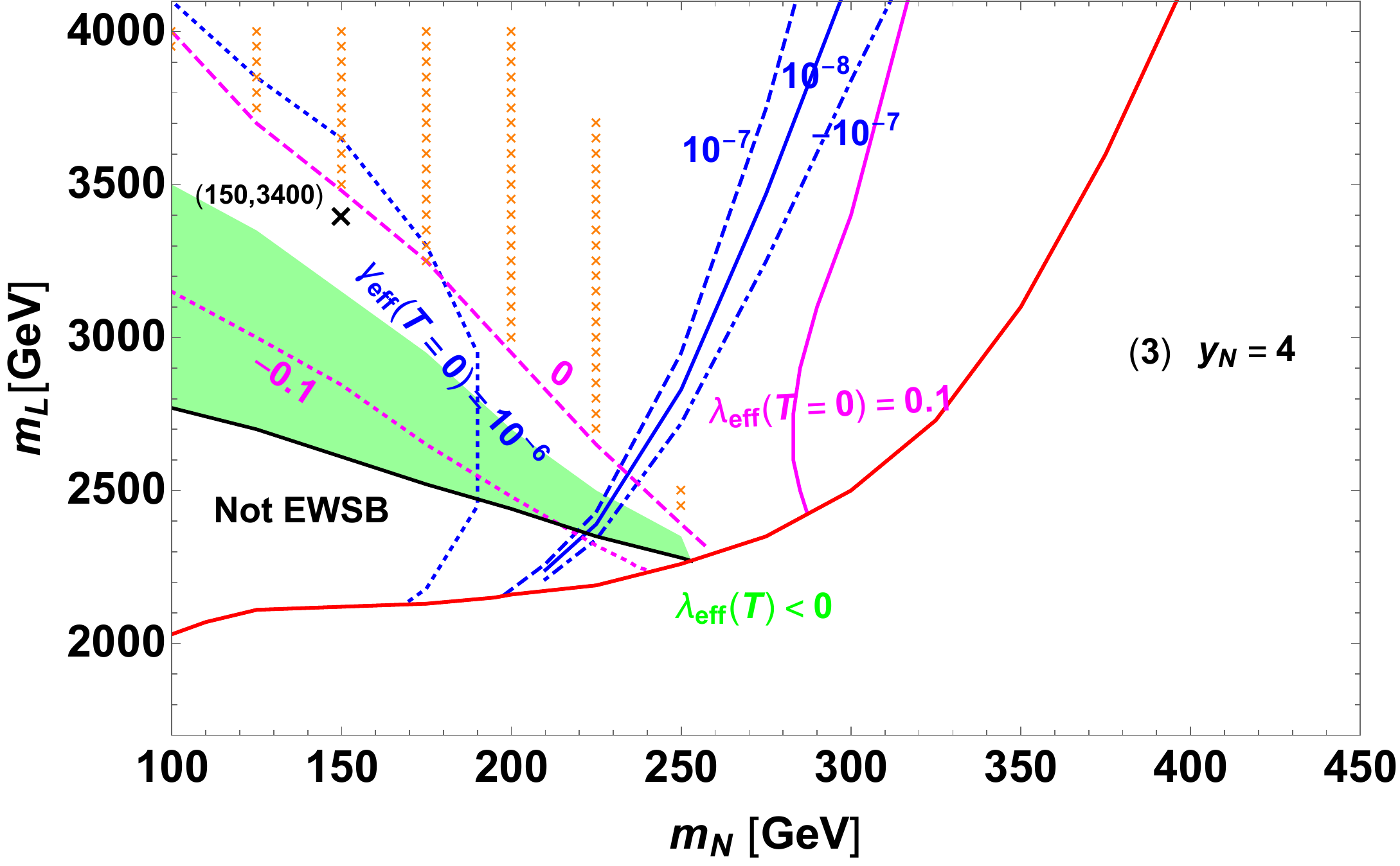}
\caption{The region represents the parameters which could generate a barrier through the scenarios (I) and (II). 
The contours are the same as Fig.~\ref{gamlamyN}.
Orange cross mark region can satisfy the conditions in Eq.~(\ref{cond1}) [scenario (I)].
In the green region, negative $\lambda_{eff}(T)$ can be achieved [scenario (II)].
Black cross mark is one of example in the vacant region between the orange mark region and the green region.
At this black cross mark, the first-order EWPT can be realized in the scenario (I) (Fig.~\ref{black}).
}
\label{12res}
  \end{center}
\end{figure}
%%%%%%%%%%%%%%%%%%%%%%%%%%%%%%%%%%%%
%%%%%%%%%%%%%%%%%%%%%%%%%%%%%%%%%%%%
%%%%%%%%%%%%%%%%%%%%%%%%%%%%%%%%%%%%
%%%%%%%%%%%%%%%%%%%%%%%%%%%%%%%%%%%%
\begin{figure}[htb]
  \begin{center}
\includegraphics[width=0.45\textwidth]{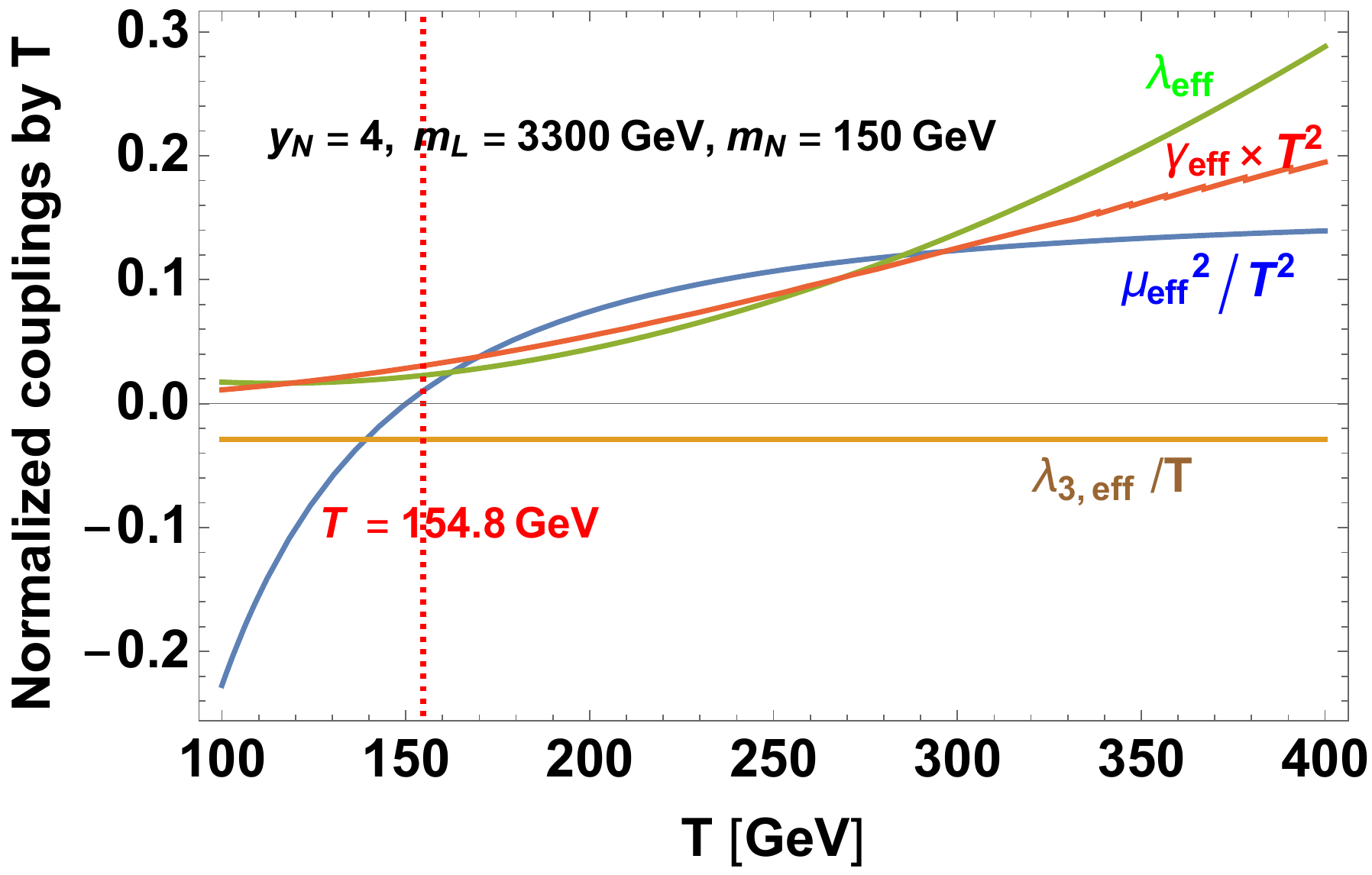}
\includegraphics[width=0.4\textwidth]{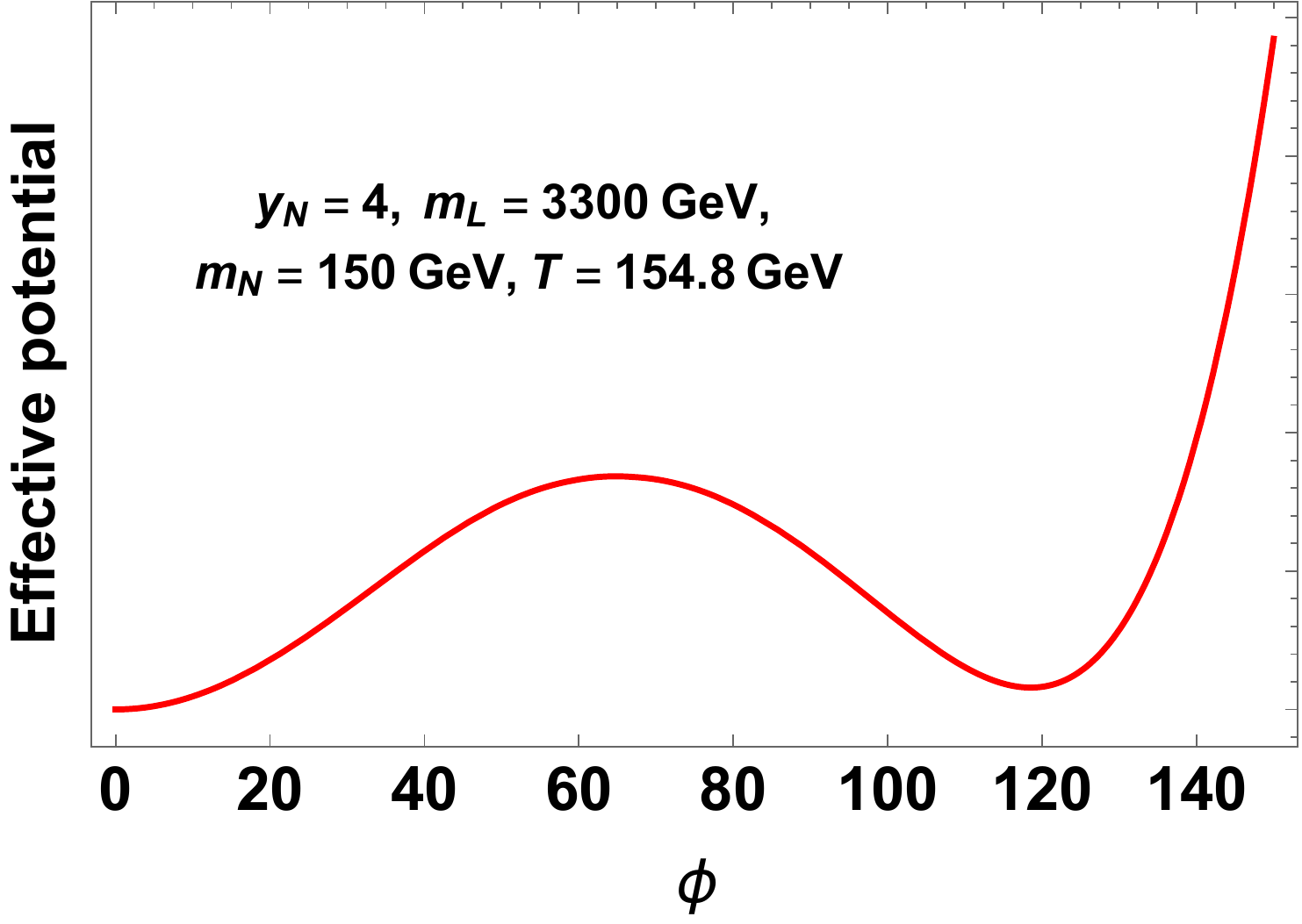}
\caption{ Left: The $T$ dependence of normalized effective couplings in the extended fermion model at the black cross mark in Fig.~\ref{12res} ($y_N=4$, $m_L=3300$ GeV and $m_N=150$ GeV).
 Right: The shape of the effective potential in the model with this benchmark point at $T=154.8$ GeV, which is a red dotted line in left figure. }
\label{black}
  \end{center}
\end{figure}
%%%%%%%%%%%%%%%%%%%%%%%%%%%%%%%%%%%%
%%%%%%%%%%%%%%%%%%%%%%%%%%%%%%%%%%%%

Now we start to scan over possible parameter regions and perform numerical calculation on the exact thermal Higgs potential. 
In practice, we adopt the following strategies: (1) in the scenario I, the quadratic term is nearly zero, while the quartic term is still positive; (2) in the scenario II, the only requirement is the negative quartic coupling near the critical temperature. 
In order to check the parameter region where a barrier is generated through the scenario (I) more precisely, we explore a temperature $T$ by the following strategy:
%%%%%%%%%%%%%%
	\begin{align} % requires amsmath; align* for no eq. number
	\label{cond1}
\frac{\mu_{eff}^2}{T^2} =0,\quad 0.7< \frac{\lambda_{eff} }{\frac{\lambda_{3, eff}}{T }}<3,\quad \lambda_{eff}^{T=0}>0.
	\end{align}
%%%%%%%%%%%%%%
From Fig.~\ref{TdepI}-(1), the scenario (I) requires $\mu_{eff}^2/T^2\sim 0$, almost the same sizes of $\lambda_{eff}$ and $\lambda_{3, eff}/T$ and positive quartic coupling $\lambda_{eff}$ to generate a sizable barrier.
Therefore we use the conditions in Eq.~(\ref{cond1}) to represent the parameter region where a sizable barrier is developed through the scenario (I).
 Also, to find the model parameters ($y_N$, $m_N$, $m_L$) for generating a barrier via the scenario (II), we numerically analyze the parameter space where negative $\lambda_{eff}(T)$ appears in $T$ = [10, 246] GeV per 10 GeV.
 For the parameter scan, we use $m_L$ = [1000, 4000] GeV per 50 GeV and $m_N$ = [100, 700] GeV per 25 GeV. 
 The results of numerical analysis for the scenarios (I) and (II) are shown in Fig.~\ref{12res}.
 The orange cross marks in this figure are parameters satisfying the condition about the scenario (I) in Eq.~(\ref{cond1}).
 The green regions represent $\lambda_{eff}(T)<0$.
 Otherwise the same as Fig.~\ref{gamlamyN}.
 There are the vacant regions between the green region and the orange mark in Fig.~\ref{12res}.
 A sizable barrier could be developed in these vacant regions, for example, Fig.~\ref{black} shows the $T$ dependence of the normalized couplings at a black cross mark in Fig.~\ref{12res}-(3). 
According to the Fig.~\ref{black}, the black cross mark could have a barrier through the scenario (I).
The different point between vacant region and orange cross marks is the sign of quartic coupling at zero temperature $\lambda_{eff}^{T=0}$, however, in the vacant region, this coupling is positive tiny amount around the critical temperature like Fig.~\ref{black}. 
 Although the quartic coupling becomes negative at zero temperature, the Higgs potential is always stable.
 The detail is discussed in appendix B.

 From Fig.~\ref{12res}, a barrier could be generated in the case that $\lambda_{eff}^{T=0}$ is small positive value or negative value, especially around magenta dashed line. 
Such a value of $\lambda_{eff}^{T=0}$ can be achieved by the large $\gamma_{eff}^{T=0}$ (second term in Eq.~(\ref{lambdaval})) or the large one-loop effect of the new light fermion (last term in Eq.~(\ref{lambdaval})). 
  The scenario (II) (green region) typically requires larger $m_N$ or smaller $m_L$ than the scenario (I) (orange crossed marks) when $y_N\leq3$.
  In the case that $y_N=4$, because the behaviour of magenta line $\lambda_{eff}^{T=0}$ could be similar to blue line $\gamma_{eff}^{T=0}$, the behaviours of orange marks and green region in Fig.~\ref{12res}-1, -2 are different from ones in Fig.~\ref{12res}-3.

  Finally, we discuss experimental constraints and testability in the parameter regions in Fig.~\ref{12res}.
  Because there is one light stable neutral fermion $N_1$ at the EW scale, there are some experimental constraints, such as the invisible $Z$ decay, the invisible $h$ decay and direct search of dark matter~\cite{Bhattacharya:2015qpa}. 
For large values of $y_N$, a sizable barrier could be easily developed in the potential, however, on the other hand, the constraint from the direct search for dark matter becomes strong.
Therefore, we consider the mixing term between the SM particle and the new fermion: $\epsilon_{N^\prime} \overline{L}_\tau H N^\prime$ and $\epsilon_{L} \overline{L} H \tau$ where $L_\tau^T=(\tau,\nu_\tau)$ and $\tau$ are related to the third generation of the SM lepton. 
The light neutral fermion $N_1$ can decay to $W^\ast \tau$ by this mixing effect.
Then a doublet fermion in the mass range of 120--790 GeV, which couples to the third generation leptons, is excluded at $95\%$ confidence level~\cite{Sirunyan:2019ofn}.
In our analyses in Fig.~\ref{12res}, we set the doublet fermion mass to $O(1$ ${\rm TeV})$ to take these experimental data into account, and we also take the mixing effects $\epsilon_{N^\prime}$ and $\epsilon_{L}$ to be tiny amounts in order to make sure the shape of effective potential does not change much and the $N_1$ particle is unstable.

\begin{figure}[t]
  \begin{center}
\includegraphics[width=0.45\textwidth]{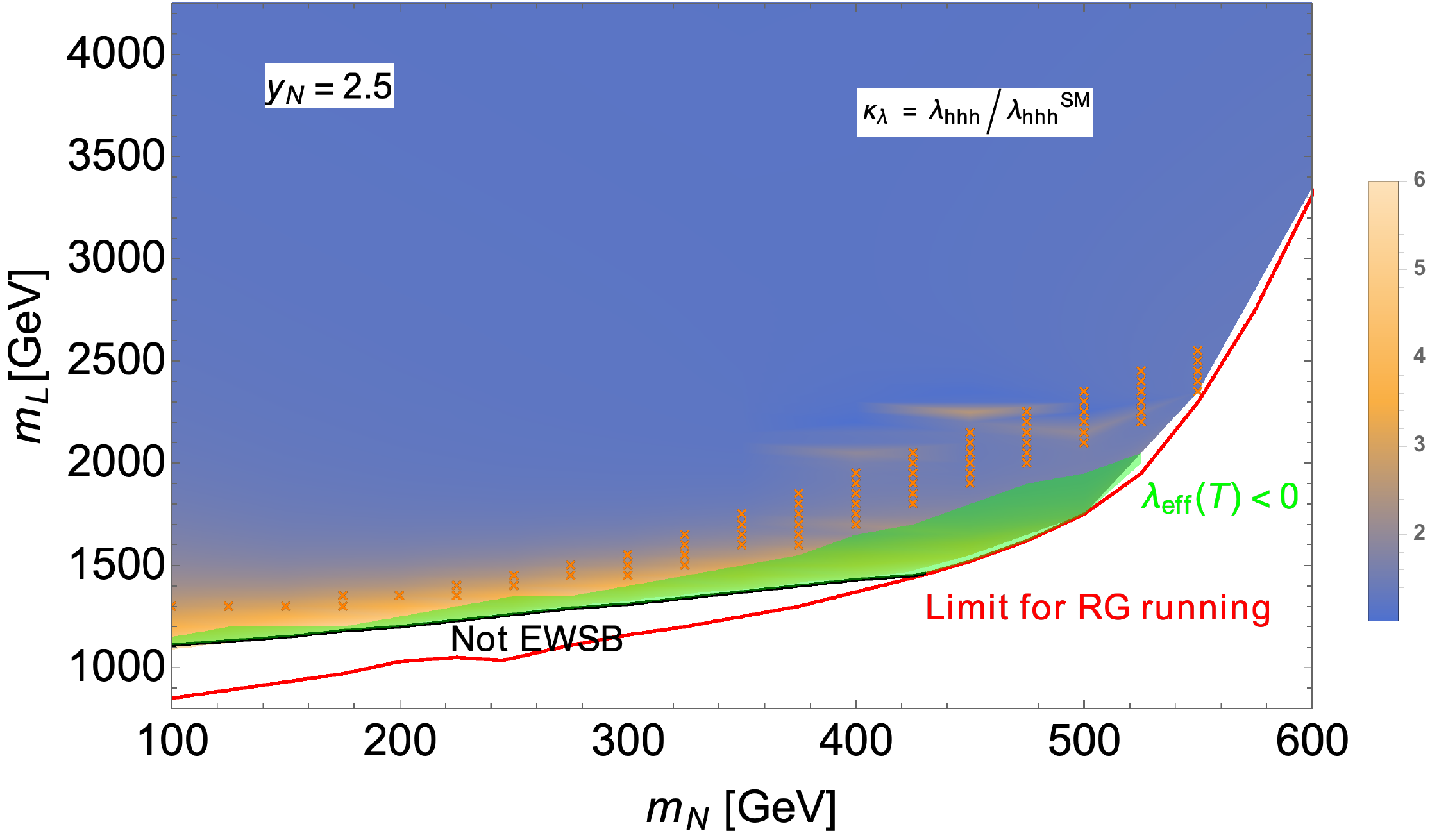}
\includegraphics[width=0.45\textwidth]{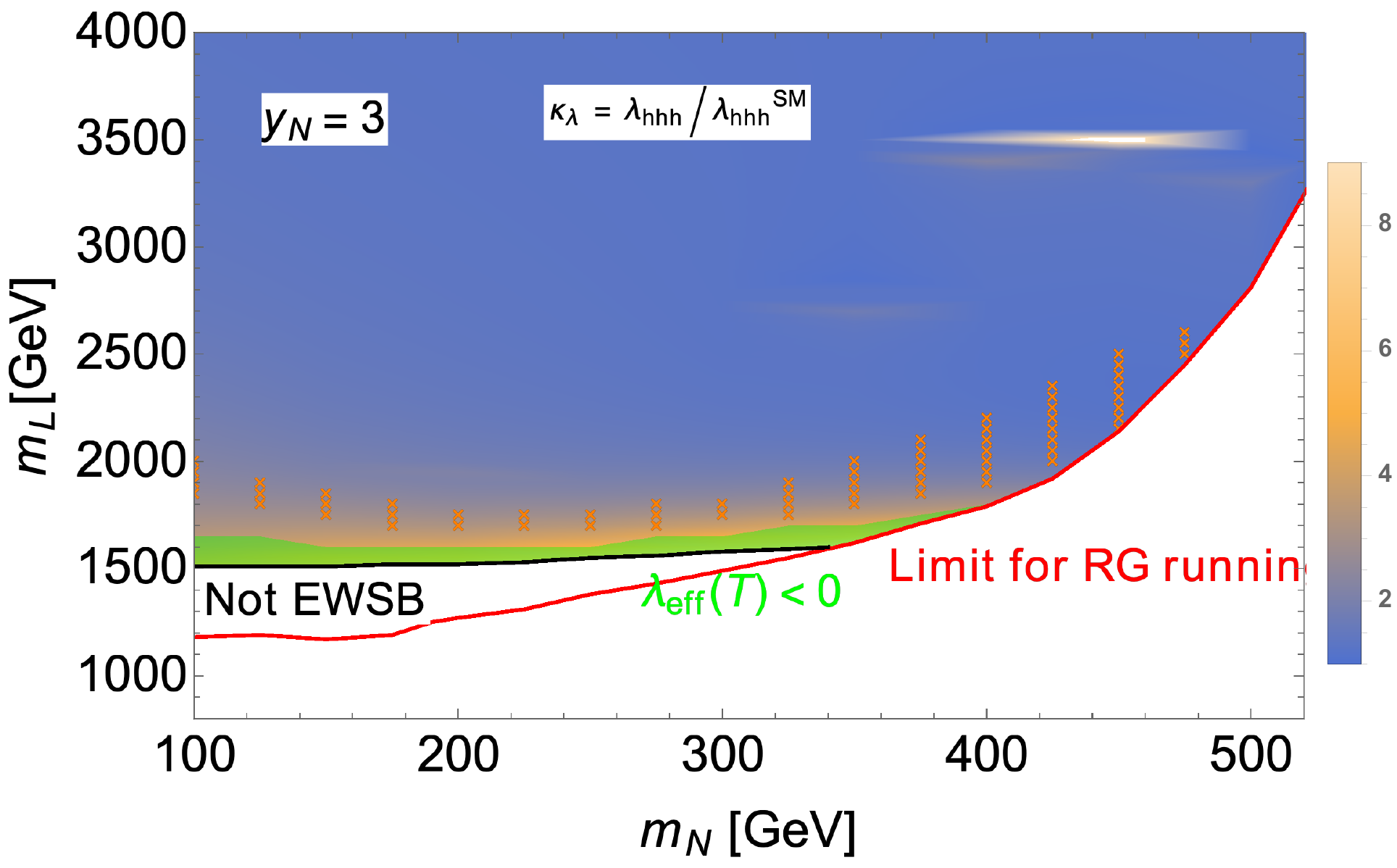}
\\
\includegraphics[width=0.45\textwidth]{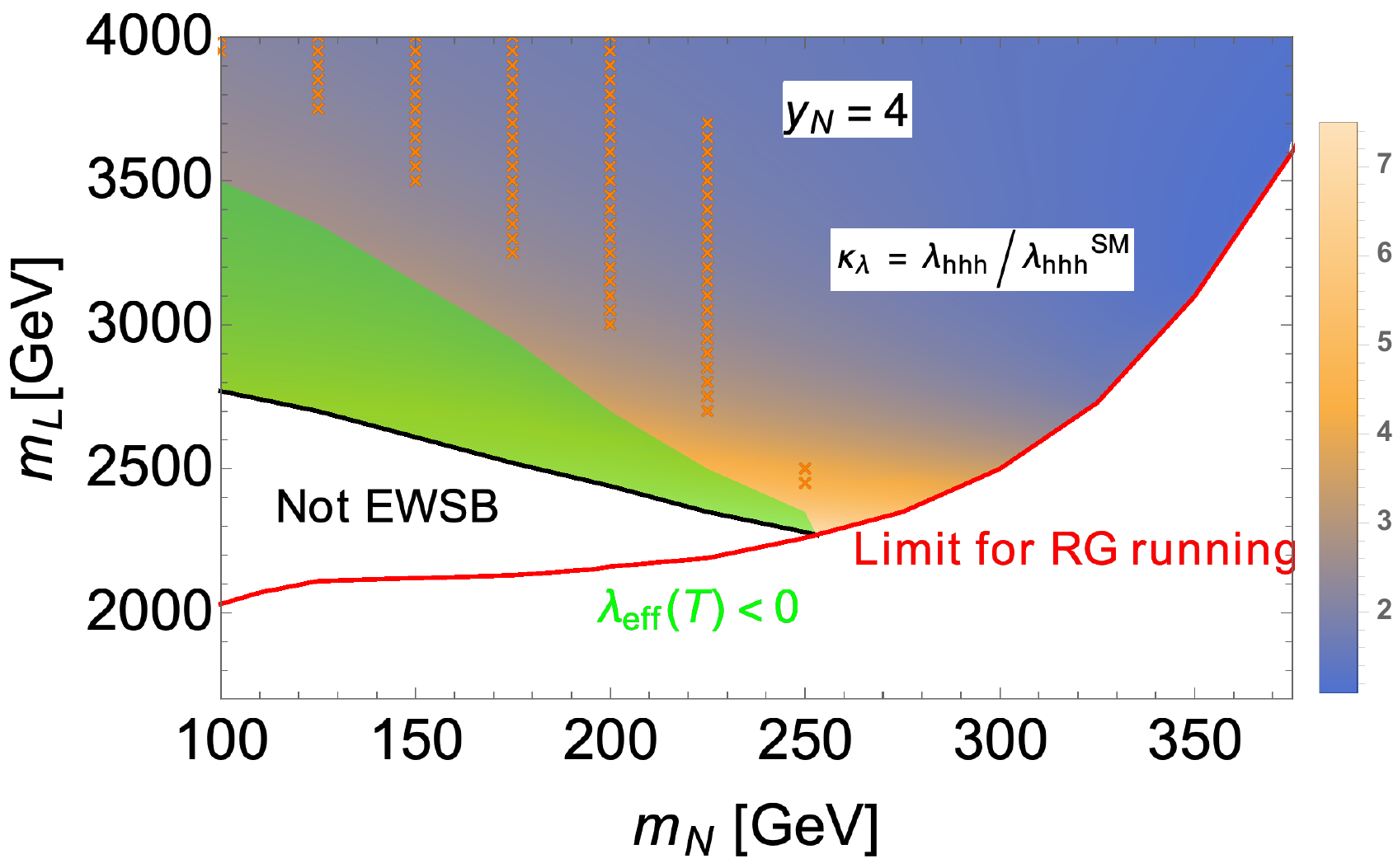}
\caption{Density plots of the ratio of triple Higgs boson coupling between the extended fermion model and the SM prediction $\kappa_\lambda = \lambda_{hhh}/\lambda_{hhh}^{SM}$. The Orange cross mark region and the green region are the same as Fig.~\ref{12res} and the black and the red lines are the same as Fig.~\ref{gamlamyN}.}
\label{hhh}
  \end{center}
\end{figure}

  The parameter regions where a barrier could be developed in Fig.~\ref{12res} could be tested by the precision measurements of the triple Higgs boson coupling $\lambda_{hhh}$.
  The ratio of $\lambda_{hhh}$ between the extended fermion model and the SM prediction is defined as 
   %%%%%%%%%%%%%%
	\begin{align} % requires amsmath; align* for no eq. number
\kappa_\lambda = \frac{\lambda_{hhh}}{\lambda_{hhh}^{SM}},
	\end{align}
%%%%%%%%%%%%%%
where $ \lambda_{hhh}$ and $\lambda_{hhh}^{SM}$ are the triple Higgs boson coupling in the extended fermion model and in the SM, which is obtained as 
   %%%%%%%%%%%%%%
	\begin{align} % requires amsmath; align* for no eq. number
	\label{hhhcalc}
 \lambda_{hhh}\equiv \left. \frac{\partial^3 V_{eff}}{\partial \varphi^3} \right|_{\varphi=v}.
	\end{align}
%%%%%%%%%%%%%%
Fig.~\ref{hhh} represents the value of $\kappa_\lambda$ for the parameter region in Fig.~\ref{12res}.
 To understanding the behaviours of density plots in the figures, we show the analytical results for new fermion contributions from Eq.~(\ref{caseBpotentialEW}) to the $hhh$ coupling.
 Based on Eqs.~(\ref{caseBpotentialEW}) and (\ref{hhhcalc}),  
 the new fermion effects in the $hhh$ coupling is roughly
   %%%%%%%%%%%%%
	\begin{align} % requires amsmath; align* for no eq. number
	\label{hhhfermion}
 \lambda_{hhh}^{\rm new \,fermion}\sim 8 \gamma v^3 + \frac{y_N^6v^3X}{\pi^2 m_L^2(1-X)^2\left(1-X - \frac{y_N^2v^2}{X m_L^2} \right)} ,
	\end{align}
%%%%%%%%%%%%%% 
and thus the value of $hhh$ coupling could be enhanced by large mass ratio $X=m_N/m_L$.
In the case of large $y_N$, the $hhh$ coupling in Eq.~\ref{hhhfermion} could be further reduced to   
   %%%%%%%%%%%%%
	\begin{align} % requires amsmath; align* for no eq. number
 \lambda_{hhh}^{\rm new \,fermion}\sim 8 \gamma v^3 - \frac{y_N^4vX^2}{\pi^2 (1-X)^2}\quad  ({\rm large\,\, } y_N).
	\end{align}
%%%%%%%%%%%%%% 
The $y_N$ dependence of second term in the $hhh$ coupling with large $y_N$ becomes $y_N^4$, which is smaller than $\gamma\sim y_N^6$ coming from Eq.~(\ref{rgamma}). 
In such a case, the behaviour of $hhh$ is similar to one of $\gamma_{eff}(T=0)$ in Eq.~(\ref{rgamma}), which is enhanced by large $m_L$ or small $m_N$. These analytical results of the $hhh$ coupling can match density plots of $\kappa_\lambda$ in Fig.~\ref{hhh}.
%%%%%%%%%%%%%%%%%%%%%%%%%%%%%%%%%%%%
%%%%%%%%%%%%%%%%%%%%%%%%%%%%%%%%%%%%
%%%%%%%%%%%%%%%%%%%%%%%%%%%%%%%%%%%%
%%%%%%%%%%%%%%%%%%%%%%%%%%%%%%%%%%%%

The color region of the ratio of $hhh$ coupling in Fig.~\ref{hhh} represents the  allowed regio by the current experimental data: $\kappa_\lambda>-2.3$ and $10.3< \kappa_\lambda$ are excluded at $95\%$ confidence level~\cite{hhh}.
 Especially, the $hhh$ coupling in the orange cross marks and green region in Fig.~\ref{12res} could be as large as  $10\%$ more than the SM prediction value.
 Such large deviation in the $hhh$ coupling can be measured at future collider experiments~\cite{Fujii:2015jha}, and thus we could check whether a first-order EWPT can be realized or not by measurements of the $hhh$ coupling.

%%%%%%%%%%%%%%%%%%%%%%%%%%%%%%%%%%%%%%%%%%%%%%%%%%%%%%%%%%%%
%%%%%%%%%%%%%%%%%%%%%%%%%%%%%%%%%%%%%%%%%%%%%%%%%%%%%%%%%%%%
%%%%%%%%%%%%%%%%%%%%%%%%%%%%%%%%%%%%%%%%%%%%%%%%%%%%%%%%%%%%
%%%%%%%%%%%%%%%%%%%%%%%%%%%%%%%%%%%%%%%%%%%%%%%%%%%%%%%%%%%%
%%%%%%%%%%%%%%%%%%%%%%%%%%%%%%%%%%%%%%%%%%%%%%%%%%%%%%%%%%%%
%%%%%%%%%%%%%%%%%%%%%%%%%%%%%%%%%%%%%%%%%%%%%%%%%%%%%%%%%%%%
\section{Summary}
%%%%%%%%%%%%%%%%%%%%%%%%%%%%%%%%%%%%%%%%%%%%%%%%%%%%%%%%%%%%
%%%%%%%%%%%%%%%%%%%%%%%%%%%%%%%%%%%%%%%%%%%%%%%%%%%%%%%%%%%%
%%%%%%%%%%%%%%%%%%%%%%%%%%%%%%%%%%%%%%%%%%%%%%%%%%%%%%%%%%%%
%%%%%%%%%%%%%%%%%%%%%%%%%%%%%%%%%%%%%%%%%%%%%%%%%%%%%%%%%%%%
%%%%%%%%%%%%%%%%%%%%%%%%%%%%%%%%%%%%%%%%%%%%%%%%%%%%%%%%%%%%
%%%%%%%%%%%%%%%%%%%%%%%%%%%%%%%%%%%%%%%%%%%%%%%%%%%%%%%%%%%%

First-order electroweak phase transition is one of the crucial ingredients to realize the baryon asymmetry of the universe in electroweak baryogenesis scenario. 
Usually it could be triggered by enhanced contribution to the cubic term in the Higgs potential from the bosonic degree of freedom.
Although fermionic degree of freedom does not contribute to cubic term in the potential, it is still possible to develop a sizable barrier in the effective potential via either decreasing quadratic/quartic terms or adding higher dimensional operators. 
In this work, we investigated the phase transition patterns in various fermionic extensions of the standard model, and realized it is necessary to introduce multi-fermions. For simplicity, we take the simplest model with more than one fermion, one doublet fermion and singlet neutral fermion.

In this model we classified the mass pattern of the fermions into three regions: (A) both fermions are at the EW scale, (B) both fermions are at the TeV scale and (C) one is at the EW scale and another is at the TeV scale. 
We found that both cases (A) and (B) could not generate a sizable barrier, while case (C) could generate one. 
Also, in case (C) we found that such a barrier can be developed if the Higgs potential satisfies two scenarios: (I) positive quadratic term $\mu^2_{eff}$, negative cubic term $\lambda_{3, eff}$ and positive quartic term $\lambda_{eff}$ or (II) positive quadratic term $\mu^2_{eff}$, negative quartic term $\lambda_{eff}$ and positive high dimensional~operator, such as dimensional~6 operator $\gamma_{eff}$.
In these scenarios, comparable size of new fermions could contribute to these effective couplings through high dimensional operators of heavy fermion and loop effect of light fermion.  
Depending on the mixing angles between new fermions and the size and sign of the quartic coupling at the zero temperature, 
some parameter regions in case (C) receive positive and negative contributions from additional fermions to $\mu^2_{eff}$ and $\lambda_{eff}$ terms, respectively. 
These effects are the source of a barrier regarding scenarios (I) or (II) and increase by large value of $m_N/m_L$. 
 
In this work, we utilize the effective field theory approach to treat the multi-scale effective potential of the two fermions model via the matching and running procedure. 
This treatment is quite general and can be extended to other new physics models with heavy fields and light fields, not limited to the fermion degree of freedom. 
Thus this general effective potential with matching and running is easy to apply to any other models with multi-scales involved in.

%====================================
	\acknowledgments
K.H. is grateful to Prof. Shinya Kanemura for helpful discussions. 
Q.H.C. is supported in part by the National Science Foundation of China under Grant Nos. 11725520, 11675002, 11635001.
Z.R. and J.H.Y. were supported by the National Science Foundation of China (NSFC) under Grants No. 12022514 and No. 11875003. J.H.Y. was also supported by National Key Research and Development Program of China Grant No. 2020YFC2201501 and the National Science Foundation of China (NSFC) under Grants No. 12047503.

%%%%%%%%%%%%%%%%%%%%%%%%%%%%%%%%%%%%%%%%%%%%%%%%%%%%%%%%%%%%
%%%%%%%%%%%%%%%%%%%%%%%%%%%%%%%%%%%%%%%%%%%%%%%%%%%%%%%%%%%%
%%%%%%%%%%%%%%%%%%%%%%%%%%%%%%%%%%%%%%%%%%%%%%%%%%%%%%%%%%%%
%%%%%%%%%%%%%%%%%%%%%%%%%%%%%%%%%%%%%%%%%%%%%%%%%%%%%%%%%%%%
%%%%%%%%%%%%%%%%%%%%%%%%%%%%%%%%%%%%%%%%%%%%%%%%%%%%%%%%%%%%
%%%%%%%%%%%%%%%%%%%%%%%%%%%%%%%%%%%%%%%%%%%%%%%%%%%%%%%%%%%%	
%%%%%%%%%%%%%%%%%%%%%%%%%%%%%%%%%%%%%%%%%%%%%%%%%%%%%%%%%%%%
%%%%%%%%%%%%%%%%%%%%%%%%%%%%%%%%%%%%%%%%%%%%%%%%%%%%%%%%%%%%
%%%%%%%%%%%%%%%%%%%%%%%%%%%%%%%%%%%%%%%%%%%%%%%%%%%%%%%%%%%%
%%%%%%%%%%%%%%%%%%%%%%%%%%%%%%%%%%%%%%%%%%%%%%%%%%%%%%%%%%%%
%%%%%%%%%%%%%%%%%%%%%%%%%%%%%%%%%%%%%%%%%%%%%%%%%%%%%%%%%%%%
%%%%%%%%%%%%%%%%%%%%%%%%%%%%%%%%%%%%%%%%%%%%%%%%%%%%%%%%%%%%
%%%%%%%%%%%%%%%%%%%%%%%%%%%%%%%%%%%%%%%%%%%%%%%%%%%%%%%%%%%%
\appendix

%%%%%%%%%%%%%%%%%%%%%%%%%%%%%%%%%%%%%%%%%%%%%%%%%%%%%%%%%%%%
%%%%%%%%%%%%%%%%%%%%%%%%%%%%%%%%%%%%%%%%%%%%%%%%%%%%%%%%%%%%
%%%%%%%%%%%%%%%%%%%%%%%%%%%%%%%%%%%%%%%%%%%%%%%%%%%%%%%%%%%%
%%%%%%%%%%%%%%%%%%%%%%%%%%%%%%%%%%%%%%%%%%%%%%%%%%%%%%%%%%%%
%%%%%%%%%%%%%%%%%%%%%%%%%%%%%%%%%%%%%%%%%%%%%%%%%%%%%%%%%%%%
%%%%%%%%%%%%%%%%%%%%%%%%%%%%%%%%%%%%%%%%%%%%%%%%%%%%%%%%%%%%
\section{Derivation of the effective potential in case B and C}
%%%%%%%%%%%%%%%%%%%%%%%%%%%%%%%%%%%%%%%%%%%%%%%%%%%%%%%%%%%%
%%%%%%%%%%%%%%%%%%%%%%%%%%%%%%%%%%%%%%%%%%%%%%%%%%%%%%%%%%%%
%%%%%%%%%%%%%%%%%%%%%%%%%%%%%%%%%%%%%%%%%%%%%%%%%%%%%%%%%%%%
%%%%%%%%%%%%%%%%%%%%%%%%%%%%%%%%%%%%%%%%%%%%%%%%%%%%%%%%%%%%
%%%%%%%%%%%%%%%%%%%%%%%%%%%%%%%%%%%%%%%%%%%%%%%%%%%%%%%%%%%%
%%%%%%%%%%%%%%%%%%%%%%%%%%%%%%%%%%%%%%%%%%%%%%%%%%%%%%%%%%%%
In this appendix, we discuss the detail of the treatment of the effective potential with multi-scales in section III. 
%%%%%%%%%%%%%%%%%%%%%%%%%%%%%%%%%%%%%%%%%%%%%%%%%%%%%%%%%%%%
%%%%%%%%%%%%%%%%%%%%%%%%%%%%%%%%%%%%%%%%%%%%%%%%%%%%%%%%%%%%
%%%%%%%%%%%%%%%%%%%%%%%%%%%%%%%%%%%%%%%%%%%%%%%%%%%%%%%%%%%%
\subsection{Case B-1: $m_L \gg m_N\gg y_N v$}
%%%%%%%%%%%%%%%%%%%%%%%%%%%%%%%%%%%%%%%%%%%%%%%%%%%%%%%%%%%%
%%%%%%%%%%%%%%%%%%%%%%%%%%%%%%%%%%%%%%%%%%%%%%%%%%%%%%%%%%%%
%%%%%%%%%%%%%%%%%%%%%%%%%%%%%%%%%%%%%%%%%%%%%%%%%%%%%%%%%%%%
  In this parameter, we consider two-step matchings: (i) very HE scale $m_L$ and HE scale $m_N$ and (ii) the HE scale and LE scale. 
 The Lagrangian at the very HE scale is Eq.~(\ref{Lagrang}).
  The field dependent masses are the same as Eq.~(\ref{fieldMASS}).
The very HE and HE effective potentials at high matching scale $Q_{M, heavy}$ are
%%%%%%%%%%%%%%%%%%%%%%%%%%%%%%%%%%%%
\begin{align}
\label{Veffallatveryheavy}
V_{eff}^{very\,HE} &= \frac{\mu^2}{2}\varphi^2 + \frac{\lambda}{4}\varphi^4  + \sum_{i=N_1, N_2, SM}\frac{n_i}{64\pi^2} M_i(\varphi)^4\left(\log\frac{M_i(\varphi)^2 }{Q_{M, heavy}^2}-c_i\right)
\nonumber\\
V_{eff}^{HE} &= \frac{(\mu^2)^{HE}}{2}\varphi^2 + \frac{\lambda^{HE}}{4}\varphi^4  + \sum_{i=N_1, SM} \frac{n_i}{64\pi^2} M_i(\varphi)^4\left(\log\frac{M_i(\varphi)^2 }{Q_{M, heavy}^2}-c_i\right)\nonumber\\
&\quad+\frac{\gamma^{HE}}{6}\varphi^6+\frac{\delta^{HE}}{8}\varphi^8+\frac{\epsilon^{HE}}{10}\varphi^{10}.
\end{align}
%%%%%%%%%%%%%%%%%%%%%%%%%%%%%%%%%%%%
By using the matching procedure in section 2, the parameters in the HE scale potential at $Q_{M, heavy}$ are given as
 %%%%%%%%%%%%%%%%%%%%%%%%%%%%%%%%%%%%
\begin{align}
\label{appendixB1}
&(\mu^2)^{HE}=\mu^2 + \frac{y_N^2m_L^3}{4\pi^2(m_L-m_N)}  ,\quad
\lambda^{HE}= \lambda - \frac{m_L^2(m_L+m_N)y_N^4}{8\pi^2(m_L-m_N)^3}, \nonumber\\
 \gamma^{HE} &=  \frac{m_Ly_N^6(m_L^2+7m_Lm_N-2m_N^2)}{16\pi^2(m_L-m_N)^5}, \quad\delta^{HE} = - \frac{y_N^8(7m_L^3+27m_L^2m_N - 4m_N^3)}{48\pi^2(m_L-m_N)^7},\nonumber\\
& \epsilon^{HE} = - \frac{y_N^{10}(107m_L^4 + 342 m_L^3m_N+ 42 m_L^2m_N^2 - 68 m_L m_N^3 - 3m_N^4)}{384\pi^2m_L(m_L-m_N)^9},
\end{align}
%%%%%%%%%%%%%%%%%%%%%%%%%%%%%%%%%%%%
 where we set $Q_{M, heavy}$ as $m_L$. 
    The RG running effect from the HE scale $Q_{HE}$ to the LE scale $Q_{LE}$ roughly is
  %%%%%%%%%%%%%%%%%%%%%%%%%%%%%%%%%%%%
\begin{align}
\label{RUNNINGofcoupling}
\beta_{n\varphi} = \frac{d \lambda_{(n)}}{d \log \mu}\quad \to\quad\lambda_{(n)}^{RG} \simeq \int^{Q_{HE}}_{Q_{LE}} \beta_{n\varphi} d\log\mu = - \beta_{n\varphi} \log\frac{Q_{HE}}{Q_{LE}},
\end{align}
%%%%%%%%%%%%%%%%%%%%%%%%%%%%%%%%%%%% 
 In order to obtain the beta function, we treat the flavor state as mass eigenstate, and then the Lagrangian of new fermion at the HE scale is given by
%%%%%%%%%%%%%%%%%%%%%%%%%%%%%%%%%%%%
\begin{align}
\label{LAGra}
-{\cal L}_{VLL}^{HE}\supset &\overline{N}_1(-i  \,/ \hspace{-0.2cm}\partial + m_{N}) N_1 - M_{2\varphi}^{N}  \overline{N}_1 N_1 \left( \varphi^2+ \varphi_z^2 + 2\varphi^+ \varphi^-\right) 
\end{align}
%%%%%%%%%%%%%%%%%%%%%%%%%%%%%%%%%%%%  
where $N_1$ is mass eigenstate of light fermion and 
 %%%%%%%%%%%%%%%%%%%%%%%%%%%%%%%%%%%%
\begin{align}
M_{2\varphi}^{N} &=  \frac{y_N^2}{2(m_L-m_N)} .
\end{align}
%%%%%%%%%%%%%%%%%%%%%%%%%%%%%%%%%%%%
 By using the Lagrangian of Eq.~(\ref{LAGra}), the beta functions of dimensional~6, 8 and 10 are
%%%%%%%%%%%%%%%%%%%%%%%%%%%%%%%%%%%%
\begin{align}
\label{betaHEscale}
\beta_{6\varphi}^{HE}&=  \frac{3m_N \left(M_{2\varphi}^{N}\right)^3}{\pi^2} -6 \frac{\gamma^{HE}}{16\pi^2} \left(-3 y_t^2  + \frac{3}{4}g_1^2+ \frac{9}{4}g_2^2 \right),\nonumber\\
\beta_{8\varphi}^{HE}&= - \frac{ \left(M_{2\varphi}^{N}\right)^4}{\pi^2}  -8 \frac{\delta^{HE}}{16\pi^2} \left(-3 y_t^2  + \frac{3}{4}g_1^2+ \frac{9}{4}g_2^2 \right), \nonumber\\ \beta_{10\varphi}^{HE}&=  -10 \frac{\epsilon^{HE}}{16\pi^2} \left(-3 y_t^2  + \frac{3}{4}g_1^2+ \frac{9}{4}g_2^2 \right).
\end{align}
%%%%%%%%%%%%%%%%%%%%%%%%%%%%%%%%%%%% 
 The second terms in the dimensional~6 and 8 operators and the right-hand side of dimensional~10 operator come from $ \gamma_\varphi $ function in Eq.~(\ref{CSE}).
Furthermore, the first terms in the dimensional~6 and 8 operators can be obtained by the first term in Eq.~(\ref{CSE}).
The SM fields, such as the gauge boson and the top quark, do not contribute to these operators through the first term in Eq.~(\ref{CSE}).
 The dimensional~6, 8 and 10 operators at low matching scale $Q_{M, LE}$, which is related to second step matching (ii) the HE scale and the LE scale, are roughly 
 %%%%%%%%%%%%%%%%%%%%%%%%%%%%%%%%%%%%
\begin{align}
\gamma^{HE, M}=  \gamma^{HE}& -\beta_{6\varphi}^{HE} \log\frac{m_L}{Q_{M, LE}},\quad
\delta^{HE, M} =  \delta^{HE} -\beta_{8\varphi}^{HE} \log\frac{m_L}{Q_{M, LE}} \nonumber\\
&\quad \epsilon^{HE, M}  = \epsilon^{HE} -\beta_{10\varphi}^{HE} \log\frac{m_L}{Q_{M, LE}} 
\end{align}
 %%%%%%%%%%%%%%%%%%%%%%%%%%%%%%%%%%%
The HE and LE effective potentials at $Q_{M, LE}$ are
%%%%%%%%%%%%%%%%%%%%%%%%%%%%%%%%%%%%
\begin{align}
V_{eff}^{HE} &= \frac{(\mu^2)^{HE, M}}{2}\varphi^2 + \frac{\lambda^{HE, M}}{4}\varphi^4  + \sum_{i=N_1, SM} \frac{n_i}{64\pi^2} M_i(\varphi)^4\left(\log\frac{M_i(\varphi)^2 }{Q_{M, LE}^2}-c_i\right)\nonumber\\
&\quad+\frac{\gamma^{HE, M}}{6}\varphi^6+\frac{\delta^{HE, M}}{8}\varphi^8+\frac{\epsilon^{HE, M}}{10}\varphi^{10}.\nonumber\\
V_{eff}^{LE} &= \frac{(\mu^2)^{LE}}{2}\varphi^2 + \frac{\lambda^{LE}}{4}\varphi^4  + \sum_{i=SM} \frac{n_i}{64\pi^2} M_i(\varphi)^4\left(\log\frac{M_i(\varphi)^2 }{Q_{M, LE}^2}-c_i\right)\nonumber\\
&\quad+\frac{\gamma^{LE}}{6}\varphi^6+\frac{\delta^{LE}}{8}\varphi^8+\frac{\epsilon^{LE}}{10}\varphi^{10}.
\end{align}
%%%%%%%%%%%%%%%%%%%%%%%%%%%%%%%%%%%%
In the following, we set $Q_{M, LE}$ as $m_N$.
The LE parameters at this scale are obtained by the matching conditions:
 %%%%%%%%%%%%%%%%%%%%%%%%%%%%%%%%%%%%
\begin{align}
\label{B1thEW}
&(\mu^2)^{LE}=(\mu^2)^{HE, M} - \frac{y_N^2m_N^3}{4\pi^2(m_L-m_N)}  ,\quad
\lambda^{LE}= \lambda^{HE, M}  + \frac{m_N^2(m_L+m_N)y_N^4}{8\pi^2(m_L-m_N)^3}, \nonumber\\
 \gamma^{LE} &= \gamma^{HE, M} + \frac{m_Ny_N^6(2m_L^2-7m_Lm_N-m_N^2)}{16\pi^2(m_L-m_N)^5}, \quad\delta^{HE} = \delta^{HE, M} - \frac{y_N^8(4m_L^3-27m_Lm_N^2 - 7m_N^3)}{48\pi^2(m_L-m_N)^7},\nonumber\\
& \epsilon^{LE} = \epsilon^{HE, M} + \frac{y_N^{10}(3m_L^4 + 68 m_L^3m_N- 42 m_L^2m_N^2 - 342 m_L m_N^3 -107m_N^4)}{384\pi^2m_N(m_L-m_N)^9}.
\end{align}
%%%%%%%%%%%%%%%%%%%%%%%%%%%%%%%%%%%%
The Lagrangian at the LE scale is the same as the SM, so the beta functions of the high dimensional~operators are
%%%%%%%%%%%%%%%%%%%%%%%%%%%%%%%%%%%%
\begin{align}
\label{betaLNheavyLE}
\beta_{6\varphi}^{LE}&=  -6 \frac{\gamma^{LE}}{16\pi^2} \left(-3 y_t^2  + \frac{3}{4}g_1^2+ \frac{9}{4}g_2^2 \right),\quad \beta_{8\varphi}^{LE}=  -8 \frac{\delta^{LE}}{16\pi^2} \left(-3 y_t^2  + \frac{3}{4}g_1^2+ \frac{9}{4}g_2^2 \right), \nonumber\\ 
&\quad\quad \beta_{10\varphi}^{LE}=  -10 \frac{\epsilon^{HE}}{16\pi^2} \left(-3 y_t^2  + \frac{3}{4}g_1^2+ \frac{9}{4}g_2^2 \right).
\end{align}
%%%%%%%%%%%%%%%%%%%%%%%%%%%%%%%%%%%% 
By using them, we can obtain the effective potential at EW scale in Eq.~(\ref{LEscalepotential}).

%%%%%%%%%%%%%%%%%%%%%%%%%%%%%%%%%%%%%%%%%%%%%%%%%%%%%%%%%%%%
%%%%%%%%%%%%%%%%%%%%%%%%%%%%%%%%%%%%%%%%%%%%%%%%%%%%%%%%%%%%
%%%%%%%%%%%%%%%%%%%%%%%%%%%%%%%%%%%%%%%%%%%%%%%%%%%%%%%%%%%%
\subsection{Case B-2: $m_L\sim m_N\gg y_N v$}
%%%%%%%%%%%%%%%%%%%%%%%%%%%%%%%%%%%%%%%%%%%%%%%%%%%%%%%%%%%%
%%%%%%%%%%%%%%%%%%%%%%%%%%%%%%%%%%%%%%%%%%%%%%%%%%%%%%%%%%%%
%%%%%%%%%%%%%%%%%%%%%%%%%%%%%%%%%%%%%%%%%%%%%%%%%%%%%%%%%%%%
In this parameter case, we need one-step matching between HE and LE scale to integrate out additional doublet and singlet fermions.
The HE and LE effective potentials at the matching scale $Q_M$ are
%%%%%%%%%%%%%%%%%%%%%%%%%%%%%%%%%%%%
\begin{align}
\label{Veffallheavy}
V_{eff}^{HE} &= \frac{\mu^2}{2}\varphi^2 + \frac{\lambda}{4}\varphi^4  + \sum_{i=N_1, N_2, SM}\frac{n_i}{64\pi^2} M_i(\varphi)^4\left(\log\frac{M_i(\varphi)^2 }{Q^2}-c_i\right)
\nonumber\\
V_{eff}^{LE} &= \frac{(\mu^2)^{LE}}{2}\varphi^2 + \frac{\lambda^{LE}}{4}\varphi^4  + \sum_{i=SM} \frac{n_i}{64\pi^2} M_i(\varphi)^4\left(\log\frac{M_i(\varphi)^2 }{Q^2}-c_i\right)\nonumber\\
&\quad+\frac{\gamma^{LE}}{6}\varphi^6+\frac{\delta^{LE}}{8}\varphi^8+\frac{\epsilon^{LE}}{10}\varphi^{10}.
\end{align}
%%%%%%%%%%%%%%%%%%%%%%%%%%%%%%%%%%%%
 The parameters of the potential at the LE scale can be obtained by the matching conditions:
 %%%%%%%%%%%%%%%%%%%%%%%%%%%%%%%%%%%%
\begin{align}
(\mu^2)^{LE}&=\mu^2 + \frac{y_N^2m_L^2}{4\pi^2}  ,\quad
\lambda^{LE}= \lambda - \frac{y_N^4}{3\pi^2}  ,\quad \gamma^{LE} =  \frac{y_N^6}{160\pi^2m_L^2}, \nonumber\\
&\quad\delta^{LE} =  \frac{y_N^8}{2240\pi^2m_L^4},\quad\epsilon^{LE} =  \frac{y_N^{10}}{16128\pi^2m_L^6}.
\end{align}
%%%%%%%%%%%%%%%%%%%%%%%%%%%%%%%%%%%%
where we set the matching scale $Q_M$ as $m_L$.
At the LE scale, the Lagrangian only has the SM fields.
Therefore the beta function of high dimensional~operators are the same as Eq.~(\ref{betaLNheavyLE}):
 %%%%%%%%%%%%%%%%%%%%%%%%%%%%%%%%%%%%
\begin{align}
\label{betaheavy}
 \beta_{6\varphi}&= -6 \frac{\gamma^{LE}}{16\pi^2} \left(-3 y_t^2  + \frac{3}{4}g_1^2+ \frac{9}{4}g_2^2 \right),\quad
\beta_{8\varphi}=  -8 \frac{\delta^{LE}}{16\pi^2} \left(-3 y_t^2  + \frac{3}{4}g_1^2+ \frac{9}{4}g_2^2 \right),\nonumber\\
 \beta_{10\varphi}&=  -10 \frac{\epsilon^{LE}}{16\pi^2} \left(-3 y_t^2  + \frac{3}{4}g_1^2+ \frac{9}{4}g_2^2 \right).
\end{align}
%%%%%%%%%%%%%%%%%%%%%%%%%%%%%%%%%%%% 
By using them, we can obtain the effective potential at EW scale in Eq.~(\ref{LEscalepotential}).

%%%%%%%%%%%%%%%%%%%%%%%%%%%%%%%%%%%%%%%%%%%%%%%%%%%%%%%%%%%%
%%%%%%%%%%%%%%%%%%%%%%%%%%%%%%%%%%%%%%%%%%%%%%%%%%%%%%%%%%%%
%%%%%%%%%%%%%%%%%%%%%%%%%%%%%%%%%%%%%%%%%%%%%%%%%%%%%%%%%%%%
\subsection{Case C: $m_L \gg m_N \sim y_N v$}
%%%%%%%%%%%%%%%%%%%%%%%%%%%%%%%%%%%%%%%%%%%%%%%%%%%%%%%%%%%%
%%%%%%%%%%%%%%%%%%%%%%%%%%%%%%%%%%%%%%%%%%%%%%%%%%%%%%%%%%%%
%%%%%%%%%%%%%%%%%%%%%%%%%%%%%%%%%%%%%%%%%%%%%%%%%%%%%%%%%%%%
In this case, $m_L$ is at the TeV scale and $m_N$ is at the EW scale.
The effective potentials at the HE and the LE scale are
%%%%%%%%%%%%%%%%%%%%%%%%%%%%%%%%%%%%
\begin{align}
V_{eff}^{HE} &= \frac{\mu^2}{2}\varphi^2 + \frac{\lambda}{4}\varphi^4  + \sum_{i=N_1, N_2, SM}\frac{n_i}{64\pi^2} M_i(\varphi)^4\left(\log\frac{M_i(\varphi)^2 }{Q^2}-c_i\right)
\nonumber\\
V_{eff}^{LE} &= \frac{(\mu^2)^{LE}}{2}\varphi^2 + \frac{\lambda^{LE}}{4}\varphi^4  + \sum_{i=N_1, SM} \frac{n_i}{64\pi^2} M_i(\varphi)^4\left(\log\frac{M_i(\varphi)^2 }{Q^2}-c_i\right)\nonumber\\
&\quad+\frac{\gamma^{LE}}{6}\varphi^6+\frac{\delta^{LE}}{8}\varphi^8+\frac{\epsilon^{LE}}{10}\varphi^{10}
\end{align}
%%%%%%%%%%%%%%%%%%%%%%%%%%%%%%%%%%%%
 and the field dependent masses are the same as Eq.~(\ref{fieldMASS}).
 At the matching scale $Q_M$, the parameters $\varphi^n$ terms at the HE scale match ones at the LE scale by the matching conditions:
 %%%%%%%%%%%%%%%%%%%%%%%%%%%%%%%%%%%%
\begin{align}
\label{thresh}
(\mu^2)^{LE}&=\mu^2 + \frac{y_N^2m_L^3}{4\pi^2(m_L-m_N)}  ,\quad
\lambda^{LE}= \lambda - \frac{y_N^4m_L^2(m_L+m_N)}{8\pi^2(m_L-m_N)^3}  , \nonumber\\
&\gamma^{LE} = \gamma^{th},\quad\delta^{LE} = \delta^{th},\quad\epsilon^{LE} = \epsilon^{th},
\end{align}
%%%%%%%%%%%%%%%%%%%%%%%%%%%%%%%%%%%%
where we set $Q_M$ as $m_L$.
The threshold effects of high dimensional~operators $\gamma^{th}$, $\delta^{th}$ and $\epsilon^{th}$ are given by 
 %%%%%%%%%%%%%%%%%%%%%%%%%%%%%%%%%%%%%%%%%%%%%%%%%%%%%%%%%%%%
	\begin{align} % requires amsmath; align* for no eq. number
	\label{darac}
\gamma^{th}&=  \frac{m_L y_N^6}{16\pi^2(m_L-m_N)^5}(m_L^2+7m_Nm_L - 2m_N^2) ,\\
 \quad \delta^{th}&=  - \frac{y_N^8}{48 \pi^2(m_L-m_N)^7}(7m_L^3+27m_Nm_L^2 - 4m_N^3),\\
 \quad \epsilon^{th}&=  \frac{y_N^{10}}{384\pi^2m_L(m_L-m_N)^9}(107m_L^4+342m_L^3m_N+42m_L^2m_N^2 - 68 m_Lm_N^3 - 3m_N^4),
	\end{align}
%%%%%%%%%%%%%%%%%%%%%%%%%%%%%%%%%%%%%%%%%%
 The threshold effects of $\gamma^{th}$ and $\delta^{th}$ are the same ones in Ref.~\cite{Davoudiasl:2012tu}.
 In order to obtain the beta functions, we use the Lagrangian in Eq.~(\ref{LAGra}). 
 The beta functions of dimensional~6, 8 and 10 operators are the same as Eq.~(\ref{betaHEscale}):
 %%%%%%%%%%%%%%%%%%%%%%%%%%%%%%%%%%%%
\begin{align}
\label{betaHE6810}
\beta_{6\varphi}&=  \frac{3m_N \left(M_{2\varphi}^{N}\right)^3}{\pi^2} -6 \frac{\gamma^{th}}{16\pi^2} \left(-3 y_t^2  + \frac{3}{4}g_1^2+ \frac{9}{4}g_2^2 \right),\nonumber\\
\beta_{8\varphi}&= - \frac{ \left(M_{2\varphi}^{N}\right)^4}{\pi^2}  -8 \frac{\delta^{th}}{16\pi^2} \left(-3 y_t^2  + \frac{3}{4}g_1^2+ \frac{9}{4}g_2^2 \right), \nonumber\\
 \beta_{10\varphi}&=  -10 \frac{\epsilon^{th}}{16\pi^2} \left(-3 y_t^2  + \frac{3}{4}g_1^2+ \frac{9}{4}g_2^2 \right).
\end{align}
%%%%%%%%%%%%%%%%%%%%%%%%%%%%%%%%%%%% 
By using them, we can obtain the effective potential at EW scale in Eq.~(\ref{LEscalepotential}).

%%%%%%%%%%%%%%%%%%%%%%%%%%%%%%%%%%%%%%%%%%%%%%%%%%%%%%%%%%%%
%%%%%%%%%%%%%%%%%%%%%%%%%%%%%%%%%%%%%%%%%%%%%%%%%%%%%%%%%%%%
%%%%%%%%%%%%%%%%%%%%%%%%%%%%%%%%%%%%%%%%%%%%%%%%%%%%%%%%%%%%
%%%%%%%%%%%%%%%%%%%%%%%%%%%%%%%%%%%%%%%%%%%%%%%%%%%%%%%%%%%%
%%%%%%%%%%%%%%%%%%%%%%%%%%%%%%%%%%%%%%%%%%%%%%%%%%%%%%%%%%%%
%%%%%%%%%%%%%%%%%%%%%%%%%%%%%%%%%%%%%%%%%%%%%%%%%%%%%%%%%%%%
\section{Vacuum stability and Landau pole}
%%%%%%%%%%%%%%%%%%%%%%%%%%%%%%%%%%%%%%%%%%%%%%%%%%%%%%%%%%%%
%%%%%%%%%%%%%%%%%%%%%%%%%%%%%%%%%%%%%%%%%%%%%%%%%%%%%%%%%%%%
%%%%%%%%%%%%%%%%%%%%%%%%%%%%%%%%%%%%%%%%%%%%%%%%%%%%%%%%%%%%
%%%%%%%%%%%%%%%%%%%%%%%%%%%%%%%%%%%%%%%%%%%%%%%%%%%%%%%%%%%%
%%%%%%%%%%%%%%%%%%%%%%%%%%%%%%%%%%%%%%%%%%%%%%%%%%%%%%%%%%%%
%%%%%%%%%%%%%%%%%%%%%%%%%%%%%%%%%%%%%%%%%%%%%%%%%%%%%%%%%%%%

 %%%%%%%%%%%%%%%%%%%%%%%%%%%%%%%%%%%%
%%%%%%%%%%%%%%%%%%%%%%%%%%%%%%%%%%%%
\begin{figure}[htb]
  \begin{center}
\includegraphics[width=0.5\textwidth]{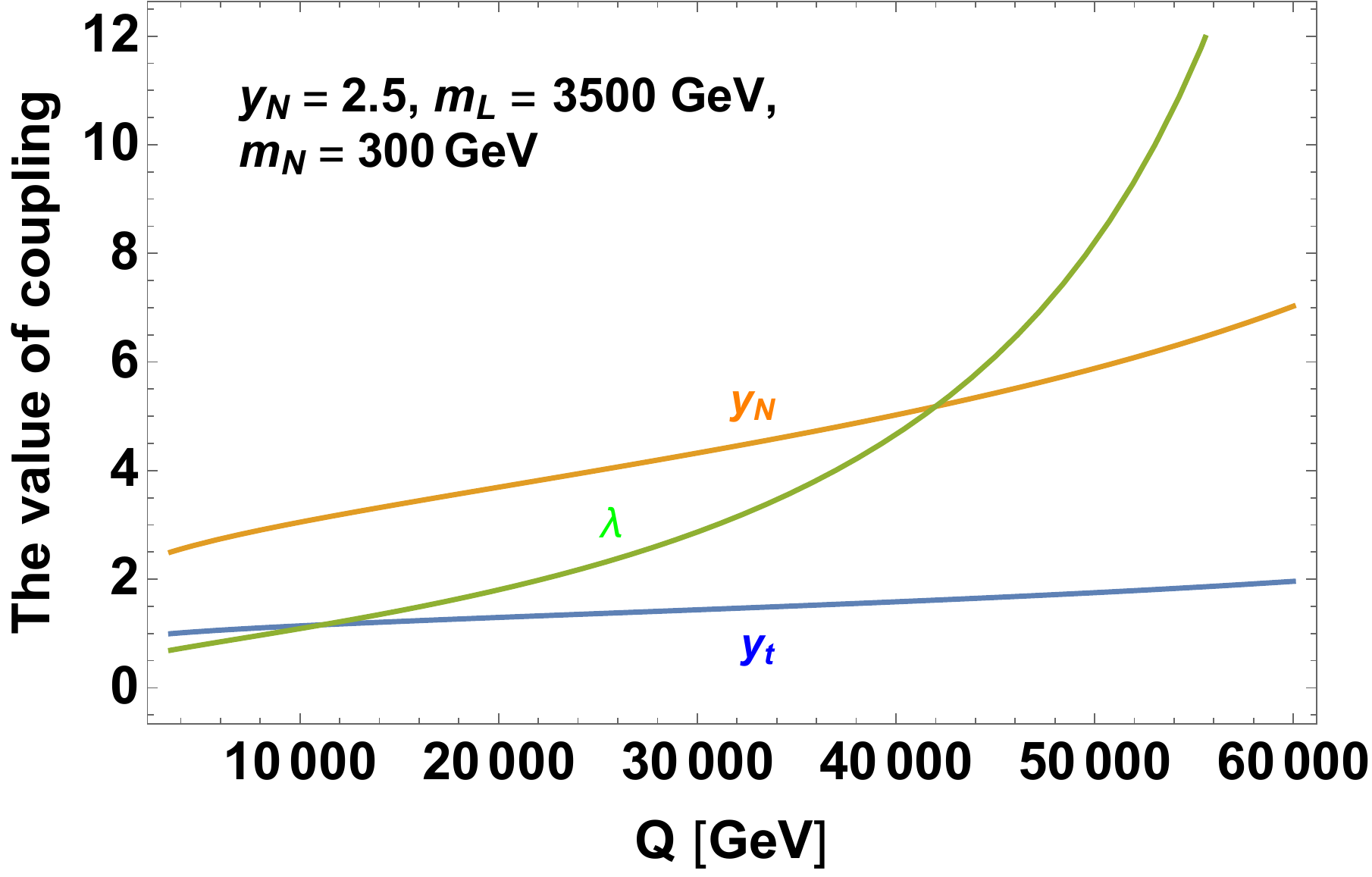}
\includegraphics[width=0.48\textwidth]{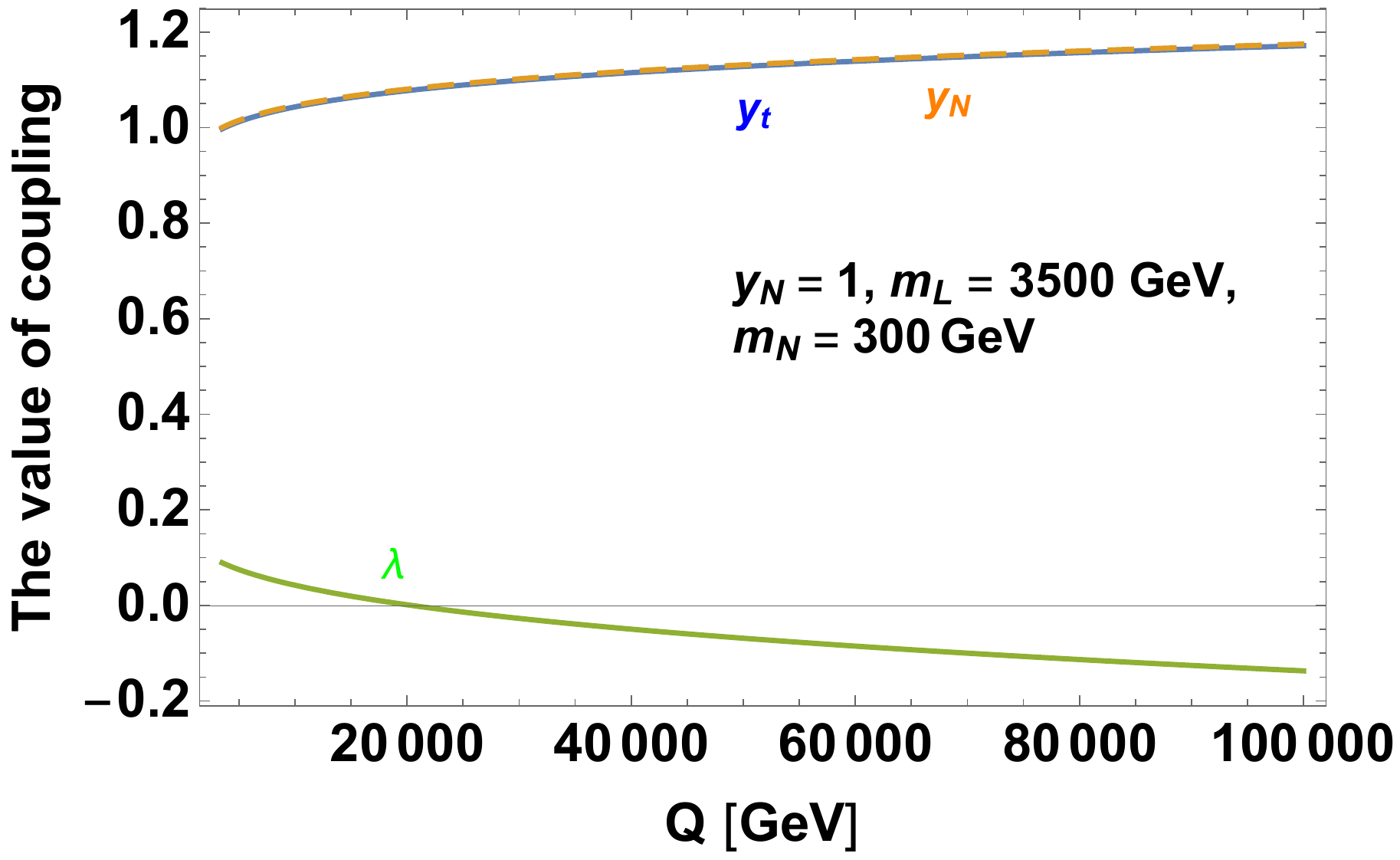}
\caption{The results of RG running fllows of the $y_t$, $y_N$ and $\lambda$ couplings in the extended fermion model with $m_N=300$ GeV, $m_L=3500$ GeV and $y_N=2.5$ (left figure) and $y_N=1$ (right figure) from matching scale $Q_M=m_L$ to 60 TeV (left figure) and to 100 TeV (right figure).
  Green line: $\lambda$ coupling, yellow line: $y_N$ coupling, blue line: $y_t$ coupling. 
  In right figure, the yellow and blue lines are the almost same and $\lambda$ coupling becomes negative. }
\label{stable}
  \end{center}
\end{figure}
%%%%%%%%%%%%%%%%%%%%%%%%%%%%%%%%%%%%
%%%%%%%%%%%%%%%%%%%%%%%%%%%%%%%%%%%%

\begin{figure}[htb]
  \begin{center}
\includegraphics[width=0.4\textwidth]{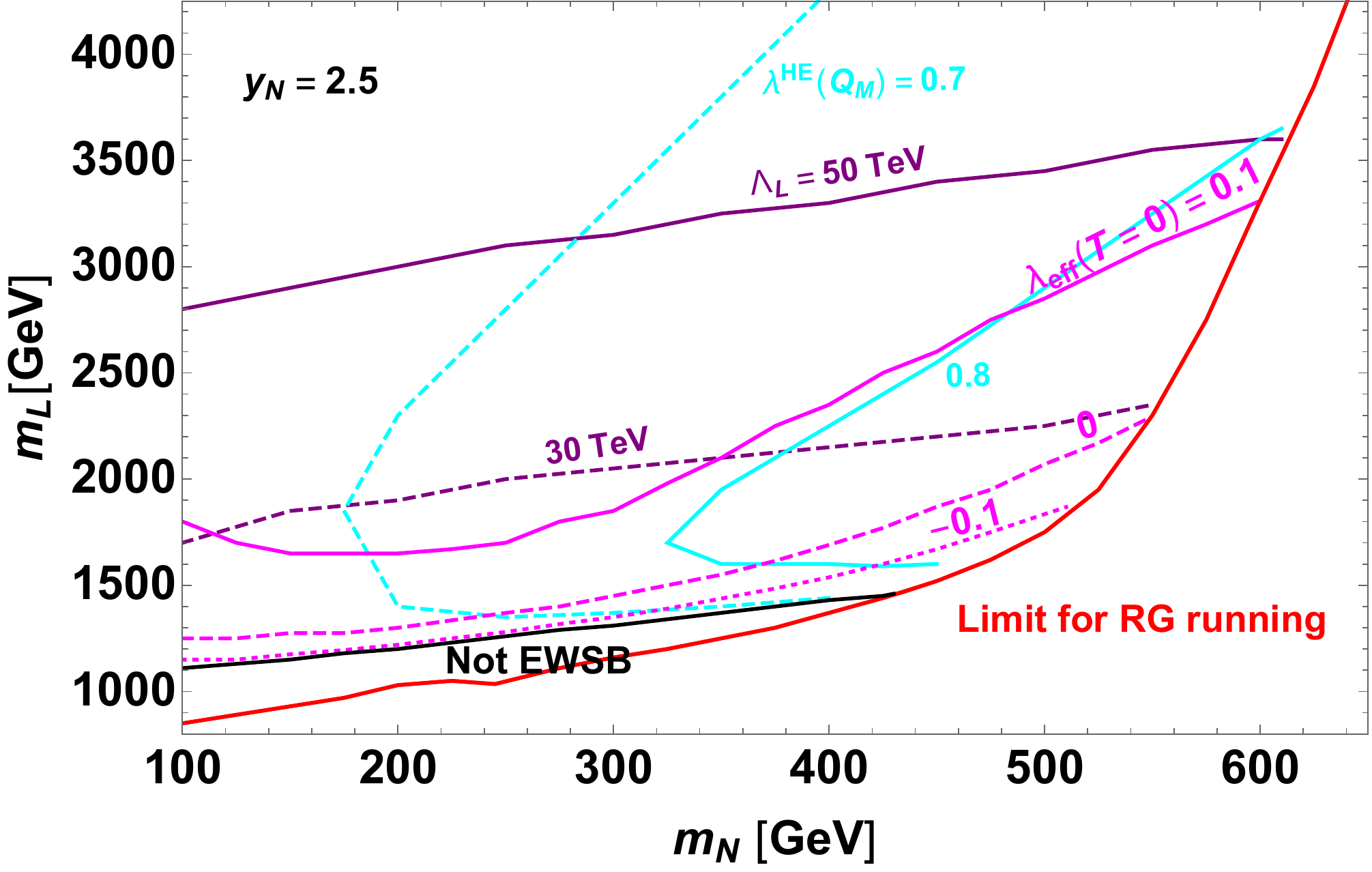}\\
\includegraphics[width=0.4\textwidth]{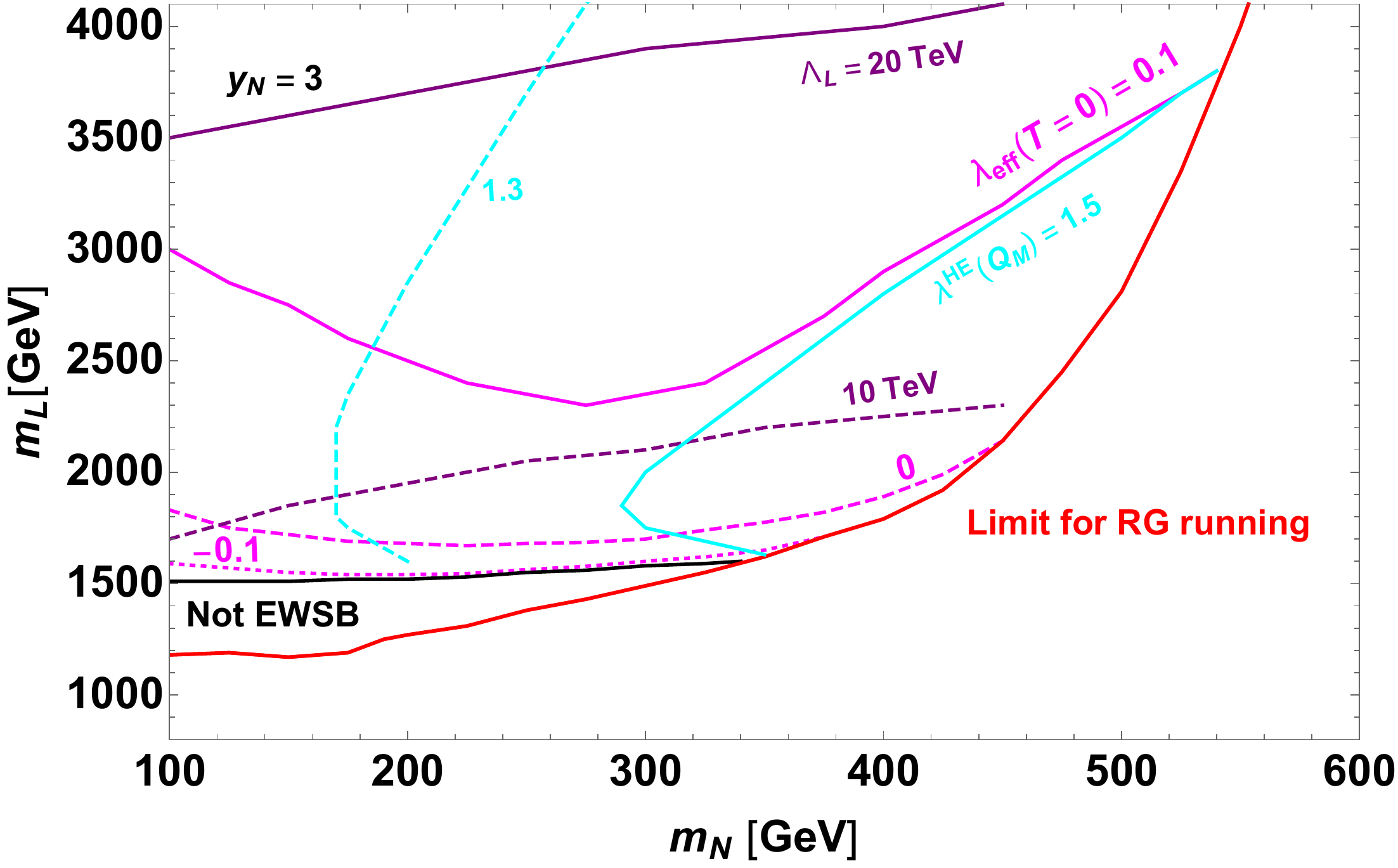}
\includegraphics[width=0.4\textwidth]{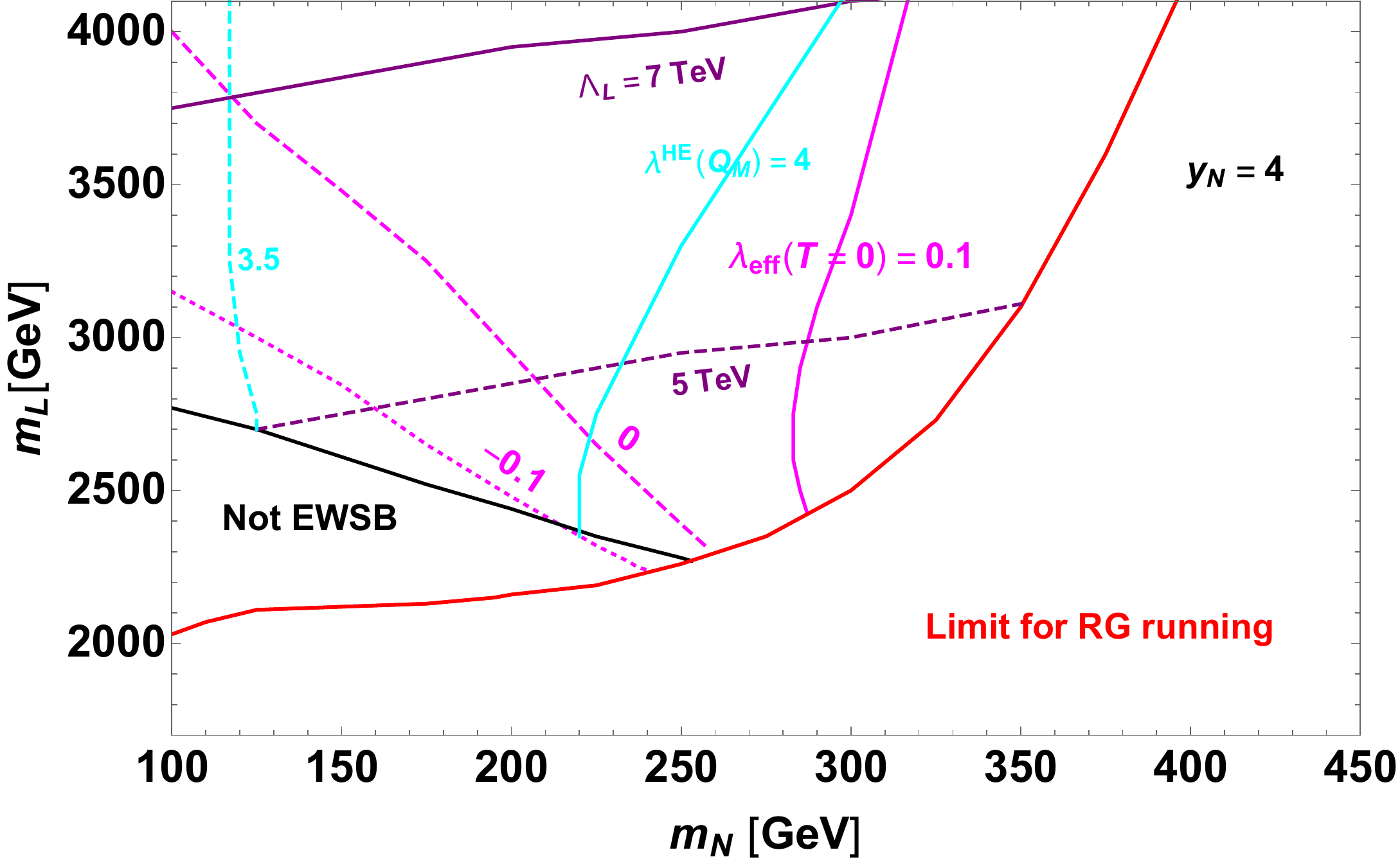}
\caption{ Cyan contours represent the value of $\lambda^{HE}$ in Eq.~(\ref{thinHElam}) at matching scale $Q_M=m_L$ and purple contours represent value of landau pole $\Lambda_L$, where one of coupling is $4\pi$ at this energy scale, in the extended fermion model with $y_N=2.5$ (upper), 3 (lower left) and 4 (lower right).
  Magenta contours, black and red lines are the same as Fig.~\ref{gamlamyN}.}
\label{landau}
  \end{center}
\end{figure}

In this section, we discuss the vacuum stability and the Landau pole $\Lambda_{L}$ in the parameter (C): $m_L \gg m_N \sim y_N v$.
 Although $\lambda_{eff}^{T=0}$ becomes negative in the parameter space in Fig.~\ref{12res}, the effective potential at the EW scale is stable by positive high dimensional~operators. 
 But, at the HE scale, the potential could become unstable, so we discuss the vacuum stability at the HE scale. 
The condition of unstable potential at the HE scale is 
%%%%%%%%%%%%%%
	\begin{align} % requires amsmath; align* for no eq. number
\lambda^{HE}(Q) < 0,
	\end{align}
%%%%%%%%%%%%%%
 where $\lambda^{HE}(Q)$ is a coefficient of $\varphi^4$ in the Higgs potential at HE scale.  
We focus on the dominant contributions coming from $y_t$, $y_N$ and $\lambda$ in order to get the energy scale $\Lambda_{stable}$: $\lambda(\Lambda_{stable}) =0$. 
The beta functions of these couplings at the HE scale are 
%%%%%%%%%%%%%%%%%%%%%%%%%%%%%%%%%%%%%%%%%%%%%%%%%%%%%%%%%%%%
	\begin{align} % requires amsmath; align* for no eq. number
 \beta^{HE}_{y_t}&=\frac{y_t}{16\pi^2}\left(\frac{9y_t^2}{2} +2y_N^2 \right),\quad
 \beta^{HE}_{y_N}=\frac{y_N}{16\pi^2}\left(\frac{7y_N^2}{2} +3y_t^2 \right)\nonumber\\
 \beta^{HE}_{\lambda}&=\frac{1}{16\pi^2}\left( -6y_t^4-2y_N^4 + 24 (\lambda^{HE})^2  -4 \left(-3 y_t^2 -2 y_N^2 \right)  \lambda^{HE} \right).
 	\end{align}
%%%%%%%%%%%%%%%%%%%%%%%%%%%%%%%%%%%%%%%%%%
We note that the quartic coupling in the HE scale $\lambda^{HE}$ at the matching scale is given by
%%%%%%%%%%%%%%%%%%%%%%%%%%%%%%%%%%%%%%%%%%%%%%%%%%%%%%%%%%%%
	\begin{align} % requires amsmath; align* for no eq. number
	\label{thinHElam}
\lambda^{HE} &= \lambda^{LE}  + \frac{y_N^4m_L^2(m_L+m_N)}{8\pi^2(m_L-m_N)^3}
 	\end{align}
%%%%%%%%%%%%%%%%%%%%%%%%%%%%%%%%%%%%%%%%%%
from Eq.~(\ref{thresh}).
The second term in the right-hand side of this equation represents the threshold effect in Eq.~(\ref{thresh}) which is positive.
If the quartic coupling at the matching scale is small positive value, such as the value of the SM case $\lambda\sim0.1$, the quartic coupling $\lambda$ becomes negative above $\Lambda_{stable}$, in other words, the shape of the potential becomes unstable.  
 Fig.~\ref{stable} shows an example of the RG running flow of the $y_t$, $y_N$ and $\lambda$ couplings.
 Left panel represents the flows in the extended fermion model with $m_N=300$ GeV, $m_L=3500$ GeV and $y_N=2.5$, while right panel has different $y_N$ value from left panel: $m_N=300$ GeV, $m_L=3500$ GeV and $y_N=1$.
  The right panel has $\Lambda_{stable}$, because the value of $\lambda^{HE}$ in the panel at the matching scale is similar to one of the SM case: $\lambda\sim0.1$.
  In this case, the negative fermion effects in the RG running are dominant and $\lambda^{HE}$ becomes negative above $\Lambda_{stable}$. 
 The behaviour of $y_N=1$ is the similar to Ref.~\cite{Gopalakrishna:2018uxn}.
 On the other hand, the quartic coupling in the left panel at the matching scale is larger than the value of SM case by additional positive contribution coming from second term of Eq.~(\ref{thinHElam}), and then the potential is stable not only at LE scale but also at HE scale. 
 The value of $\lambda^{HE}$ at the matching scale $Q_M$ represents the cyan lines in Fig.~\ref{landau}. 
 From this figure, the $\lambda^{HE}(Q_M)$ coupling in the parameter region of our analysis is larger than the SM value and the potential is always stable. 
%%%%%%%%%%%%%%%%%%%%%%%%%%%%%%%%%%%%
%%%%%%%%%%%%%%%%%%%%%%%%%%%%%%%%%%%%
%%%%%%%%%%%%%%%%%%%%%%%%%%%%%%%%%%%%
%%%%%%%%%%%%%%%%%%%%%%%%%%%%%%%%%%%%

Furthermore, we analyze the Landau pole $\Lambda_L$ at which a coupling is $4\pi$. 
The landau pole represents purple lines in Fig.~\ref{landau}.  
When the value of $m_L$ is small, the Landau pole $\Lambda_L$ becomes small.
  The value of $\Lambda_L$ is always larger than the matching scale $Q_M=m_L$.

\bibliographystyle{JHEP}

\end{document}